\newcommand\hst{{\it HST}}
\documentclass{mn2e}
\usepackage{epsfig}
\usepackage{epsf}
\usepackage{graphicx}

\title[HH Jets in Carina]{HST/ACS H$\alpha$ Imaging of the Carina
  Nebula: Outflow Activity Traced by Irradiated Herbig-Haro
  Jets\thanks{Based on observations made with the NASA/ESA {\it Hubble
      Space Telescope}, obtained at the Space Telescope Science
    Institute, which is operated by the Association of Universities
    for Research in Astronomy, Inc., under NASA contract NAS5-26555.}}

\author[N.\ Smith et al.]{Nathan Smith$^1$\thanks{Email:
    nathans@astro.berkeley.edu}, John Bally$^2$, \& Nolan R.\
  Walborn$^3$ \\ $^1$Department of Astronomy, University of
  California, 601 Campbell Hall, Berkeley, CA 94720, USA \\ $^2$Center
  for Astrophysics and Space Astronomy, University of Colorado, 389
  UCB, Boulder, CO 80309, USA \\ $^3$Space Telescope Science
  Institute, 3700 San Martin Dr., Baltimore, MD 21218, USA}

\begin{document}
\date{Accepted 0000, Received 0000, in original form 0000}
\pagerange{\pageref{firstpage}--\pageref{lastpage}} \pubyear{2002}
\def\arcdeg{\degr}
\maketitle
\label{firstpage}

\begin{abstract}

  We report the discovery of new Herbig-Haro (HH) jets in the Carina
  Nebula, and we discuss the protostellar outflow activity of a young
  OB association.  These are the first results of an H$\alpha$ imaging
  survey of Carina conducted with the {\it Hubble Space
    Telescope}/Advanced Camera for Surveys.  Adding to the one
  previously known example (HH~666), we detect 21 new HH jets, plus 17
  new candidate jets, ranging in length from 0.005 to 3 pc.  Using the
  H$\alpha$ emission measure to estimate jet densities, we derive jet
  mass-loss rates ranging from 8$\times$10$^{-9}$ to
  $\sim$10$^{-6}$ $M_{\odot}$ yr$^{-1}$, but a comparison to the
  distribution of jet mass-loss rates in Orion suggests that we may be
  missing a large fraction of the jets below 10$^{-8}$ $M_{\odot}$
  yr$^{-1}$.  A key qualitative result is that even some of the
  smallest dark globules with sizes of $\la$1\arcsec\ (0.01 pc) are
  active sites of ongoing star formation because we see HH jets
  emerging from them, and that these offer potential analogs to the
  cradle of our Solar System because of their proximity to dozens of
  imminent supernovae that will enrich them with radioactive nuclides
  like $^{60}$Fe.  Whereas most proplyd candidates identified from
  ground-based data are dark cometary globules, {\it HST} images now
  reveal proplyd structures in the core of the Tr~14 cluster, only
  0.1--0.2 pc from several extreme O-type stars.  Throughout Carina,
  some HH jets have axes bent away from nearby massive stars, while
  others show no bend, and still others are bent {\it toward} the
  massive stars.  These jet morphologies serve as ``wind socks'';
  strong photoevaporative flows can shape the jets, competing with the
  direct winds and radiation from massive stars.  We find no clear
  tendency for jets to be aligned perpendicular to the axes of dust
  pillars.  Finally, even allowing for a large number of jets that may
  escape detection, we find that HH jets are negligible to the global
  turbulence of the surrounding region, which is driven by massive
  star feedback.

\end{abstract}


\begin{figure*}\begin{center}
\includegraphics[width=5.9in]{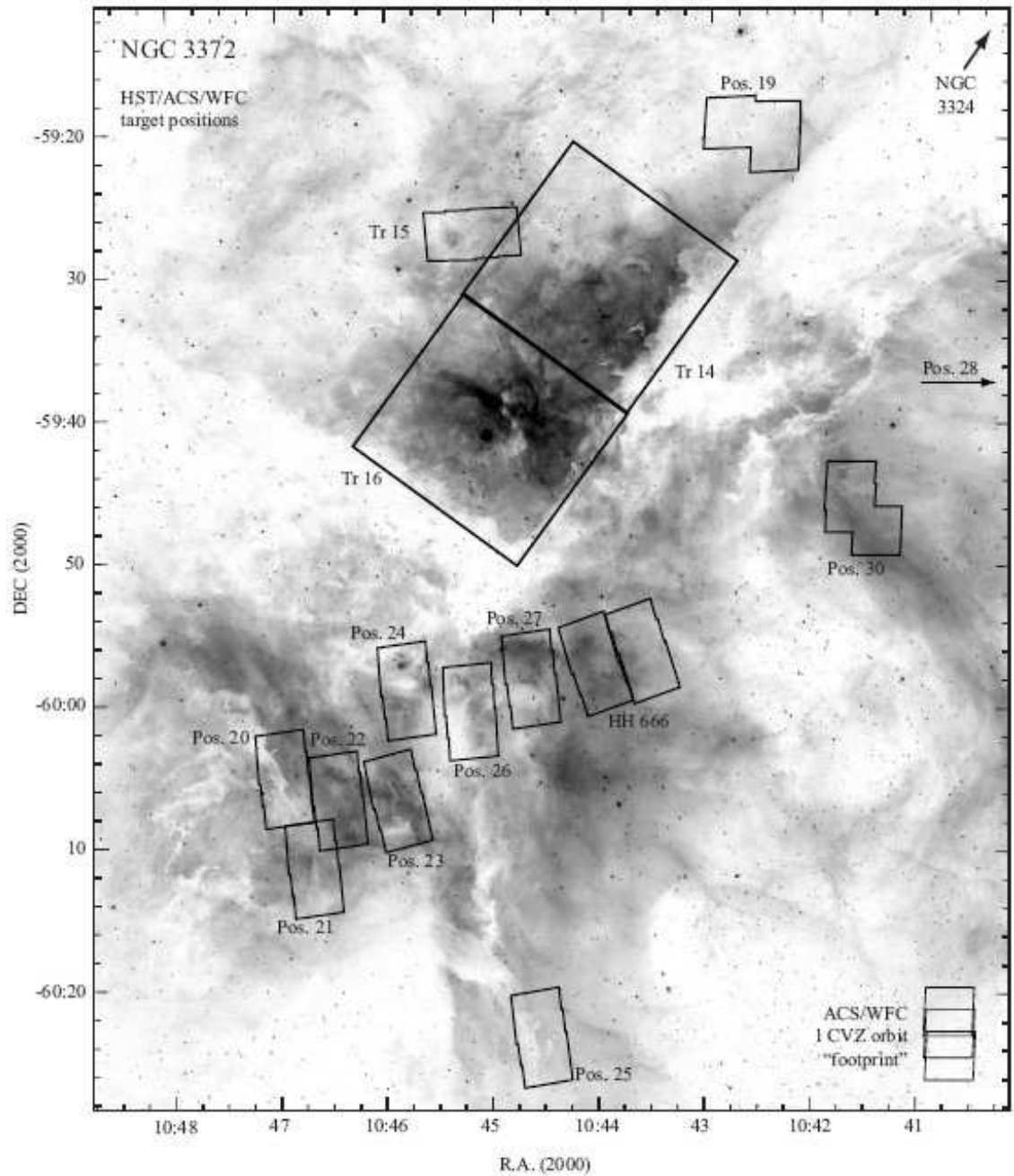}
\end{center}
\caption{A large-scale H$\alpha$ image of the Carina Nebula obtained
  with the Curtis-Schmidt telescope at CTIO (from Smith 2002).
  Positions of \hst\ observations are indicated.  Most target
  positions used a ``footprint'' composed of three individual dithered
  exposures in a single CVZ orbit, shown at the lower right corner.
  In a few cases, this pattern was modified to suit the particular
  target (Positions 19 and 30).  In the central region, two large
  mosaics of the regions around Tr14 and Tr16 were composed of eight
  of these footprints each to yield a large contiguous mosaic of the
  inner nebula.  Two pointings are not shown here: one in the
  neighboring H~{\sc ii} region NGC~3324, and one in the far-western
  part of Carina (Pos.\ 28).}\label{fig:map}
\end{figure*}

\begin{table*}\begin{minipage}{5.4in}
\caption{ACS Observation Log}\scriptsize
\begin{tabular}{@{}lcclrcl}\hline\hline
Target &$\alpha_{2000}$ &$\delta_{2000}$ &Date &P.A.($\arcdeg$) &Orbits &Comment \\ \hline
N3324   &10:36:55  &--58:38:05  &2006 Mar 7   &9.7    &1 &edge-on I.F.    \\
Tr 14   &10:44:00  &--59:30:50  &2005 Jul 17  &144.0  &8 &edge-on I.F., dust pillars, globules \\
Tr 15   &10:45:10  &--59:27:00  &2006 Nov 26  &95.0   &1 &globules near Tr15    \\
Tr 16   &10:44:55  &--59:41:10  &2005 Jul 18  &144.0  &8 &Keyhole, $\eta$ Car, globules, etc.  \\
HH 666  &10:43:50  &--59:56:40  &2005 Mar 30  &20.0   &2 &south pillars, Axis of Evil  \\
Pos 19  &10:42:30  &--59:19:50  &2006 Mar 8   &8.9    &1 &globules        \\
Pos 20  &10:47:00  &--60:05:00  &2006 Mar 2   &8.0    &1 &dark globules, Bo 11   \\
Pos 21  &10:46:40  &--60:12:00  &2006 Mar 4   &8.0    &1 &south pillars   \\
Pos 22  &10:46:25  &--60:06:40  &2006 Mar 4   &8.0    &1 &globules        \\
Pos 23  &10:45:53  &--60:06:40  &2005 Mar 28  &15.0   &1 &south pillars   \\
Pos 24  &10:45:49  &--59:59:00  &2006 Mar 2   &8.0    &1 &south pillars, Treasure Chest cluster  \\
Pos 25  &10:44:31  &--60:23:10  &2005 Sep 10  &10.0   &1 &globules        \\
Pos 26  &10:45:13  &--60:00:20  &2006 Mar 2   &5.8    &1 &south pillars   \\
Pos 27  &10:44:40  &--59:58:00  &2006 Mar 7   &8.3    &1 &south pillars   \\
Pos 28  &10:38:52  &--59:35:50  &2006 Apr 20  &75.0   &1 &dust pillars, west   \\
Pos 30  &10:41:33  &--59:44:00  &2005 Sep 11  &10.7   &1 &dust pillars, west edge  \\
\hline
\end{tabular}
\end{minipage}
\end{table*}

\section{INTRODUCTION}

The Carina Nebula (NGC~3372) provides us with a rich and unique
laboratory in which to perform detailed study of massive-star feedback
on a surrounding second generation of newly formed stars.  It offers
our best available trade-off between an extreme population of the most
massive O-type stars known, weighed against its nearby distance and
relatively low interstellar extinction.  The central star clusters
Tr~14 and Tr~16 (e.g., Massey \& Johnson 1993) are famous for their
collection of extremely massive stars, including $\eta$ Car (Davidson
\& Humphreys 1997), three WNH stars (see Smith \& Conti 2008, and
references therein), several of the earliest O-type stars known
including the O2 supergiant HD~93129A (Walborn et al.\ 2002), and
dozens of additional O-type stars.  See Walborn (2009) and Smith
(2006a) for recent reviews of Carina's remarkable massive-star content
and its collective energy input, and see Smith \& Brooks (2007) for a
recent census of the global nebulosity and integrated mass/energy
budgets of the region.

Surrounding the bright central clusters, one finds evidence for active
ongoing star formation, particularly in the collection of dust pillars
to the south (Smith et al.\ 2000; Smith \& Brooks 2007; Megeath et
al.\ 1996; Rathborne et al.\ 2004).  Smith \& Brooks (2007) pointed
out, however, that the current star formation activity appears less
vigorous than the first generation of star formation that yielded
Tr~14 and Tr~16.  They conjectured that this is because most of the
present-day nebular mass budget consists of warm atomic gas in
photodissociation regions (PDRs) and not in cold molecular cloud
cores, so that most of the gas mass reservoir is not currently able to
participate in new star formation. Star formation appears to be
``percolating'' in many smaller clouds in Carina, inhibited by the
strong UV radiation from O stars in the region.  These $\sim$70 O
stars are destined to explode as supernovae (SNe) soon, which will rob
the region of its UV radiation and will sweep these clouds into dense
clumps.  In this way, Smith \& Brooks (2007) speculated that a new
wave of star formation may occur in the near future.

The imminent explosion of several dozen SNe has implications for the
new generation of stars currently forming in cometary clouds and dust
pillars only 5--30 pc away.  Namely, the clouds from which these young
stars form will soon be pelted by ejecta from numerous SNe over a
relatively short time period.  SN ejecta contain radioactive nuclides
such as $^{60}$Fe, $^{26}$Al, and other materials found in
Solar-System meteorites, which require that a nearby SN injected
processed material into the presolar cloud or disk (Lee et al.\ 1976;
Cameron \& Truran 1977; Tachibana \& Huss 2003; Hester et al.\ 2004;
Desch \& Ouellete 2005; Boss et al.\ 2008; Vanhala \& Boss 2000,
2002).  Thus, Smith \& Brooks (2007) argued that regions like the
South Pillars may provide a close analog to the cradle of our own
Solar System.

Whether or not SN explosions have already occurred in Carina is still
a topic of current research.  SNe are not required for the global
energy budget (Smith \& Brooks 2007), but high velocity gas seen in
absorption (Walborn et al.\ 2007; and references therein) and the
recent discovery of a neutron star in the field (Hamaguchi et al.\
2009) imply that a recent SN occurred here.  However, this SN may be
projected along our line-of-sight to Carina rather than {\it inside}
it (Walborn et al.\ 2007).  In any case, Carina provides a laboratory
in which to study a new generation of young stars exposed to feedback
from an extreme population of massive stars.  Here, we focus on the
outflows that are a signpost of embedded protostars caught in an
active accretion phase.

\begin{table*}\begin{minipage}{5.4in}
\caption{HH Jets in Carina and NGC~3324}\scriptsize
\begin{tabular}{@{}lcccll}\hline\hline
HH &$\alpha_{2000}$ &$\delta_{2000}$ &P.A.($\arcdeg$) &Pointing &Comment \\ \hline
666   &10:43:51.3  &--59:55:21  &293.5 &HH 666 &Previously discovered \\
900   &10:45:19.3  &--59:44:23  &242  &Tr 16    &small glob., points to $\eta$ Car \\
901   &10:44:03.5  &--59:31:02  &279  &Tr 14    &pillar head, points to Tr14 \\
902   &10:44:01.7  &--59:30:32  &258  &Tr 14    &pillar head, points to Tr14 \\
903   &10:45:56.6  &--60:06:08  &278  &Pos 23   &side of large pillar \\
1002  &10:36:57.1  &--58:37:26  &106  &NGC 3324 &edge-on I.F. \\
1003A &10:36:53.6  &--58:38:09  &171  &NGC 3324 &jet body, near I.F. \\
1003C &10:36:54.8  &--58:39:09  &171  &NGC 3324 &bow shock \\
1004  &10:46:44.8  &--60:10:20  &247  &Pos 21   &pillar head, bipolar \\
1005  &10:46:44.2  &--60:10:35  &112  &Pos 21   &same pillar as HH~1004 \\
1006  &10:46:33.0  &--60:03:54  &352  &Pos 22   &proplyd cand., bipolar \\
1007  &10:44:29.5  &--60:23:05  &270  &Pos 25   &irregular \\
1008  &10:44:47.0  &--59:57:25  &170  &Pos 27   &large bow shock, from pillar head? \\
1009  &10:44:39.5  &--59:58:26  &?    &Pos 27   &irregular \\
1010  &10:41:48.7  &--59:43:38  &215  &Pos 30   &dark pillar head, bipolar \\
1011  &10:45:04.9  &--59:26:59  &260  &Tr 15    &small glob., one-sided jet \\
1012  &10:44:38.6  &--59:30:07  &140  &Tr 14    &LL Ori object, microjet \\
1013  &10:44:19.2  &--59:26:14  &35   &Tr 14    &proplyd, bow shock, microjet \\
1014  &10:45:45.9  &--59:41:06  &290  &Tr 16    &pillar head, toward $\eta$ Car \\
1015  &10:44:27.9  &--60:22:57  &315  &Pos 25   &pillar head \\
1016  &10:45:53.2  &--59:56:05  &320  &Pos 24   &pillar head, Treasure Chest \\
1017  &10:44:41.5  &--59:33:57  &25   &Tr14     &jet bends away from Tr~14 \\
1018  &10:44:52.9  &--59:45:26  &332  &Tr16     &microjet and shock \\
\hline
\end{tabular}
\end{minipage}
\end{table*}

\begin{table*}\begin{minipage}{5.6in}
\caption{Candidate Jets in Carina and NGC~3324}\scriptsize
\begin{tabular}{@{}lcccll}\hline\hline
HH &$\alpha_{2000}$ &$\delta_{2000}$ &P.A.($\arcdeg$) &Pointing &Comment \\ \hline
c-1 (3324) &10:36:55.9 &--58:36:38  &90   &NGC~3324 &edge-on I.F. \\
c-2 (3324) &10:37:01.1 &--58:38:37  &?    &NGC~3324 &shocks, unclear source \\
c-1   &10:44:05.4  &--59:29:40  &300  &Tr~14    &proplyd cand. \\
c-2   &10:36:55.9  &--58:36:38  &5    &Pos 27   &pillar head, unclear morph. \\
c-3   &10:45:04.6  &--60:03:02  &?    &Pos 26   &small glob., unclear morph. \\
c-4   &10:45:08.3  &--60:02:31  &?    &Pos 26   &bow shock? \\
c-5   &10:45:09.3  &--60:01:59  &298  &Pos 26 &pillar head, coll.\ chain of knots \\
c-6   &10:45:09.3  &--60:02:26  &2    &Pos 26 &pillar side, bipolar, par.\ to I.F. \\
c-7   &10:45:13.4  &--60:02:55  &165? &Pos 26   &bow shock? unclear source \\
c-8   &10:45:12.2  &--60:03:09  &?    &Pos 26   &LL Ori object?  \\
c-9   &10:45:08.0  &--59:31:03  &275? &Tr~14    &large bow shock, unclear source \\
c-10  &10:45:57.2  &--60:05:32  &230  &Pos 23   &pillar head, bow shocks? \\
c-11  &10:46:36.0  &--60:07:12  &225  &Pos 22   &bow shock, unclear source \\
c-12  &10:45:07.8  &--60:03:23  &?    &Pos 26   &shock, unclear source \\
c-13  &10:44:00.6  &--59:54:27  &169  &HH~666   &microjet from star near HH~666 \\
c-14  &10:45:43.7  &--59:41:39  &?    &Tr~16    &curved shocks near HH~1014 \\
c-15  &10:43:54.6  &--59:32.46  &80   &Tr~14    &core of Tr~14, microjet \\
\hline
\end{tabular}
\end{minipage}
\end{table*}

\begin{figure*}\begin{center}
\includegraphics[width=6.5in]{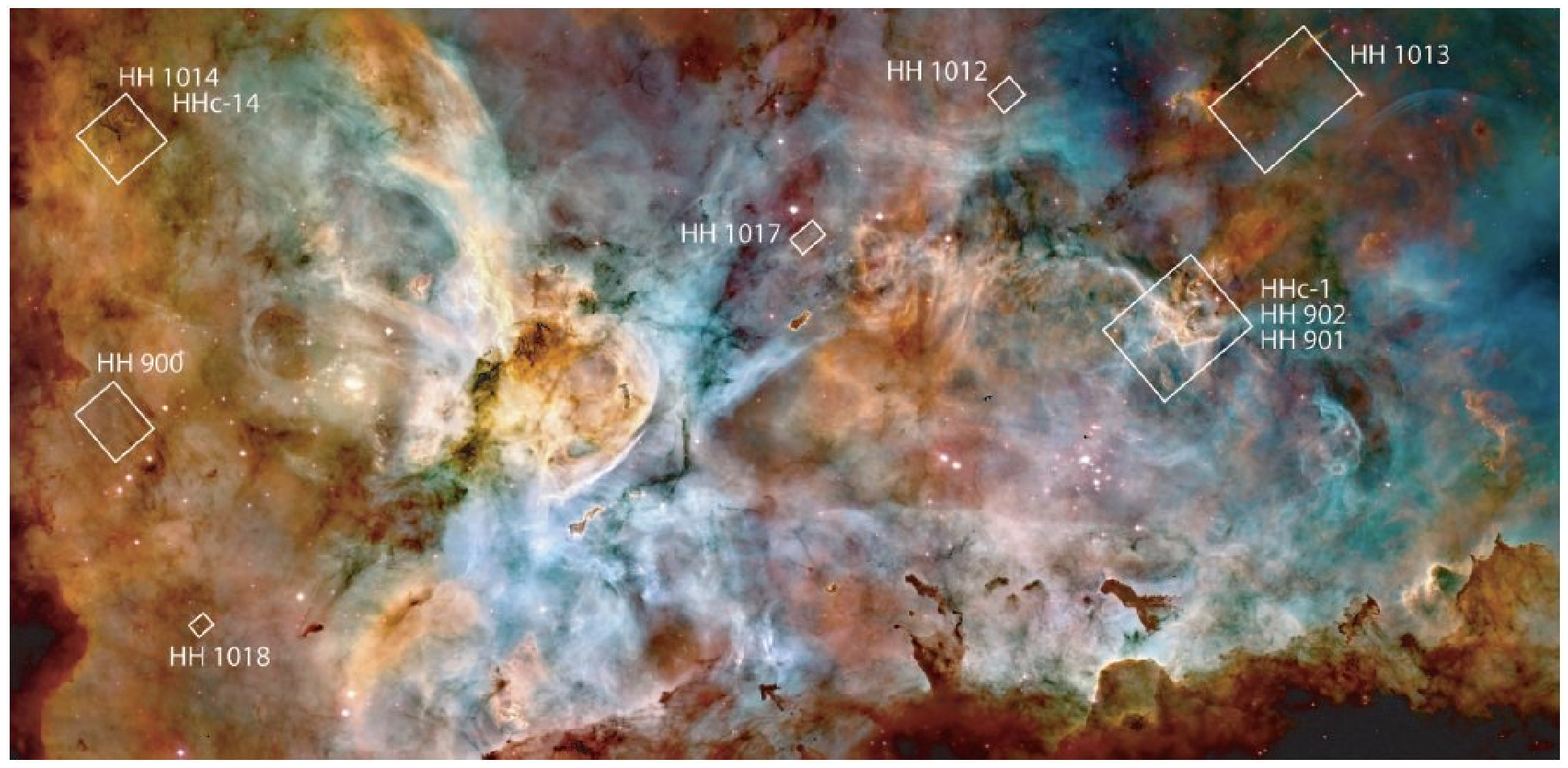}
\end{center}
\caption{The color Hubble Heritage image of the inner Carina Nebula
  including the Tr~14 and Tr~16 clusters, the Keyhole, and
  $\eta$~Carinae, plus several newly discovered HH~jets.  This is made
  from a large ACS H$\alpha$ mosaic used for the intensity scale in
  the image, while the color coding is taken from ground-based
  narrow-band images obtained with the MOSIAC camera on the 4m Blanco
  telescope at CTIO (from Smith et al.\ 2003), with [O~{\sc iii}] in
  blue, H$\alpha$ in green, and [S~{\sc ii}] in red.  This
  ground-based image was also used to patch small gaps in the ACS
  mosaic.  See the Hubble Heritage webpage for more information about
  the image processing for this large mosaic, where color images for
  individual jets and other structures can be found as well
  (http://heritage.stsci.edu).  The full field shown here is roughly
  12$\arcmin\times$25$\arcmin$.  The rectangular boxes show the
  detailed fields of view for various HH jets included in the large
  mosaic.  North is to the upper right; the small boxes around
  individual HH jets are aligned with R.A. and DEC directions.}
\label{fig:color}
\end{figure*}


Herbig-Haro (HH) objects are nebulosities associated with protostellar
outflows that are identified at visual wavelengths, usually by their
shock-excited emission in lines like H$\alpha$ and [S~{\sc ii}] (see
Reipurth \& Bally 2001 for a review).  Since they arise in
protostellar outflows, HH~jets are signposts of actively accreting
young stars embedded within dark clouds.  When their physical
properties can be measured accurately, they can be used to estimate
the jet mass, momentum, kinetic energy, and mass-loss rate, and hence,
the injection of energy and turbulence into the surrounding
interstellar medium (ISM).  They also provide a way to constrain the
recent jet mass-loss history and to infer the past accretion rate of
the underlying star.  Large-format array detectors applied to nearby
outflows have revealed HH objects from single flows more than a parsec
in length (e.g., Reipurth et al.\ 1997b; Devine et al.\ 1997;
Eisl\"{o}ffel \& Mundt 1997), providing a record of the accretion
history of the driving star over the previous $\sim$10$^4$ yr --- a
significant fraction of the most active accretion phase.

In most cases of nearby jets in low-mass star-forming regions, the
nebular emission arises from collisional excitation in shocks, so
observations can only trace the jet material that was very recently
heated by passage through a shock.  Much of the remaining jet material
may therefore be invisible, making estimates of the physical
parameters difficult.  In the presence of a strong external UV
radiation field provided by nearby OB stars, however, even unshocked
jet material is rendered visible when it is photoionized.  Such
irradiated HH jets (Reipurth et al.\ 1998; Bally \& Reipurth 2001)
emit visual-wavelength spectra characteristic of photoionized gas,
making it easier to constrain the physical properties of the jet using
standard photoionization theory.  Even in the absence of spectra, the
H$\alpha$ surface brightness in images can be used to infer the
emission measure $EM \propto \int n_e^2 \ dl$, and hence, the electron
density of the jet $n_e$, providing that the emitting path length $dl$
can be estimated from the spatially resolved jet geometry in
high-resolution images such as those attainable with the {\it Hubble
  Space Telescope} ({\it HST}).


With its active ongoing star formation and strong UV radiation field,
Carina provides fertile hunting ground in which to identify a large
population of irradiated HH jets.  So far, the only HH jet to have
been discovered in Carina is HH~666 (Smith et al.\ 2004c), which is a
parsec-scale bipolar jet emerging from one of the South Pillars.  The
embedded infrared (IR) source that drives the outflow was also
identified (Smith et al.\ 2004c).  Its discovery in ground-based
images was enabled by the fact that it is such a large and particuarly
bright jet.  One might expect that many other HH jets are lurking in
such a rich region, but that they are difficult to see at ground-based
angular resolution because their thin filaments and shocks are drowned
in the bright background nebular light.

In this paper we show that HH~666 was the tip of the iceberg, with
many more HH jets and candidates seen in our {\it HST} images.
Throughout, we adopt the precise distance to Carina of
2300($\pm$50)~pc derived from the expansion kinematics of the
Homunculus Nebula around $\eta$ Carinae by Smith (2006b), which is
known to reside within the center of the Carina Nebula rather than
being projected along the same sightline (Allen 1979; Lopez \& Meaburn
1986; Walborn \& Liller 1977).  We describe the {\it HST} observations
in \S 2.  In \S 3 we provide an overview of the new HH jets and HH jet
candidates; \S 3 is detailed and largely descriptive, so readers may
wish to only skim this section and refer back to it when interested in
a specific object.  \S 4 takes a closer look at three remarkable
bipolar jets, while \S 5 investigates the distribution of mass-loss
rates.  We discuss main points in \S 6, and we provide a short
conclusion in \S 7.

\section{OBSERVATIONS: THE ACS H$\alpha$ SURVEY OF 
THE CARINA NEBULA}

We present images covering a large portion of the Carina Nebula,
comparable to our previous H$\alpha$ mapping of the Orion Nebula with
{\it HST} (Bally et al.\ 2005, 2006; Smith et al.\ 2005b).  With a
similar setup, we used the Wide-Field Channel (WFC) of \hst's Advanced
Camera for Surveys (ACS) under joint programs GO-10241 and GO-10475
(PI: Smith) conducted in Cycles 13 and 14.

Because of the large area of the Carina Nebula (several square
degrees), it was impractical to make a single contiguous map of the
entire region with \hst \ as we had done in Orion.  Instead, we
conducted the survey in two phases.  First, during Cycle 13 we made a
large (13$\farcm$6$\times$27$\arcmin$) contiguous mosaic image of the
brightest inner parts of Carina surrounding the clusters Tr14 and
Tr16, including the Keyhole Nebula and $\eta$ Carinae.  Second, we
selected several individual pointings at various positions in the
nebula.  These targets were chosen because they were deemed to have
more potential to yield interesting results related to ongoing star
formation (protoplanetary disks, globules, outflows, embedded sources,
young star clusters) based on their structure in our previous
ground-based narrow-band survey of the entire Carina region with the
Cerro Tololo Inter-American Observatory (CTIO) 4m telescope (e.g.,
Smith et al.\ 2003).

The ACS/WFC aperture positions we targeted are shown in
Figure~\ref{fig:map}, and the individual target positions and
observation dates are listed in Table~1.  The target positions in
Table~1 are the approximate centers of each group of offset pointings
defining one ``visit''.  We were able to maximize the efficiency of
these observations (area covered per orbit) because Carina is far
enough south that all orbits were conducted in continuous viewing zone
(CVZ) mode.  Given the restrictions of the ACS buffer, this allowed
for a maximum of 6 individual exposures per orbit.  The observations
during each orbit were identical, consisting of 3 pairs of {\sc
  cr-split} exposures to correct for cosmic rays and hot pixels.  We
used the F658N filter (transmitting both H$\alpha$ and [N~{\sc ii}]
$\lambda$6583), with a total exposure time of 1000 s at each position
(500 s per individual exposure).  For each CVZ orbit, we adopted a
3-position linear offset pattern that maximized the area covered on
the sky while also filling-in the inter-chip gap on the ACS detector.
This typical ACS 1-orbit ``footprint'' is shown in the lower right
corner of Figure~\ref{fig:map}, covering a rectangular
205\arcsec$\times$400\arcsec\ region.  The large square regions
representing the contiguous maps of the Tr14 and Tr16 regions
(Figure~\ref{fig:color}) each consisted of 8 of these footprints.  In
two cases, positions 19 and 30, this offset pattern was modified
slightly to include interesting targets while compensating for
roll-angle and scheduling restrictions; see Fig.~\ref{fig:map}.  Two
positions, the adjacent region NGC~3324 and Position 28, are not shown
in Figure~\ref{fig:map} because they are too far outside the field of
view.

The total area on the sky covered by our ACS imaging was roughly 706
arcmin$^2$.  This program includes the first \hst \ imaging of the
Carina Nebula for $\sim$98\% of the area we surveyed.  Prior to this
program, only small regions around $\eta$ Car (e.g., Morse et al.\
1998; Smith et al.\ 2005a; Currie et al.\ 1996) and the
Keyhole\footnote{The Hubble Heritage Image of the Keyhole Nebula is
  available from {\tt http://oposite.stsci.edu/pubinfo/pr/2000/06/};
  see also Smith et al.\ 2004a.}  had been imaged with the Wide Field
Planetary Camera 2 (WFPC2), and very small regions in the core of the
Tr14 cluster and $\eta$ Car itself were imaged with ACS's High
Resolution Channel (ACS/HRC) (e.g., Smith et al.\ 2004b; Ma\'{i}z
Apell\'{a}niz et al.\ 2005).  The ACS/WFC images were reduced and
combined using the PYRAF routine MULTIDRIZZLE, with manually
determined pixel shifts using the pipeline-calibrated, flat-fielded
individual exposures.  MULTIDRIZZLE combines exposures, generates
static bad-pixel masks, corrects for image distortions, and removes
cosmic rays using the {\sc cr-split} pairs.  In this paper we report
the discovery of several new HH jets in Carina.  Other new results
from this {\it HST} imaging program will be presented in future
papers.

\begin{figure*}\begin{center}
\includegraphics[width=6.2in]{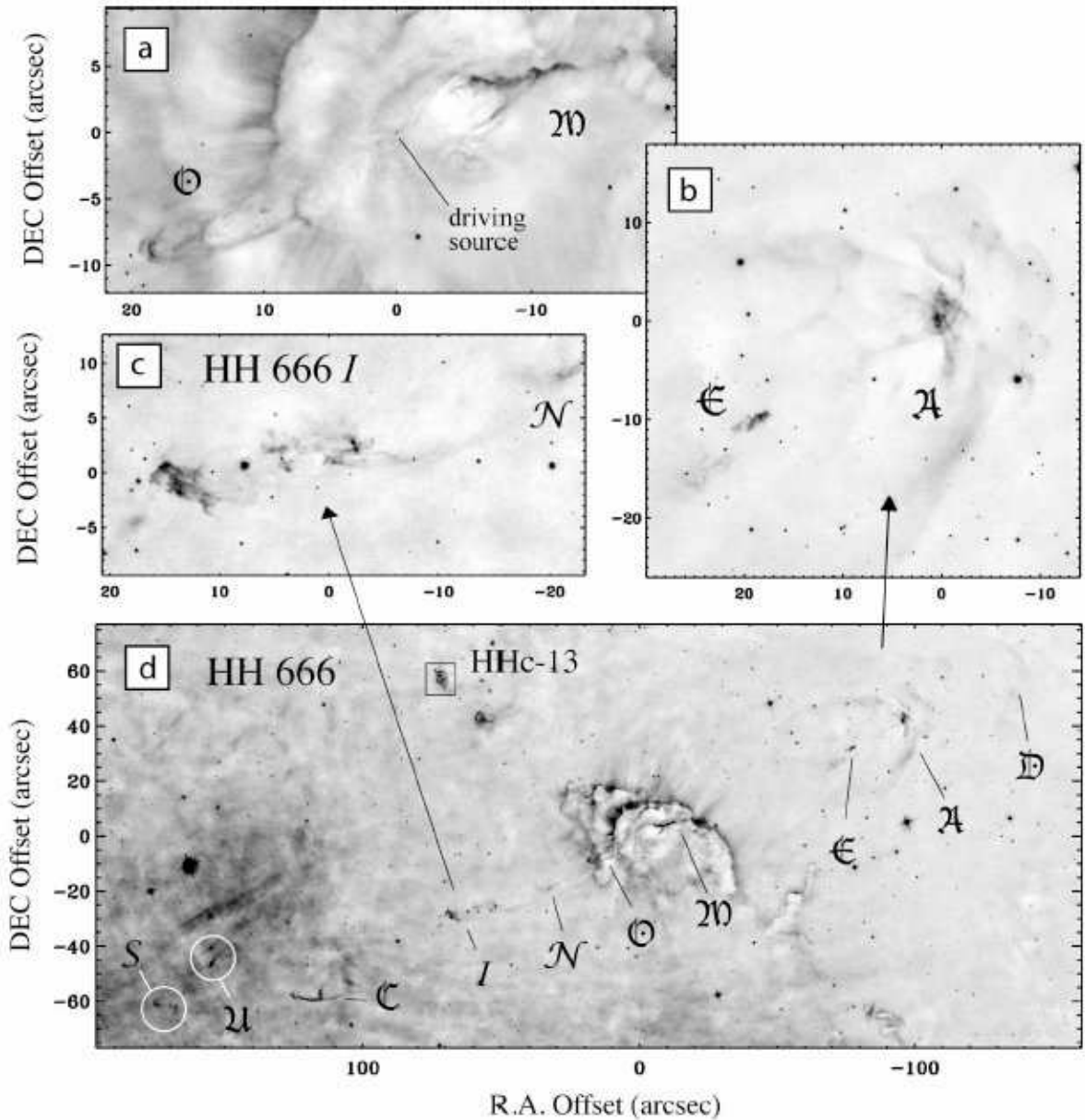}
\end{center}
\caption{H$\alpha$ images of the field including the Axis of Evil,
  HH~666.  Panels (a, b, c) show details of features within the flow,
  whereas Panel (d) shows the full bipolar jet.  Panel (a) is the
  center of the flow including the putative driving source,
  HH~666~IRS1, and the collimated inner jet features HH~666~M and O.
  Panel (b) shows the main bow shock HH~666~A and an internal shock in
  the flow, HH~666~E.  Panel (c) shows shock structures within the
  flow, HH~666~N and I.  The full bipolar jet in Panel (d) includes
  two new shocks seen in the ACS images, HH~666~U and S (lower left),
  which were not seen in ground-based images (Smith et al.\ 2004c).
  The location of HHc-13 is also identified (see \S 3.4.9).}
\label{fig:hh666}
\end{figure*}

\section{RESULTS}

\subsection{New HH Jets and Candidate Jets}

We carefully searched the full area covered by all frames in our
ACS/WFC survey of the Carina Nebula and NGC~3324 in order to identify
protostellar outflows visible in H$\alpha$ emission.  This
identification process is necessarily subjective without multi-epoch
\hst \ images to detect proper motions (e.g., Bally et al.\ 2006;
Smith et al.\ 2005b) and without high-resolution spectra of all
sources to separate Doppler-shifted jet emission from the background.
(These techniques will be used in the future to confirm or reject
candidate sources and to identify new flows.)  HH jets were identified
by nebular features showing well-structured parabolic bow-shocks and
Mach disks, resembling the morphology seen in objects such as HH~34,
HH~47, and HH~111 (Reipurth et al.\ 1997b, 2002; Heathcote et al.\
1996; Hartigan et al.\ 1999, 2001; Morse et al.\ 1992, 1994), or by
high-contrast emission from the body of a collimated flow, and in some
cases the likely counter flow.  A list of new HH jet sources is given
in Table 2, and a list of HH jet candidates for which the jet nature
is less clear is given in Table 3.  HH~666 has been previously
discovered in ground-based data (Smith et al.\ 2004c), but is included
in Table~2 for completeness.  The discovery of all other HH jets and
candidate sources is reported here for the first time.  Individual
sources are discussed below.

\subsection{Comments on individual Jets}

\subsubsection{HH 666}

HH~666, the so-called ``Axis of Evil'', is the only jet in Carina that
could be confidently identified in ground-based data.  Smith et al.\
(2004c) discovered this remarkable jet and presented a detailed
analysis of ground-based images and spectra.  It is a linear bipolar
jet emerging from the head of a dark dust pillar that points toward
$\eta$~Carinae.  It has a projected length of 4$\farcm$5 on the sky,
and with a total length of over 3 pc, HH~666 is one of the longest HH
jets known (see Devine et al.\ 1997; Eisl\"{o}ffel \& Mundt 1997).
Optical images show a chain of dense knots along the jet axis and a
clear bow shock/Mach disk pair (HH~666~A) at the main terminal working
surface.  Along the jet axis, near-IR images reveal an embedded
Class~I protostar inside the globule that is the presumed driving
source of the flow, with $L \simeq 210 L_{\odot}$.  Narrow-band
near-IR images show bright [Fe~{\sc ii}] $\lambda$16435 emission from
the obscured parts of the jet inside the globule, connecting the
parsec-scale outflow seen at visual wavelengths to the embedded inner
IR point source.  Long-slit echelle spectra revealed a coherent
outflow pattern, with the south-east part of the flow being redshifted
and the north-west flow being blueshifted.  These spectra revealed
Doppler shifts of more than $\pm$200 km s$^{-1}$, and a velocity
pattern consistent with a sequence of individual Hubble-like flows
along the jet axis.  This provided suggestive evidence that the jet
was formed by a series of discrete, episodic mass-loss events rather
than a continuous steady outflow, perhaps related to the class of
FU~Ori outbursts (Hartmann \& Kenyon 1985).

Our new ACS/WFC image of HH~666 is shown in Figure~\ref{fig:hh666}.
The entire flow is shown in Figure~\ref{fig:hh666}d, while
Figures~\ref{fig:hh666}a, \ref{fig:hh666}b, and \ref{fig:hh666}c
zoom-in on the base of the jet where it breaks out of the globule, the
main working surface HH~666~A, and an internal working surface
HH~666~I, respectively.  The morphology seen in the new {\it HST}
H$\alpha$ image confirms that the features HH~666~D, A, E, M, O, N, I,
and C are shock structures in the HH~666 flow along the jet axis, and
not small irradiated clouds.  The {\it HST} image also reveals two new
shock structures that we denote HH~666~U and HH~666~S at the eastern
end of the flow.  Together, HH~666~C, U, and S may compose a larger
fragmented curved bow shock that is the eastern flow's analog of
HH~666~A.  A terminus of the eastern flow was not apparent in our
ground-based study, but fainter structures are hard to detect here
because the background nebula is brighter than near HH~666~A.

Some of the most intriguing new results from the {\it HST} images of
HH~666 concern the inner structure of the jet near the driving source,
detailed in Figure~\ref{fig:hh666}a. A faint star is detected along
the jet axis at roughly the same position as the IR source that we
proposed to be the driving source of the HH~666 jet (Smith et al.\
2004c).  This faint star exhibits low-level extended emission to the
west along the jet axis, so we consider it likely that this is indeed
the driving source of the outflow seen in the H$\alpha$ filter despite
the considerable extinction to the central star.  HH~666~M and 666~O
are the blueshifted and redshifted components, respectively, of the
inner collimated jet that is also seen to be very bright in [Fe~{\sc
  ii}] $\lambda$16435 traced back to the IR driving source (Smith et
al.\ 2004c).

The detailed structure of HH~666~M is intriguing
(Figure~\ref{fig:hh666}a). At about 12\arcsec$-$15\arcsec\ northwest
of the star, the jet body is well-collimated along the jet axis.  As
we head eastward, back toward the driving source along the jet, the
brightest emission bends to the north off the jet axis before it can
reach all the way back to the driving source, seeming to curve around
a prominent dark patch located immediately north-west of the driving
source.  This is true for the brightest line emission in the IR as
well (Smith et al.\ 2004c), so it is not merely an obscuration effect
as the jet passes behind the dark globule, for example.  The
surprising structure revealed by {\it HST} shows additional
filamentary emission structures at other positions around the dark
globule that {\it also} seem to merge with the collimated jet axis by
the time they reach $\sim$10\arcsec\ north-west of the driving source.
It is almost as if the jet body has been divided into several streams
as it is diverted around the surface of a dark object in its path.  On
the other hand, the dark oval along the flow axis may not be an
obscuring globule, but rather, a cavity that is dark because a hole is
punched into a dark cloud.

In any case, the jet flow then converges again beyond this dark
feature to form a narrow jet.  While there are many known cases of
variability in collimation, mass-loss rate, speed, and orientation
along a given flow (e.g., Reipurth \& Bally 2001), HH~666~M is one of
the clearest and most dramatic examples of recollimation in a
protostellar jet. Mundt et al. (1991) discussed other possible
examples such as HH~24G.  The specific cause of the recollimation is
not clear; it may indicate variable collimation with time, the
apparenly poor collimation may be caused by transverse motions as fast
material overtakes slower material in the jet, or it may be a
hydrodynamic refocussing of the jet as in a ``Cant\'{o} nozzle''
(Cant\'{o} et al.\ 1981) in objects like HH~526 in Orion (Bally et
al.\ 2000).

\begin{figure*}\begin{center}
\includegraphics[width=6.3in]{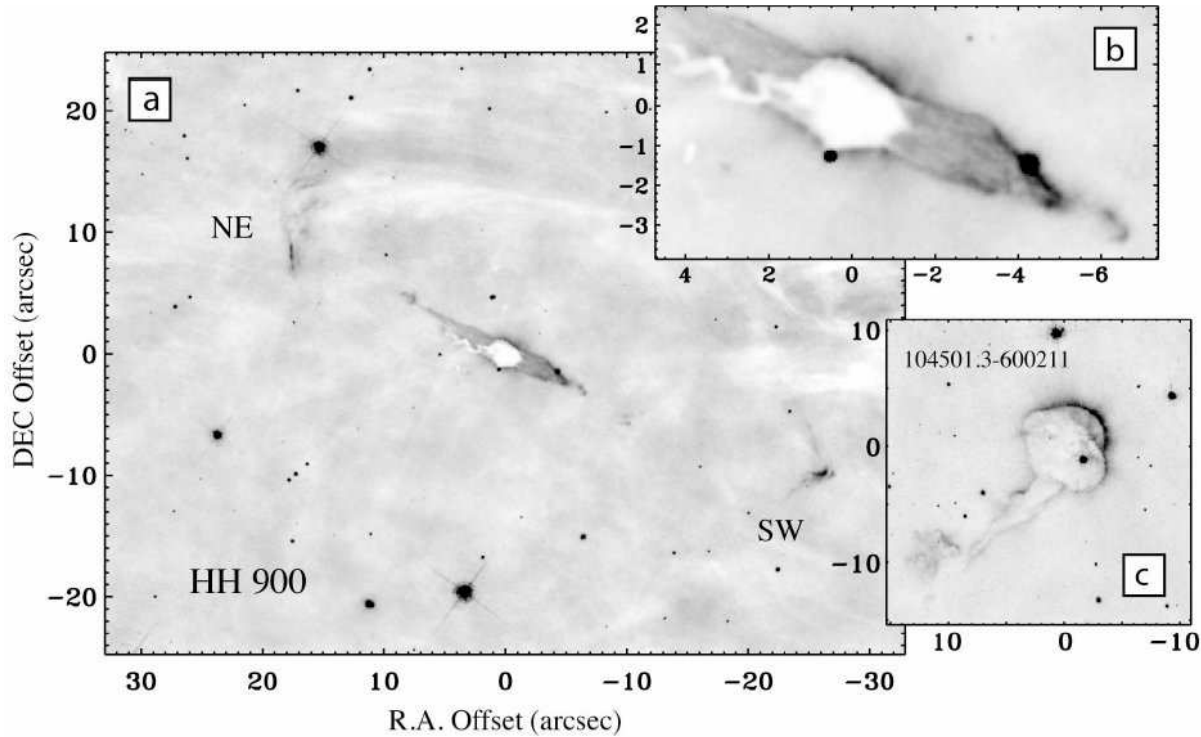}
\end{center}
\caption{(a) H$\alpha$ image of the bipolar jet HH~900 near Tr~16 (see
  Fig.~\ref{fig:color}).  North is up and east is to the left.  (b) A
  detail of the inner region of the jet and globule. (c) A different
  small tadpole-shaped dark globule (104501.3-600211; Smith et al.\
  2003) near Tr14 that does not show evidence for a jet.}
\label{fig:hh900}
\end{figure*}

The main terminal bow shock of the blueshifted north-western flow,
HH~666~A, is shown in Figure~\ref{fig:hh666}b.  This field also
includes the internal working surface HH~666~E along the jet axis.
HH~666~A is a huge bow shock extending over $\sim$40\arcsec\ and
spreading $\sim$15\arcsec\ on either side of the jet axis.  While
HH~666~A maintains its overall parabolic shape seen earlier, at {\it
  HST} resolution it exhibits a more fragmented structure implying a
series of overlapping shock arcs running into a mildly inhomogeneous
medium. This irregular shock front is reminiscent of the complicated
structure seen in HH~1 \& 2 (Bally et al.\ 2002) or HH~168 (Hartigan
et al.\ 2000). The peak emission is still dominated by what appears to
be a Mach disk or reverse shock along the jet axis, which shows
complex small scale structure including multiple dense knots and
filaments, and resembles reverse shocks in well-studied nearby HH jets
such as HH~47 (Heathcote et al.\ 1996; Morse et al. 1994) and HH~34
(Reipurth et al.\ 2002; Morse et al.\ 1992).  Part of the forward
shock along the jet axis extends ahead of the putative Mach disk by
$\sim$10\arcsec\ (0.1 pc).  Multi-filter imaging or spatially resolved
spectroscopy with {\it HST} could resolve the post-shock cooling zone.

The complex structure of an internal working surface in the flow is
exemplified by HH~666~I, shown in Figure~\ref{fig:hh666}c.  It is
perhaps not surprising that compared to ground-based images (Smith et
al.\ 2004c), the H$\alpha$ emission structure seen with {\it HST}
breaks up into a series of smaller clumps, filaments, and numerous
thin arcs.  These multiple shock structures indicate that even at
large distances of $\sim$1 pc from where it was launched, the
structure of the jet body is still quite inhomogeneous on small size
scales of $\la$10$^{15}$ cm.  Thinner filaments produce the same flux
in a smaller area on the sky (compared to fuzzy structures in
ground-based images), so they have higher surface brightness,
$I_{H\alpha}$, and they also have a smaller emitting layer thickness
$L$ along the line of sight.  When deriving the electron density from
the emission measure (see section 5), we have $n_e \propto
(I_{H\alpha}/L)^{1/2}$, so the narrower structures imply higher
electron densities.  This is important, because it resolves a
discrepancy from our previous study (Smith et al.\ 2004c) where
electron densities derived from the H$\alpha$ emission measure were
systematically lower by factors of 2--3 than electron densities from
spectral diagnostics like the red [S~{\sc ii}] doublet ratio,
primarily because the individual filaments were not fully resolved.

\begin{figure*}\begin{center}
\includegraphics[width=5.5in]{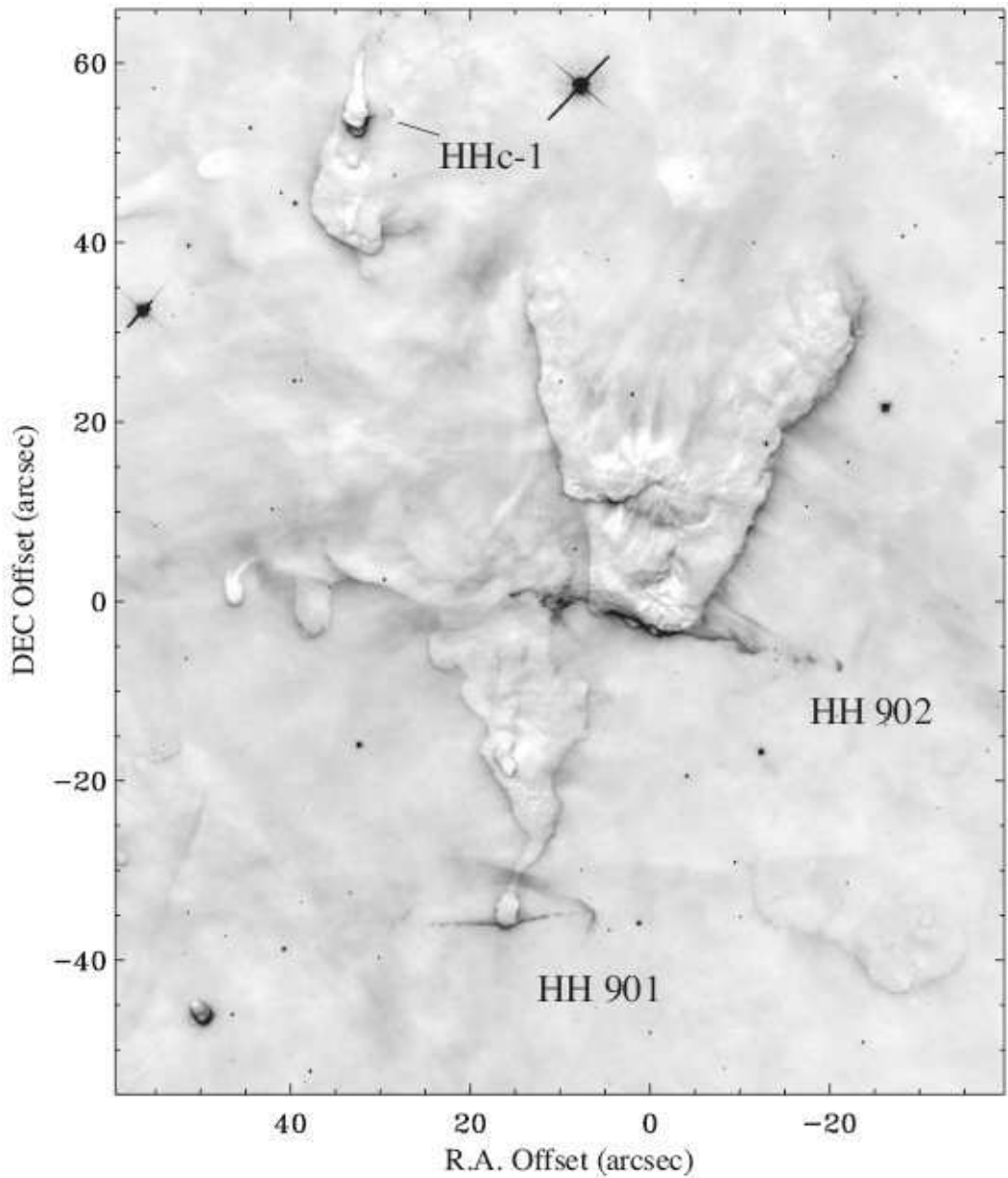}
\end{center}
\caption{H$\alpha$ image of the field including the two bipolar jets
  HH~901 and HH~902, north of Tr~14 (the Tr~14 cluster is located to
  the south, off the bottom of this figure).  The jet candidate HHc-1
  is also included in the field.  North is up and east is to the left,
  and the offsets are relative to an arbitrary point near the origin
  of the HH~902 jet.}
\label{fig:hh901}
\end{figure*}

\subsubsection{HH 900}

HH~900, shown in Figure~\ref{fig:hh900}a, is a stunning object.  This
bipolar jet explodes from a tiny ($\sim$1\arcsec\ diameter) dark
globule that was one of the original proplyd candidates seen in
ground-based images of Carina (Smith et al.\ 2003).  It resides within
the bright inner region of NGC~3372 in Tr~16, but near the edge of the
dark molecular ridge that bisects the Carina Nebula (see
Figure~\ref{fig:color}).  It is found only 2-3 pc from $\eta$ Carinae
and several massive O-type stars in the Tr~16 cluster, so the jet
material outside the globule is bathed in a strong UV radiation field.

The morphology revealed in {\it HST} images is quite unusual.  The
small dark globule has a wiggly dark tail extending to the east, with
peculiar and very angular kinks that make it look like a swimming
tadpole or spermatozoon.  On larger scales, one can see qualitatively
similar wiggles in the tails of cometary clouds or in thin cords
connecting the heads of dust pillars to a more distant clouds
throughout Carina, as in HH~901 and HH~1008 (see below).  The origin
of these wiggles is unclear; they may arise from hydrodynamic
instabilities as these cometary clouds are shocked and ablated (e.g.,
Pittard et al.\ 2009; Marcolini et al.\ 2005; Klein et al.\ 1994), or
they may point to time-variable radiation fields in the region.  An
alternative explanation in the case of HH~900 is that two competing
effects influence the strange tail morphology.  On the one hand, the
cometary tail pointing to the east must have been formed as a result
of strong UV radiation or winds from the massive Tr~16 stars 2--3 pc
to the west (from the right in Figure~\ref{fig:hh900}).\footnote{Note,
  however, that HH~900's tail does not point exactly away from $\eta$
  Car itself, which is to the northwest (upper right) in
  Figure~\ref{fig:hh900}, rather than due west.}  On the other hand,
HH~900 is even closer ($\sim$0.5 pc) to the edge of a dark molecular
ridge (see Figure~\ref{fig:color}), which appears as a very bright
edge-on PDR at IR wavelengths (Smith et al.\ 2000; Smith \& Brooks
2007).  This PDR probably has a strong photoevaporative flow from the
illuminated surface of the cloud, and this flow would form a pervasive
westerly wind in the vicinity of HH~900.  This photoevaporative wind
may counteract the UV radiation force, and may act to compress the
tail.  Hydrodynamic models of such a scenario would be interesting.
This same photoevaporative wind may also be needed to explain the jet
bend toward the massive stars discussed below.

The bipolar jet emerging from the head of this tiny globule shows
bright emission from both the main body of its collimated jet and from
a pair of bow shocks at the ends of the flow.  The full extent of the
jet is about 41\arcsec\ between the two bow shocks, or 0.46 pc, with
the flow axis along P.A.$\simeq$242\arcdeg.  The line connecting the
apexes of the two bow shocks, however, passes about 3\arcsec\ NW of
the dark globule from which the jet emerges, indicating that the jet
flow axis has been bent.  Interestingly, however, the jet ends are
bent {\it toward} the direction of Tr~16, rather than away from it as
one would expect if the jet were bent by radiation pressure, the
rocket effect, or a side wind as in LL~Ori objects (see Bally et al.\
2006).  Bally \& Reipurth (2001) noted several jets in NGC~1333 that
bent toward the center of star formation activity as well, but in
those cases this was thought to result because their driving
protostars were dynamically ejected from the cluster.  This is not the
case for HH~900, where the cometary tail points in the wrong
direction.

The fact that the dark tail is seen in silhouette in front of the
eastern part of the jet suggests that the eastern flow is redshifted
and the western flow is blueshifted; Doppler velocities from spectra
are not yet available to confirm this conjecture.  The morphology in
the pair of bow shocks conforms to the bow shock and Mach disk
morphology typically seen in the terminal shocks of HH jets.  The
morphology of the collimated jet body, however, is somewhat bizarre
and difficult to understand.

Zooming-in on the point where the western part of the jet flow first
emerges from the dark globule, one can see remarkable structure, shown
in the detail panel of Figure~\ref{fig:hh900}b.  The jet body is quite
wide, with 3-pronged elongated filaments seen in absorption.  The
central dark filament may correspond to dust entrained in the narrow
body of the collimated jet itself seen in silhouette against the
bright screen of the background H~{\sc ii} region.  The two remaining
dark filaments may be limb-darkened by dust in the side walls of the
flow cavity, pushed aside and entrained as the jet coccoon exits the
globule.  Similar structures were seen in HH~280 in L1451 at the point
where it first breaks out of its natal globule (Walawender et al.\
2004).

About 4$\farcs$5 west from the center of the dark globule, along the
jet body, there is a star that draws attention to itself
(Figure~\ref{fig:hh900}b).  It resides precisely at the center of a
straight 2$\farcs$5-long H$\alpha$ emission filament oriented at
P.A.$\simeq$45\arcdeg, and the ends of this filament meet the boundary
of the western part of the HH~900 jet.  Is this simply a chance
superposition of this star along the straight filament, or is this
filament actually a bipolar microjet originating from that star?  If
the former is true, then it is unclear why the wide jet body appears
to recollimate and then coincidentally end at the position of the star
and putative microjet.  If the latter is correct, then the
morphological relationship between this putative microjet and the rest
of the outflow gives the impression that this star and its associated
outflow have recently been dynamically ejected from the dark globule.
Although the image is intriguing, such conjecture is quite speculative
without detailed spectroscopic information to provide kinematics of
the star and jet material.

The HH~900 jet is significant in a broader context as well, because it
confirms that even the smallest ($\sim$1\arcsec) dark globules in the
interior of the H~{\sc ii} region can harbor actively-accreting
protostars, and that these dark globules are still important sites of
active star formation within evolving H~{\sc ii} regions.  Many small
dark globules without associated HH jets, such as the object in
Figure~\ref{fig:hh900}c, may not be caught in this brief active
outflow phase but may nevertheless harbor young stars or protostars.
In such close proximity to $\eta$~Carinae, HH~900 may --- in the
not-too-distant future --- be polluted by nuclear-processed ejecta
when that star dies in a violent supernova explosion.  In that sense,
the star that drives HH~900 provides an observed example of precisely
the sort of conditions adopted in the models of Boss et al.\ (2008),
where SN ejecta collide with a small cloud, imparting radioactive
nuclides. These, in turn, are presumably then incorporated into
chondrites within 10$^6$~yr after the SN, when the cloud material
collapses into a disk, as is thought to have occurred in our Solar
System.  With a diameter of $\sim$2\arcsec\ ($\la$10$^{17}$ cm), the
head of the dark globule is about half the {\it initial} size of the
$\sim$1 $M_{\odot}$ cloud in the simulation discussed by Boss et al.\
(2008).  At a distance of 2--3 pc from $\eta$ Car, the dark globule
has a cross section that would intercept a fraction $\sim$10$^{-5}$ of
$\eta$ Car's SN ejecta, if it were to explode.  Thus, if $\eta$ Car
ejects at least 10 $M_{\odot}$ of processed ejecta in the explosion,
then HH~900 would intercept 10$^{-4}$ $M_{\odot}$ of SN material,
comparable to the amount required to account for the inferred
$^{36}$Al and $^{60}$Fe abundance in the Solar nebula (e.g., Foster \&
Boss 1997).  HH~900, then, provides a glimpse of a young accreting
protostar in the specific sort of environment where the Sun may have
formed.

\begin{figure*}\begin{center}
\includegraphics[width=6.2in]{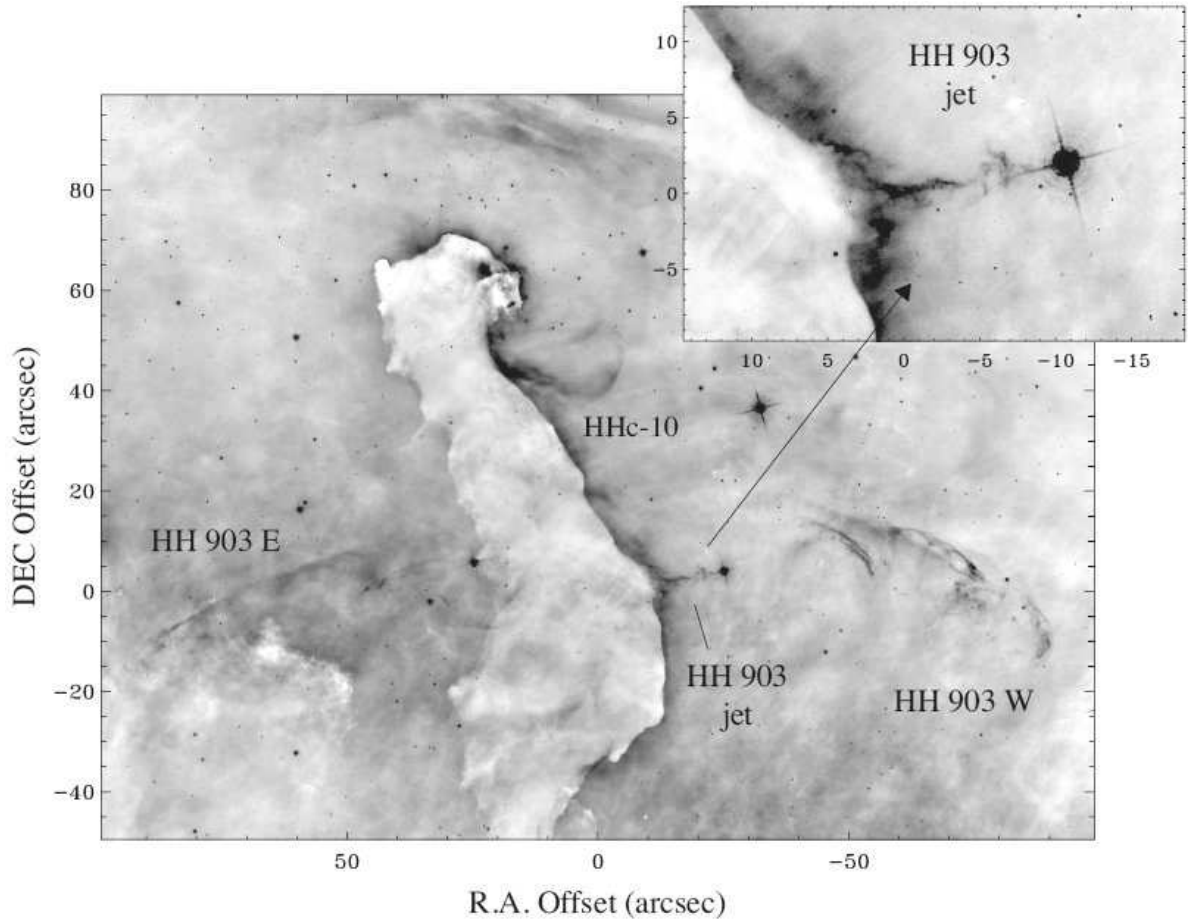}
\end{center}
\caption{ACS H$\alpha$ image of the large dust pillar (G287.88-0.93)
  in the South Pillars of Carina.  The field contains the parsec-scale
  bipolar outflow HH~903, which flows east and west to either side of
  the middle of the pillar and is over 2 pc in length. The field also
  contains the candidate jet HHc-10 associated with the pillar head
  (discussed later).  North is up and east is to the left.}
\label{fig:hh903}
\end{figure*}

\begin{figure*}
\begin{center}
\includegraphics[width=5.8in]{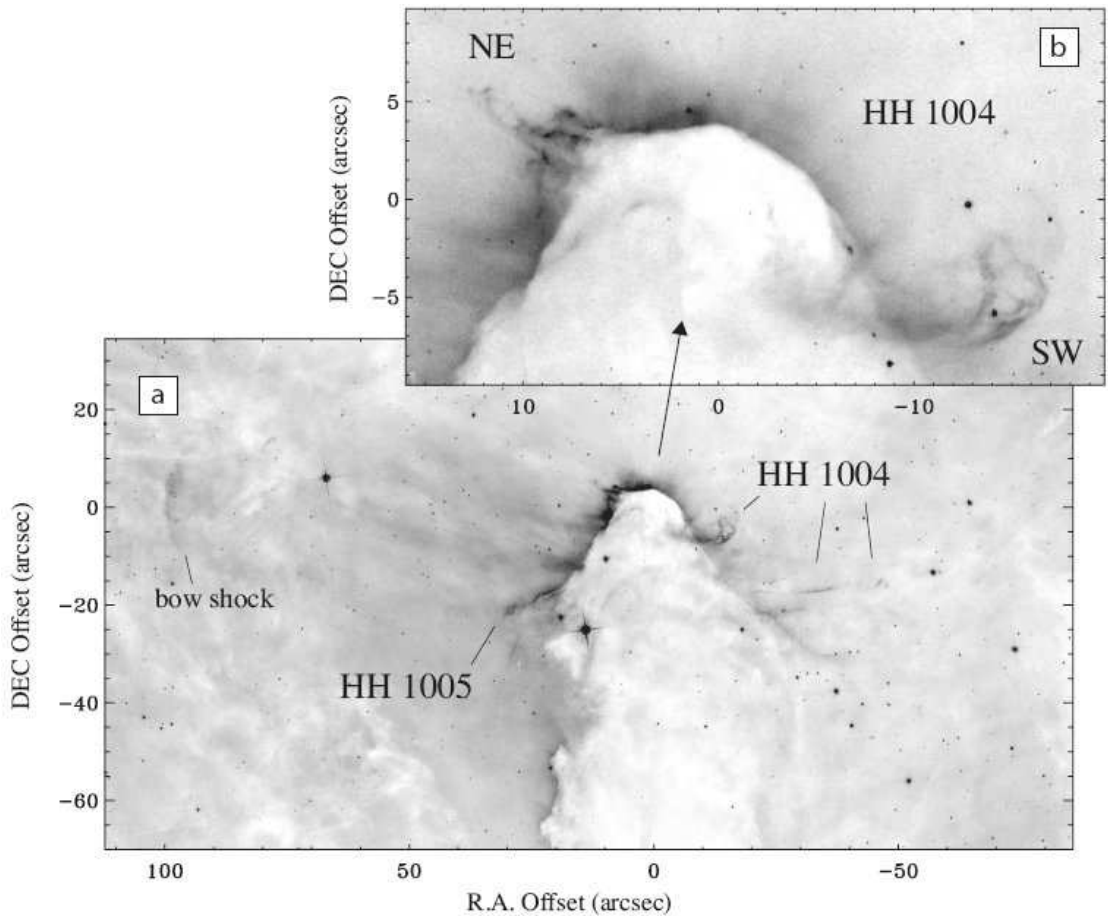}
\end{center}
\caption{H$\alpha$ image of the tip of another large dust pillar in
  the South Pillars, which contains the bipolar jet HH~1004 closest to
  the tip of the pillar, and the one-sided jet HH~1005, which appears
  to be a separate flow.  There is also a large bow shock at left,
  probably associated with HH~1004.}
\label{fig:hh1004}
\end{figure*}

\subsubsection{HH 901 and HH 902}

The pair of outflows HH~901 and HH~902 are spectacular examples of
irradiated bipolar jets (Figure~\ref{fig:hh901}).\footnote{These were
  included in the large mosaic shown in Figure~\ref{fig:color}, and
  color images of HH~901 and HH~902 can be found at {\tt
    http://heritage.stsci.edu}.} They are extrememly bright, and in
hindsight, they can be seen on high-quality optical ground-based
images.  Unambiguous recognition of these features as jets, however,
required the angular resolution of {\it HST}.  Their high surface
brightness of (2--6)$\times$10$^{-14}$ erg s$^{-1}$ cm$^{-2}$
arcsec$^{-2}$ is due to high jet densities and strong UV irradiation
by Tr~14, located only $\sim$1\arcmin\ (0.7 pc) to the south.  The
pillars from which HH~901 and 902 emerge do not point exactly at
Tr~14, however.  Instead, their axes point between Tr~14 and the
massive O stars HD~93160/93161, thought to be members of Tr~16 (the
bright pair of stars $\sim$2\arcmin\ SE of Tr~14; see Walborn 2009).

HH~901 emerges from within $\sim$1\arcsec\ (3$\times$10$^{16}$ cm) of
the very tip of a dark cometary dust cloud that is almost cut off from
its natal dust pillar, with only a thin twisted cord connecting them.
It is likely that it will soon appear as a bipolar jet from an
isolated tiny cometary cloud, much like HH~900.  HH~902, on the other
hand, emerges from the tip of a thick dust pillar that is still
solidly connected to its parent cloud.  Several other cometary clouds
without jets can be seen in the region, as well as one candidate
bipolar jet from a tiny cloud (HHc-1, located in the upper left of
Figure~\ref{fig:hh901}, and discussed more below).

The well-defined jet body of HH~901 extends to $\pm$8\arcsec\ (0.09
pc) on either side of the globule, consisting of a highly collimated
chain of H$\alpha$-emitting knots implicating a time-variable outflow.
Detailed inspection of the images shows that the jet body has a higher
surface brightness on its southern edge facing Tr~14.  The average
orientation of the bipolar flow is along
P.A.$\simeq$95\arcdeg(275\arcdeg), although there is a slight bend in
the jet, such that the eastern portion runs along
P.A.$\simeq$92\arcdeg, while the western side is angled at
P.A.$\simeq$279\arcdeg, although this may be partly due to the
non-uniform illumination.  The outflow terminates in a pair of
parabolic bow shocks to the east and west (although the western bow is
more clearly defined), marking a total end-to-end length of the
outflow of $\sim$23\arcsec\ (0.26 pc).

HH~902 is similar, oriented at P.A.$\simeq$258\arcdeg, with total jet
length of over 38\arcsec\ (0.42 pc).  It also shows clear bow shock
structures at its ends, although the structure along the jet is more
chaotic than HH~901.  It has a chain of dense knots, suggestive of
variable outflow, with several twists and turns that implicate shaping
by sideward radiation pressure, the rocket effect, or winds from the
OB stars below in Tr~14.  If the driving source were located halfway
along the jet, it would reside within a curved protrusion in the dark
dust pillar (an R.A.\ offset of $-$2\arcsec\ in
Figure~\ref{fig:hh901}).  HH~902's dust pillar has a strong
photoevaporative flow normal to the pillar surface, through which
HH~902 must burrow.

Like many jets studied here, the orientations of HH~901 and HH~902 are
such that their outflow axes run roughly perpendicular to the axis of
symmetry in their parent cometary cloud or dust pillar.  Because they
are extraordinarily bright and their bipolar jet structure is
well-defined, we pursue a more detailed and more quantitative case
study of the properties of these two jets in \S 4.

\subsubsection{HH 903}

HH~903 is a remarkable bipolar jet in the South Pillars that emerges
from the sides of a large dust pillar (G287.88-0.93), shown in
Figure~\ref{fig:hh903}.  This pillar marks the relatively narrow top
of an even larger structure that is the most massive concentration of
molecular gas in the region, known as the ``Giant Pillar''
(G287.93-0.99; Smith et al.\ 2000), both of which point roughly toward
$\eta$ Carinae.  HH~903 is an east/west flow oriented roughly along
P.A.$\simeq$278\arcdeg, with a total length of about 180\arcsec\ (2
pc).  It is therefore among the longest protostellar ouflows known, in
the elite class of parsec-scale outflows (e.g., Bally \& Devine 1994;
Devine et al.\ 1997; Eisl\"{o}ffel \& Mundt 1997; Marti et al.\ 1993;
Ogura 1995; Poetzel et al.\ 1989; Reipurth et al.\ 1997b; Smith et
al.\ 2004c).

We identify three main components to the HH~903 flow: the large bow
shock structures to the east and west, HH~903~E and H~903~W,
respectively, and the main body of the jet (see Fig.~\ref{fig:hh903}).
The bow shocks are asymmetric, with their curved shock structures
appearing much brighter on their northern sides.  If these two bow
shocks are density enhancements along parabolic cones, their
``top-lighted'' appearance in Figure~\ref{fig:hh903} probably arises
because they are illuminated from above by OB stars in Tr~16 far to
the north.  This suggests that the southern surfaces of the bow shocks
are partly shielded because neutral gas within them or at their
northern surfaces absorbs most of the ionizing radiation.
Alternatively, the northern parts of HH~903E and 903W may be brighter
because of higher density, having been compressed by the ram pressure
of the large-scale flow of plasma descending from above in
Figure~\ref{fig:hh903}.

The jet body of HH~903 erupts from the western side of the dust
pillar, with a chaotic stream of condensations about 12\arcsec\ (0.13
pc) in length. Since the detectable jet body terminates long before
reaching the bow shock, it suggests that the HH~903 flow has been
intermittent.

HH~903 is similar to many other jets in Carina in that its axis
appears to be roughly perpendicular to the axis of the elongated dust
pillar or cometary cloud from which it emerges, but it is unusual in
that the driving source is not located at the apex of the dust pillar.
Instead, the driving source is located halfway down the dust
pillar. Judging from the collimated morphology of the jet, its driving
source is likely to be close to the west ionization front.

\subsubsection{HH 1004 and HH 1005}

The two outflows HH~1004 and HH~1005 both emerge from the head of the
same dark cloud in the South Pillars, and both are shown in
Figure~\ref{fig:hh1004}a.  This is one of the more prominent dust
pillars in the region, but it has not been discussed before in the
literature.  Like its cousins, the pillar points north in the general
direction of Tr~16, although its symmetry axis actually points
$\sim$15\arcdeg\ east of $\eta$ Car.

\begin{figure}\begin{center}
\includegraphics[width=2.7in]{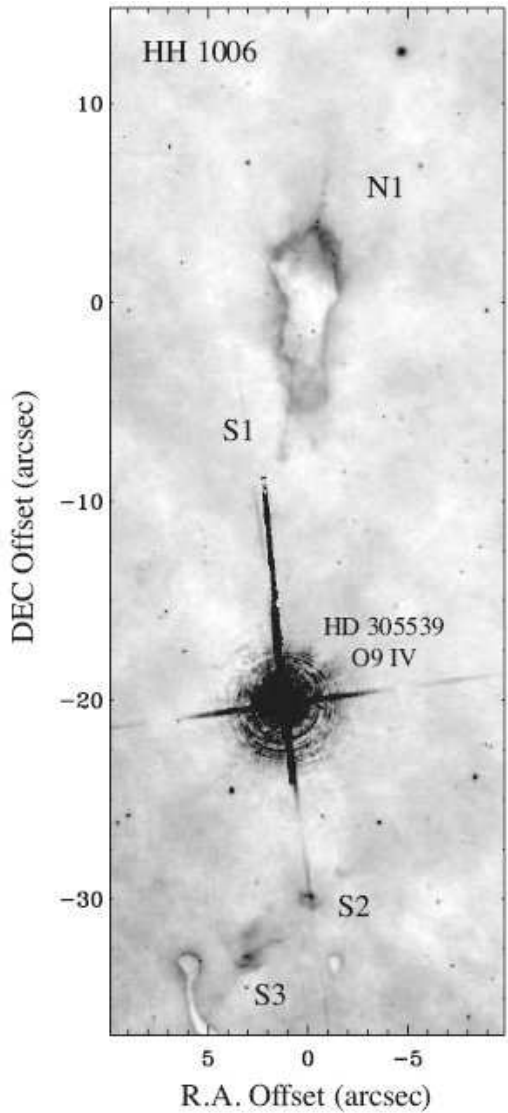}
\end{center}
\caption{ACS H$\alpha$ image of the highly collimated bipolar jet
  HH~1006, emerging from a dark cometary globule.  The north (N1) and
  south (S1) parts of the collimated bipolar jet flow are identified,
  as well as two more distant shock structures to the south (S2 and
  S3) that are probably part of the same outflow.  The bright star in
  the center is the O9 IV star HD~305539.}
\label{fig:hh1006}
\end{figure}

HH~1004 is clearly a bipolar jet with flows seen emerging to the NE
and SW from the very tip of the dark pillar in
Figure~\ref{fig:hh1004}b.  HH~1004~NE has a somewhat chaotic
structure, perhaps arising as it plows through a dense
photoevaporative flow.  HH~1004~SW has a large and more well-defined
bow shock structure that is brighter on its southern side, opposite
the direction facing the ionizing stars but toward the denser
photoevaporative flow.  The full length of the inner part of HH~1004's
bipolar jet is roughly 32\arcsec\ (0.36 pc), with its axis along
P.A.$\simeq$247\arcdeg.  However, there also appear to be more distant
shock structures associated with the HH~1004 flow.  Continuing to the
southwest from the bow-shock structure HH~1004~SW noted above, there
are several curved filaments along the same axis (identified in
Figure~\ref{fig:hh1004}a) that probably represent a more distant
terminal shock along the same flow.  Similarly, far to the east there
is a large bow shock structure, which is labeled in
Figure~\ref{fig:hh1004}a.  This bow shock is not along the presumed
axis of HH~1004, but it seems likely that this is a distant bow shock
in a curved flow that is part of HH~1004 rather than HH~1005, because
HH~1005 points in an altogether different direction.  Including these
distant shock structures, the total length of the HH~1004 jet is
roughly 150\arcsec\ (1.7 pc), placing it among the longest
parsec-scale HH jets known.

HH~1005 emerges from the tip of the same dust pillar, but only one
side of the flow is seen, directed in the SE direction.  In HH~1005,
H$\alpha$ emission traces a collimated jet body, but no distant bow
shocks are seen along a projection of the same axis.  The body of this
jet is likely to be quite dense, because its northern face is much
brighter, suggesting that the jet body is optically thick to ionizing
photons, which therefore cannot penetrate all the way through it (as
is the case for clearer examples such as HH~901 and 902, or jets in
Orion discussed by Bally et al.\ 2006).  The length of the emitting
portion of the jet that we detect is roughly 15\arcsec\ (0.17 pc),
with an axis along P.A.$\simeq$112\arcdeg.  Projecting this axis into
the cloud, the IR driving source should be located a few arcseconds
south of HH~1004's protostar.

\subsubsection{HH 1006}

HH~1006, shown in Figure~\ref{fig:hh1006}, is a narrow and straight
collimated bipolar jet.  It emerges from an embedded star inside a
dark cometary globule that is one of the proplyd candidates from Smith
et al.\ (2003), although here it resembles a bright-rimmed cometary
cloud rather than Orion's proplyds.  The putative embedded star is not
detected in our ACS H$\alpha$ image due to extinction within the
cloud.  The small cometary globule is located amid the South Pillars,
and it points north toward $\eta$ Carinae.

The narrow HH~1006 jet is a chain of faint knots, stretching over a
full jet length of roughly 16\arcsec\ (0.18 pc).  The jet points along
roughly the same axis as the globule, and the bipolar jet does not
deviate from its axis by more than about 1\arcdeg.  Farther south one
sees two additional shocks, likely to be distant terminal shocks in
the same outflow, denoted S2 and S3 in Figure~\ref{fig:hh1006}.
Including these, the total length of HH~1006 is $\sim$40\arcsec, or
over 0.45 pc.  The bright O9 IV star HD~305539 is seen along the
southern jet axis, but this may be a chance alignment.

\begin{figure*}\begin{center}
\includegraphics[width=4.4in]{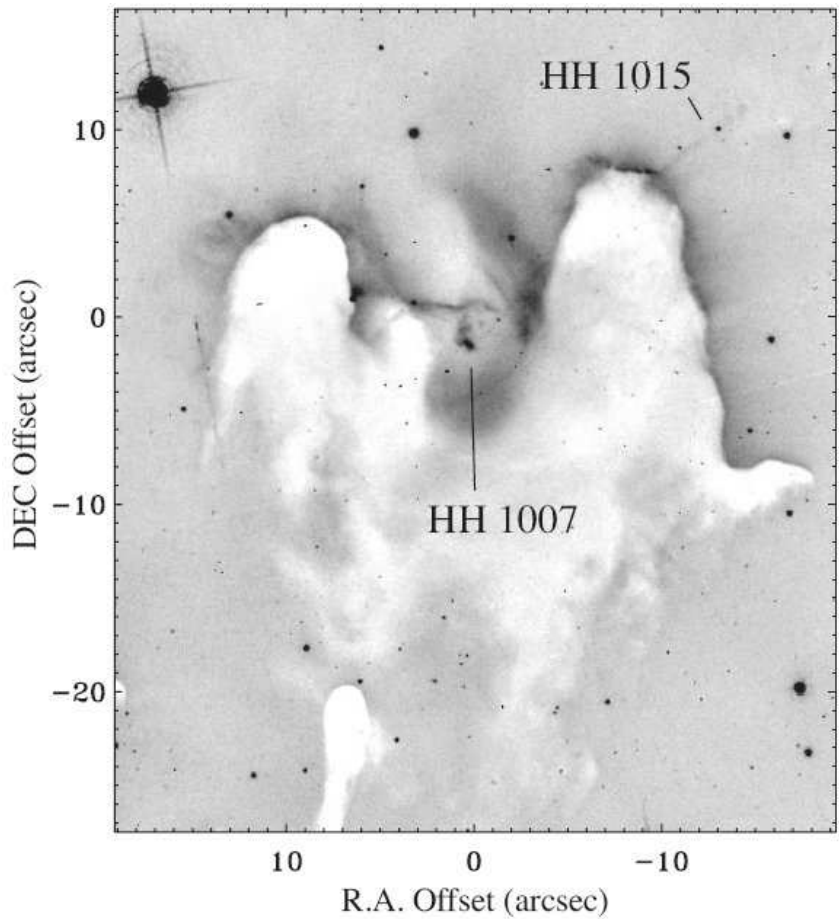}
\end{center}
\caption{ACS H$\alpha$ image of a dark globule amid the South Pillars
  in Carina (Pos.\ 25 in Figure~\ref{fig:map}), containing the rather
  mangled one-sided jet HH~1007 (center) discussed in \S 3.2.7, as
  well as the collimated jet HH~1015 (upper right) also discussed in
  \S 3.2.7.}
\label{fig:hh1007}
\end{figure*}

\begin{figure*}\begin{center}
\includegraphics[width=4.6in]{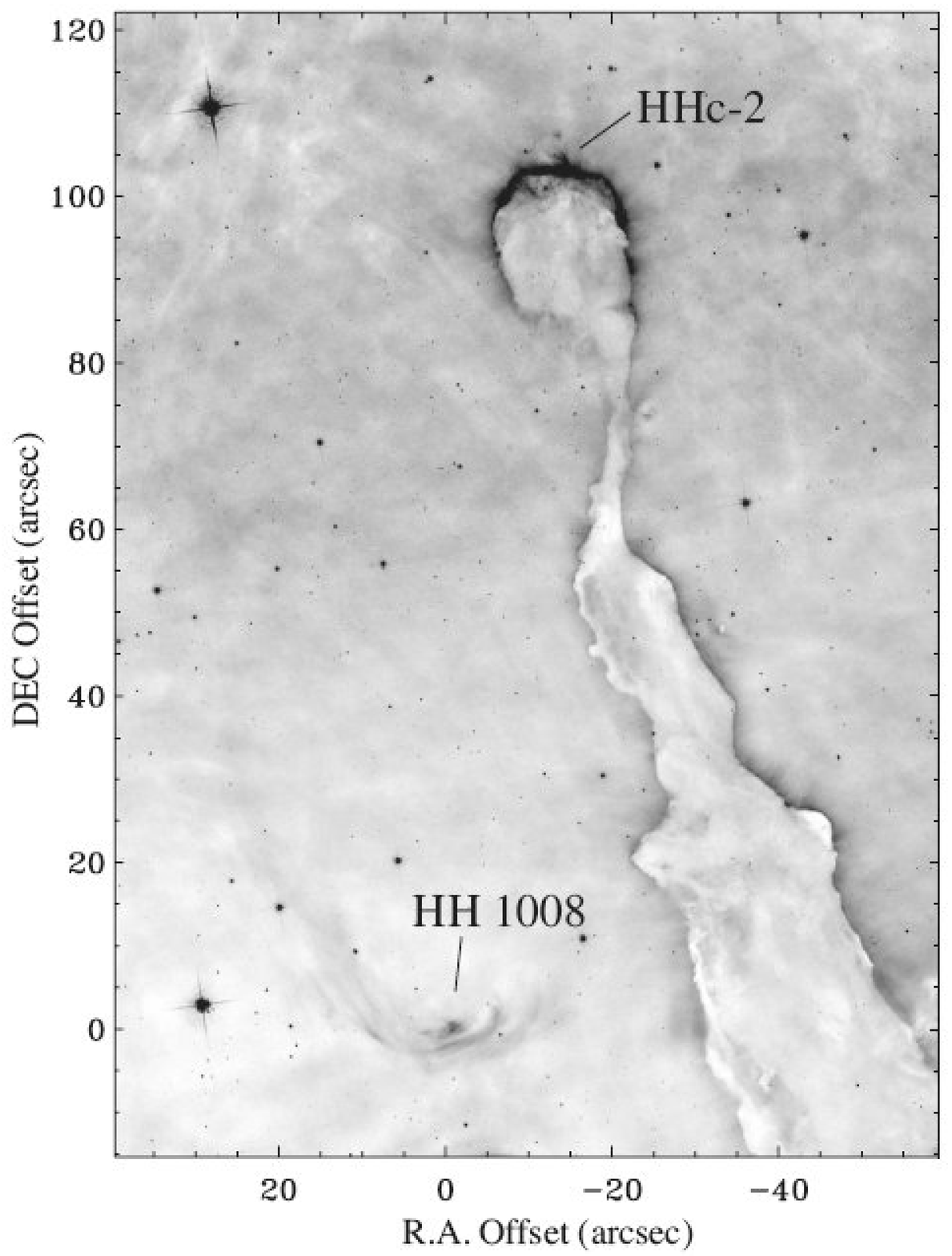}
\end{center}
\caption{H$\alpha$ image of the giant bow shock HH~1008, the wings of
  which stretch across roughly 45\arcsec (0.5 pc).  The tadpole shaped
  structure is the tip of a long wriggly dust pillar associated with
  the cloud G287.73-0.92.  Although no collimated jet body is seen,
  the driving source of the jet is clearly to the north of HH~1008 and
  may be located inside the head of this pillar.  The candidate jet
  HHc-2 emerges from the north end of this same cometary cloud and may
  be the counter-jet in a bipolar flow.}
\label{fig:hh1008}
\end{figure*}

\subsubsection{HH 1007 and HH 1015}

HH~1007 in Figure~\ref{fig:hh1007} is a dense collimated jet body
associated with a small complex of dark clouds in the far southern
region of the South Pillars (Pos.\ 25 in Fig.~\ref{fig:map}).  The
overlapping cometary clouds in this dark cloud complex all point north
toward $\eta$ Carinae, Tr~16, and Tr~14.  A large curved structure may
indicate a bow shock, but the morphology is complicated because the
jet appears to be impacting another cloud or plowing through its dense
photoevaporative flow.  The jet projects eastward to the left-most of
the dark clouds in Figure~\ref{fig:hh1007}, suggesting that this is
the location of its driving source.  The observed jet length is
roughly 8\arcsec\ (0.09 pc) running roughly due west at
P.A.$\simeq$270\arcdeg.

HH~1015, also shown in Figure~\ref{fig:hh1007}, is a chain of very
faint but well-collimated emission knots that is $\sim$6\arcsec\ (0.07
pc) long, with its axis at P.A.$\simeq$315\arcdeg.  It is a one-sided
jet emerging to the northwest from the head of a dark cometary cloud,
and appears similar to HHc-5 and HH~1014.  The dark globule from which
HH~1015 emerges is part of the same complex of overlapping cometary
clouds that launches HH~1007.

\subsubsection{HH 1008}

Figure~\ref{fig:hh1008} shows the field around HH~1008, which is a
giant and elegant bow shock structure spanning more than 45\arcsec\
(0.5 pc).  The H$\alpha$ image also shows a dense knot at the apex of
the parabolic shock, which is likely to be its Mach disk where the
collimated jet is decelerated, although the jet body itself is not
detected.

The long, thin, and twisted dark cloud in Figure~\ref{fig:hh1008} is
G287.73-0.92, which terminates with a bright tadpole-shaped cometary
cloud that points north to $\eta$ Carinae.  Although the {\it HST}
H$\alpha$ image does not reveal a collimated jet body associated with
HH~1008, the parabolic bow shock has an axis of symmetry that points
north along its open end, up toward this cometary cloud, so this is
the most likely location of its driving source.  Indeed, there is a
partially obscured star at the apex of this cometary cloud that is
seen in the H$\alpha$ image.  If this is the driving source, then the
southern flow in HH~1008 has a length of roughly 105\arcsec\ (1.2 pc)
The candidate flow HHc-2 emerges to the north from this source, and it
may be the counter-flow to HH~1008.

\subsubsection{HH 1009}

\begin{figure}\begin{center}
\includegraphics[width=2.9in]{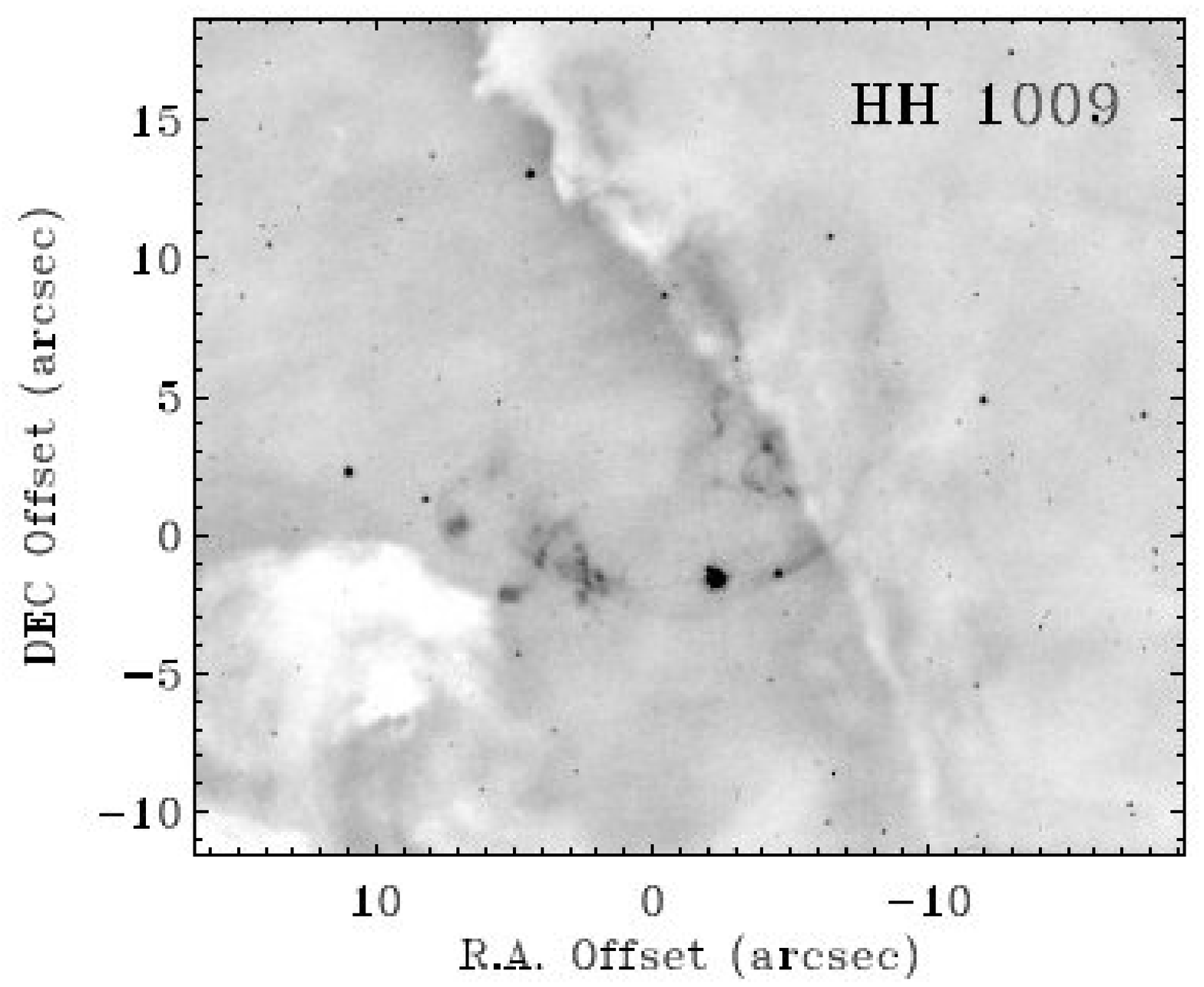}
\end{center}
\caption{H$\alpha$ image showing HH~1009, an outflow emerging from the
  side of the same pillar associated with HH~1008.  This feature is
  located off the bottom of Figure~\ref{fig:hh1008}.}
\label{fig:hh1009}
\end{figure}

HH~1009 is an unusual case, seen in Figure~\ref{fig:hh1009}.  The
dense filaments are distinct from other wisps and filaments in the
background H~{\sc ii} region, and their morphology clearly resembles
shock structures in HH jets.  (In this sense it is similar to HH~1007,
discussed above.)  One can see thin bow shocks surrounding clumps of
emission that are probably Mach disks or other working surfaces in the
flow.  However, images do not reveal a collimated jet body, so it is
unclear exactly where the jet originates.  Its likely origin is within
the base of the large dark dust pillar to its west, which is the base
of the pillar associated with HH~1008.

\subsubsection{HH 1010}

\begin{figure}\begin{center}
\includegraphics[width=2.9in]{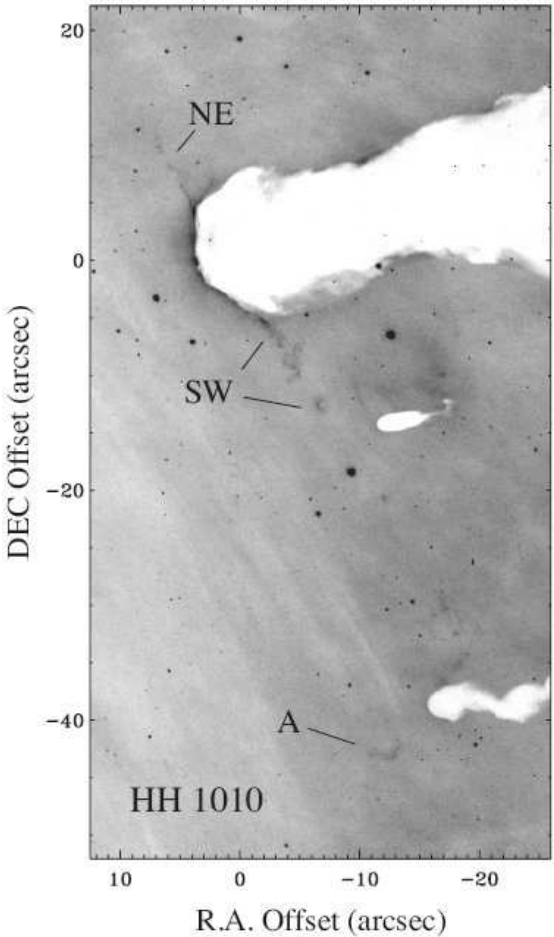}
\end{center}
\caption{H$\alpha$ image of HH~1010, a highly collimated bipolar jet
  emerging to the northeast (NE) and southwest (SW) from the very tip
  of a dust pillar near the western edge of Carina.  Several internal
  shocks and a bow shock in the flow (A) can be seen.}
\label{fig:hh1010}
\end{figure}

HH~1010 is a highly collimated bipolar jet emerging from the very tip
of a large and dark dust pillar at the western edge of the southern
Carina Nebula (Figure~\ref{fig:hh1010}).  It is associated with the
large edge-on western ionization front in the southern polar lobe of
the nebua; see Smith et al.\ (2000).  The symmetric parts of the inner
bipolar jet extend $\pm$15\arcsec\ to the northeast (NE) and southwest
(SW) from the presumed driving source at the head of the jet.
However, additional shock structures along the same axis are seen
farther to the southwest (a possible bow shock is marked as HH~1010~A
in Fig.~\ref{fig:hh1010}), making the full length of the jet at least
60\arcsec\ (0.7 pc) along roughly P.A.$\simeq$216\arcdeg.  With the
driving source located along the jet axis inside the dark dust pillar,
the driving source must be within $\sim$2$\farcs$5 ($\la$10$^{17}$ cm)
from the ionization front.  There is a faint optical point source
within the boundaries of the dark pillar and along the jet axis closer
to the NE part of the flow, but it is not clear yet if this is
actually the driving source.

This jet provides an illustrative and striking example of what seems
to be a fairly common phenomenon -- collimated bipolar jets that
emerge from the heads of dust pillars, with an axis orientation
roughly perpendicular to the long axis of the parent dust pillar or
cometary cloud.  Several examples are seen in our study of Carina
(HH~666, 900, 901, 902, 1004, HHc-1, etc.), although some counter
examples also exist where the jet emerges at apparently random angles
(e.g., HH~1011, 1013, 1014, HHc-5).  Another good example of a jet
perpendicular to its cometary globule outside Carina is the remarkable
jet system in L1451 in Perseus, consiting of the chain of HH objects
HH~280, 317, 492, and 493 (Walawender et al.\ 2004).  This is
discussed further in \S 6.3.  The large dark dust pillar from which
HH~1010 emerges points eastward, in the general direction of $\eta$
Car, Tr~16, and the upper parts of Cr~228 (see Smith 2006; Walborn
2002), as do neighboring cometary globules to the south in
Figure~\ref{fig:hh1010}.

\begin{figure}\begin{center}
\includegraphics[width=3.05in]{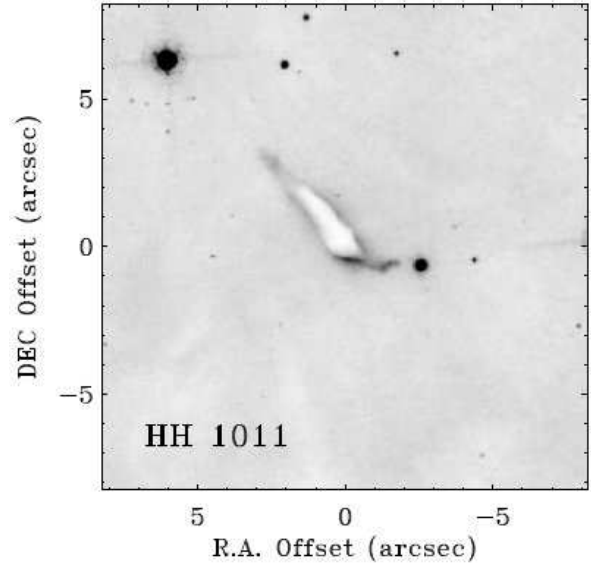}
\end{center}
\caption{ACS/WFC H$\alpha$ image of HH~1011, a microjet from a small
  dark cometary cloud located near Tr~15.}
\label{fig:hh1011}
\end{figure}

\begin{figure}\begin{center}
\includegraphics[width=3.05in]{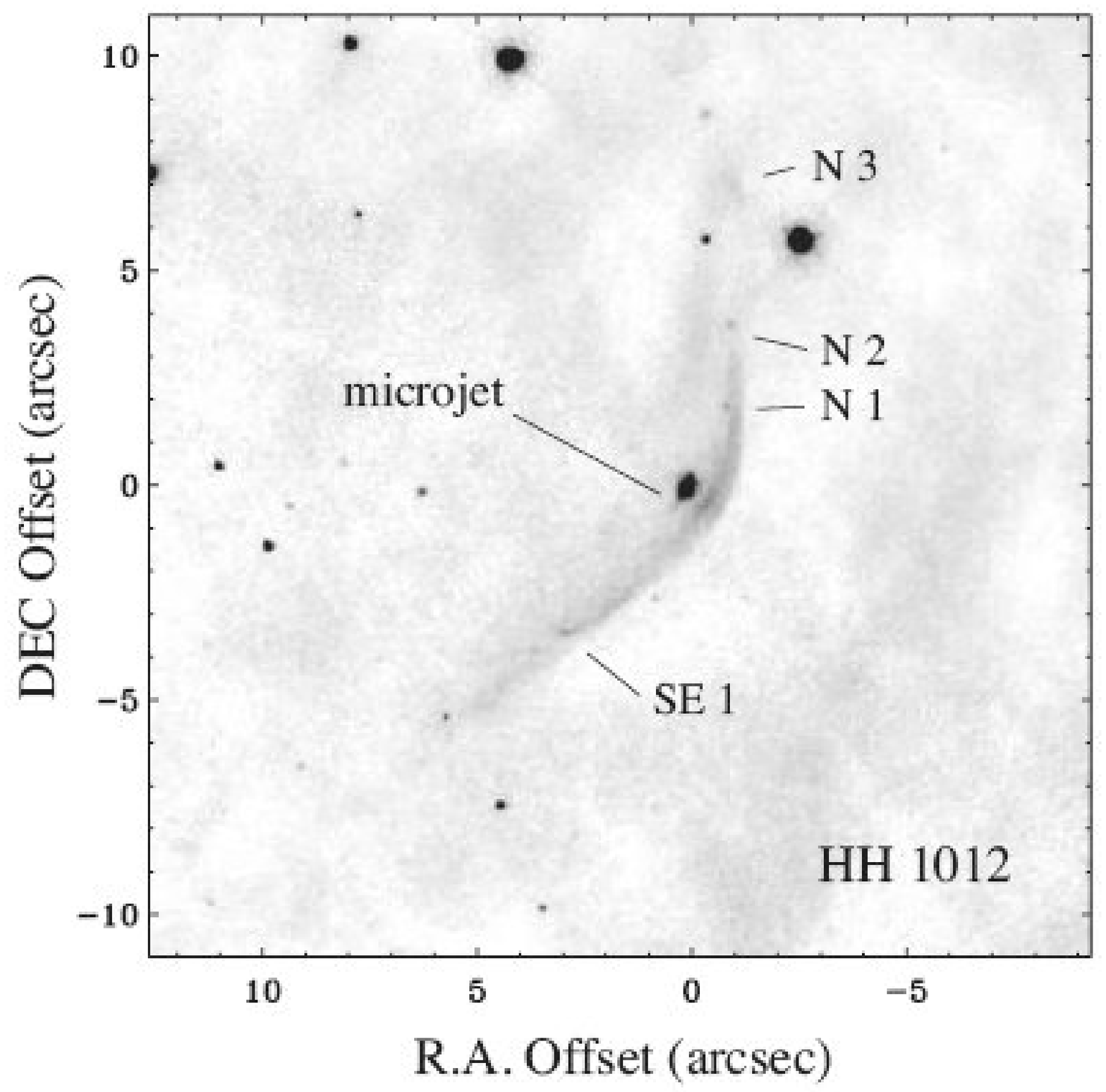}
\end{center}
\caption{The LL~Ori-like object HH~1012 in an H$\alpha$ ACS image.  In
  addition to the larger curved shock structure, one can see shocks
  within the jet and a tiny microjet from the central star.}
\label{fig:hh1012}
\end{figure}

\subsubsection{HH 1011}

HH~1011, shown in Figure~\ref{fig:hh1011}, is a one-sided jet that is
only 2\arcsec\ long (0.02 pc).  It emerges from a small cometary
globule located near the Tr~15 cluster, but like a few dozen other
small cometary clouds in the vicinity (not shown), the head of the
cometary cloud points to the Tr~14 cluster.  The H$\alpha$ images do
not show clear evidence for a bow shock further along the jet axis, nor
do they reveal any evidence for a counter-jet in the opposite
direction.

\subsubsection{HH 1012}

HH~1012, shown in Figure~\ref{fig:hh1012}, is a clear example of the
class of ``LL~Orionis objects'', which exhibit parabolic or C-shaped
shock fronts surrounding visible stars (see Bally \& Reipurth 2001;
Bally et al.\ 2001, 2006).  These parabolic fronts tend to be
associated with jets that are deflected by large-scale bulk flows of
plasma away from the core of an H~{\sc ii} region like Orion
(Masciadri \& Raga 2001), although the jet bending may also be caused
by radiation pressure or the rocket effect if the jet core is neutral
(Bally et al.\ 2006).  In the case of HH~1012, the wings of the
parabolic shock point away from the Tr~14 cluster, located a few
arcminutes southwest, which would be the dominant source of photons or
a bulk flow from its collective stellar winds.

Bally et al.\ (2006) provided a detailed analysis of LL~Ori objects
and the kinematics of their jets, such as the exemplary case of HH~505
in Orion.  HH~1012 has all the hallmark properties of Orion's
prototypical LL~Ori objects like HH~505.  There are clumps within the
parabolic front that resemble internal working shocks in the jets of
other LL~Ori objects, and HH~1012 even appears to harbor a tiny
0$\farcs$5 (0.005 pc) microjet from its central star, at
P.A.$\simeq$140\arcdeg.  The total length of the visible jet in
HH~1012 is $\sim$16\arcsec\ (0.18 pc).

\subsubsection{HH 1013}

\begin{figure*}\begin{center}
\includegraphics[width=5.2in]{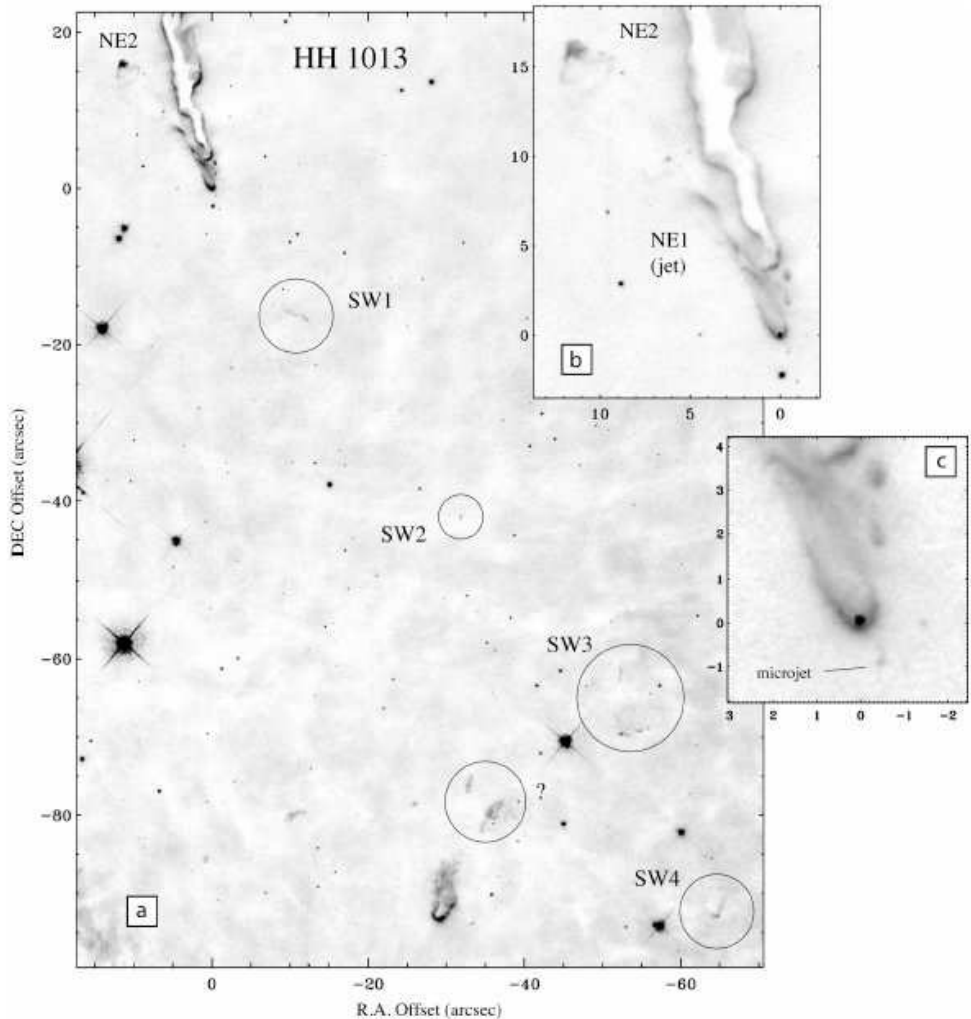}
\end{center}
\caption{(a) This ACS H$\alpha$ image captures one of the long tail
  structures noted by Smith et al.\ (2003), and the long bipolar
  HH~1013 jet associated with it.  The NE outflow of HH~1013 has a
  clear bow shock (NE2) and jet body (NE1) emerging from the bright
  central star in the proplyd at the southern end of the elongated
  cloud structure in the upper left.  Several individual shocks in the
  flow can be seen far to the southwest (SW1-4).  SW4 appears to be
  the terminal bow shock.  The one circled feature marked with a ``?''
  may be a bow shock as well, but it may also be an illuminated cloud
  like many of the other structures in the lower left portion of the
  image.  (b) A detailed view of the proplyd at the end of the cloud,
  containing the star that is the likely driving source of HH~1013, as
  well as the NE flow.  (c) An even more detailed view of the possible
  counter-jet emerging from the tip of the proplyd (labeled as
  ``microjet'').}
\label{fig:hh1013}
\end{figure*}

The outflow HH~1013, shown in Figure~\ref{fig:hh1013}, is an extremely
long jet stretching over 130\arcsec\ or almost 1.5 pc.  The source of
HH~1013 is a proplyd, found near the tip of a long and twisted
cometary cloud structure that is perhaps in a more advanced stage of
evaporation than the similarly twisted dust pillar near HH~1008.  It
is located near the Tr~14 cluster (see Fig.~\ref{fig:color}), and the
long tail points away from Tr~14, which is located a few arcminutes to
the south, off the bottom of Figure~\ref{fig:hh1013}a.  This elongated
cloud was one of the long tail structures found in ground-based images
by Smith et al.\ (2003), along with several proplyd candidates.

\begin{figure*}\begin{center}
\includegraphics[width=4.9in]{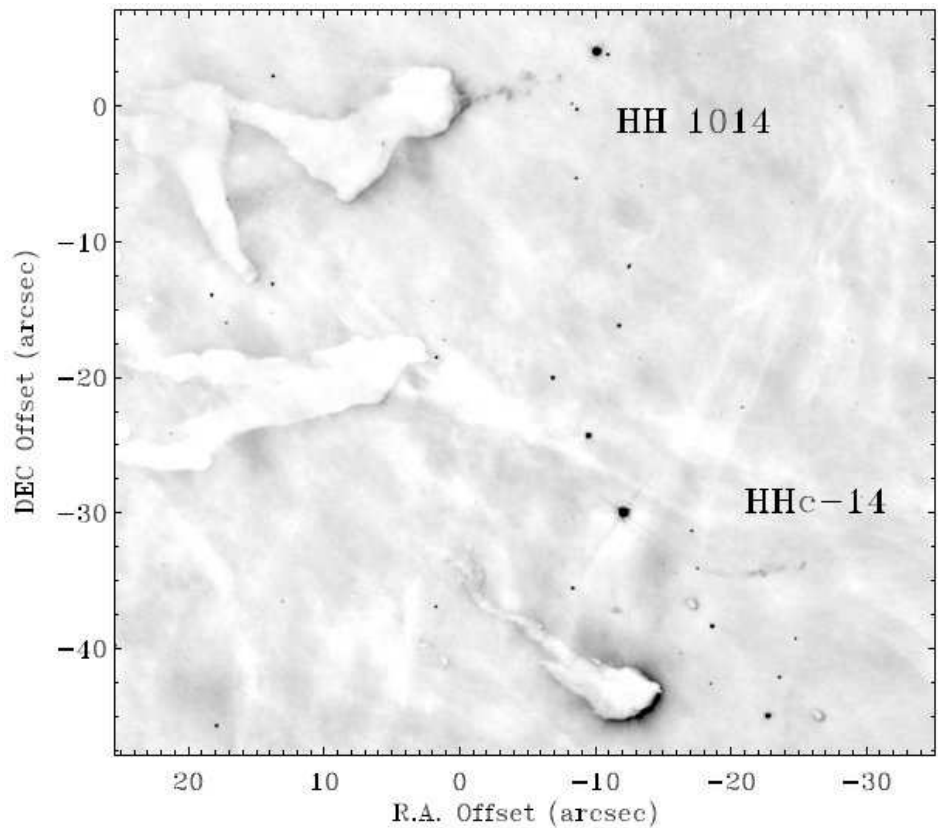}
\end{center}
\caption{HH~1014, located at the eastern edge of the inner Carina
  nebula, is within the Tr~16 mosaic field and emerges to the NW from
  a small pillar head.  The pillar head and the one-sided HH~1014 jet
  point roughly toward $\eta$ Carinae and Tr~16.  The candidate jet
  shocks labeled HHc-14 are also in the field to the lower right, but
  do not lie along the HH~1014 jet axis.}
\label{fig:hh1014}
\end{figure*}

HH~1013 is a bipolar jet that flows to the NE and SW from the proplyd
located at (0\arcsec,0\arcsec) in Figure~\ref{fig:hh1013}.  To the
northeast it is comprised of a clear parabolic bow shock
(HH~1013~NE2), and a collimated but meandering body of the jet (NE1),
shown most clearly in Figure~\ref{fig:hh1013}b. The detailed
morphology of the parabolic bow shock HH~1013~NE2 resembles other
well-studied examples of bow shocks with Mach disks, and is consistent
with the jet direction implied by the collimated jet body to its lower
right.  The driving source of the jet is almost certainly the star
within the proplyd at the head of the cometary cloud.  The length of
the north-east arm of the jet is $\sim$21\arcsec (0.23 pc), with an
axis at roughly P.A.$\simeq$35\arcdeg.  There appears to be a slight
bend in the axis of the NE jet at 7--9\arcsec\ from its origin,
perhaps due to the influence of a photoevaporative flow off the side
of the elongated globule.  In the opposite direction, a chain of
shocks stretches $\sim$110\arcsec\ (1.23 pc) to the southwest, labeled
as SW1-4.  SW4 is probably the terminal bow shock in the flow.
Several other nebular features are seen in the field, in the lower
left and center of Figure~\ref{fig:hh1013}a, but upon close inspection
most of these appear to be irradiated clouds.

The driving source of HH~1013 is significant, because it is an
unambiguous example of a true analog of the proplyds in the Orion
Nebula.  Most structures that resemble the proplyds in Carina and
other H~{\sc ii} regions appear to be dark cometary clouds upon closer
inspection.  (Some of these nevertheless appear to contain embedded
stars with active accretion and outflows, as described elsewhere in
this paper.)  A detail of this proplyd is shown in
Figure~\ref{fig:hh1013}b.  Although we do not resolve a small dark
silhouette disk inside the proplyd, its tadpole shape arises because
of a proplyd outflow rather than a from an opaque cometary cloud,
because the internal star is seen clearly.

Like many of Orion's proplyds, this one associated with HH~1013
appears to have a tiny microjet only 1\arcsec\ ($\sim$0.01) long,
which protrudes from the head of the proplyd in the direction opposite
from of the flow that terminates with HH~1013~NE2 (see
Figure~\ref{fig:hh1013}).  This microjet is the start of the
counterflow that produces the SW1--4 structures.

\subsubsection{HH 1014}

The one-sided jet HH~1014 is a collimated chain of emission knots
emerging from the head of a dark globule, seen in the ACS/WFC
H$\alpha$ image in Figure~\ref{fig:hh1014}.  These emission knots are
likely photoionized condensations arising from internal working
surfaces in the flow.  No clear bow shock structure can be identified,
nor do the images reveal any obvious counter-jet.  No driving source
is seen at visual wavelengths; it is probably embedded within the dark
cloud.

The jet stretches roughly 8\arcsec\ (0.09 pc) westward along
P.A.$\simeq$290\arcdeg.  This is only a few degrees different from the
axis of the elongated dust pillar from which it emerges, and this is
also roughly the direction toward $\eta$ Carinae.  HH~1014 is found at
the eastern edge of the bright inner part of the Carina Nebula, within
the Tr~16 mosaic (see Fig.~\ref{fig:color}) and along the dark
molecular ridge not far from HH~900.  $\eta$~Car is located a few
arcminutes to the right.  The nearby shock structure HHc-14
(Figure~\ref{fig:hh1014}) is probably part of a different flow because
it is not along the axis of HH~1014.

\subsubsection{HH 1015}

See \S 3.2.7.

\subsubsection{HH 1016}

\begin{figure}\begin{center}
\includegraphics[width=3.1in]{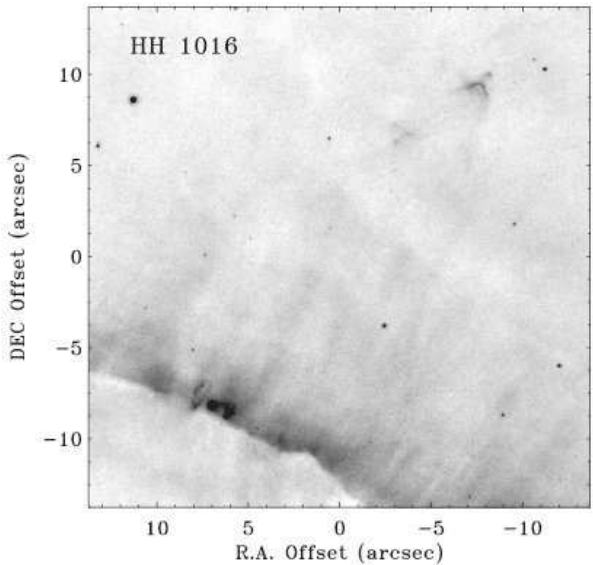}
\end{center}
\caption{An H$\alpha$ image of HH~1016, located about 30\arcsec\ north
  of the Treasure Chest cluster (Smith et al.\ 2005c), and emerging
  from the top of the same dust pillar that contains the cluster.
  There is a parabolic bow shock in the upper right of this image, and
  one can see filaments in the lower left where the jet breaks out
  through the ionization front.  The broad band of emission running
  diagonally through the image is the photoevaporative flow from the
  surface of the dust pillar.}
\label{fig:hh1016}
\end{figure}

HH~1016 (Figure~\ref{fig:hh1016}) is a one sided jet about 25\arcsec\
(0.3 pc) in length, emerging along P.A.$\simeq$320\arcdeg\ from the
top of the dust pillar that harbors the Treasure Chest cluster (Smith
et al.\ 2005c).  The dust pillar is the most luminous mid-IR source in
the South Pillars (Smith et al.\ 2000), while the Treasure Chest
cluster itself is extremely young ($\sim$10$^5$ yr) with a very high
YSO disk excess fraction (Smith et al.\ 2005).  The embedded source
that drives HH~1016 is probably a member of this young cluster.  The
counter jet is invisible, probably because it burrows into the dust
pillar.

The bow shock of HH~1016 is seen in the upper right corner of
Figure~\ref{fig:hh1016}.  It has clear parabolic bow-shock structure.
The bright Mach disk at its apex also emits strong [Fe~{\sc ii}] 1.644
$\mu$m emisson, which can be seen in hindsight in narrowband near-IR
images (Smith et al.\ 2005c).

The driving source of the jet cannot be clearly seen, and there is no
clear feature that one can attribute to the collimated inner jet of
this flow.  However, one can see bright filaments of emission at the
point where the jet breaks through the ionization front, at roughly
(6\arcsec,--8\arcsec) and (8\arcsec,--7\arcsec) in
Figure~\ref{fig:hh1016}.  These breakout filaments resemble similar
features where other HH jets emerge from their embedded sources, as in
HH~900, HH~903, HH~1003~A, HH~1004, HHc-2, etc.  The two filaments are
on either side of a faint star, which may or may not be the driving
source of the outflow.

\begin{figure}\begin{center}
\includegraphics[width=3.1in]{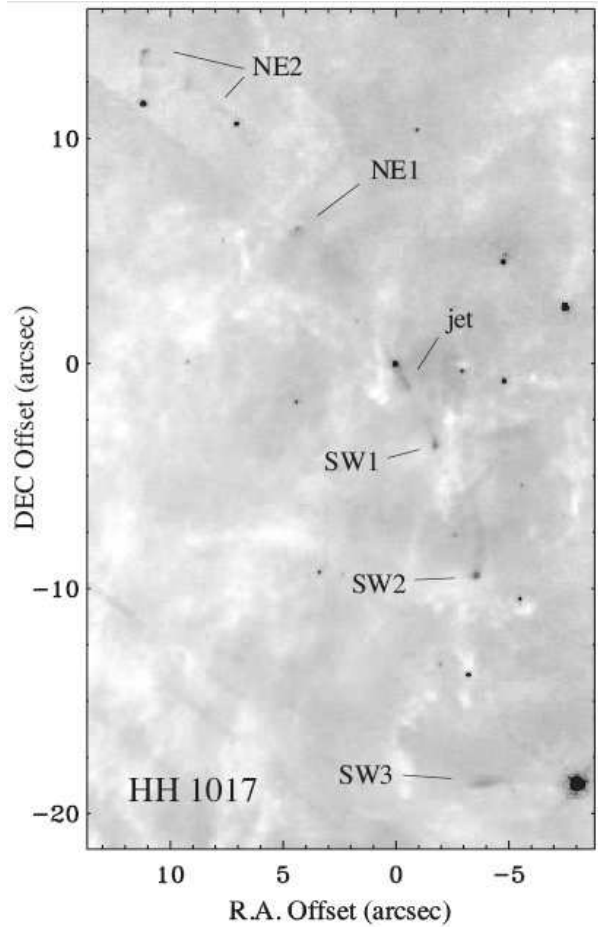}
\end{center}
\caption{The long bent jet HH~1017 in H$\alpha$.  The source of the
  jet is the star at (0,0).  A collimated jet body is seen emanating
  from this star toward the south-west.  Several shocks to the
  north-east and south-west can also be seen.}
\label{fig:hh1017}
\end{figure}
\begin{figure}\begin{center}
\includegraphics[width=3.1in]{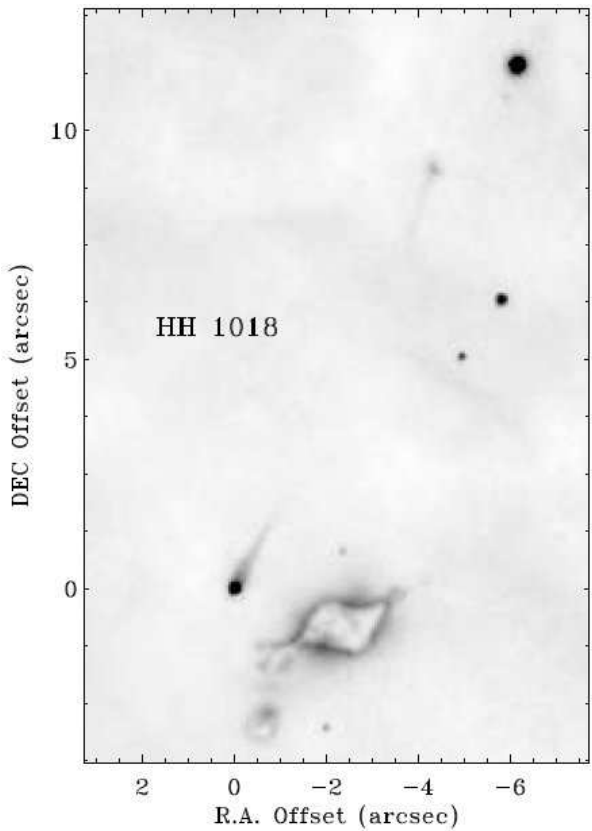}
\end{center}
\caption{The microjet HH~1018 in H$\alpha$, located south of $\eta$
  Car in Tr~16.  The source of the jet is the star at (0,0).  A
  collimated jet body is seen emanating from this star.  A shock
  structure farther along the same axis can also be seen.}
\label{fig:hh1018}
\end{figure}

\subsubsection{HH 1017}

HH~1017, shown in Figure~\ref{fig:hh1017}, is a bent bipolar
jet located east of the Tr~14 cluster (see Figure~\ref{fig:color}).
Tr~14 is located off the right side of Figure~\ref{fig:hh1017}, and
HH~1017 bends away from Tr~14, as one might expect of radiation and
winds from the extreme O stars in Tr~14 are pushing the jet.

Unlike most HH jets we have found in Carina, HH~1017 does not emerge
from an invisible source embedded deep within a dark dust pillar, but
instead comes from an exposed star with no dark clouds nearby.  A
collimated jet body runs about 4\arcsec\ (0.05 pc) to the southwest
from this driving source, while several other bow shocks in the flow
can be seen.  These shocks trace out a gracefully bending jet with a
full length over 35\arcsec\ (0.4 pc).  HH~1017 is reminiscent of
several graceful bent jets in Orion, especially HH~502 (Bally et al.\
2006).  The average direction of the flow is P.A.$\sim$25\arcdeg, but
obviously the flow direction changes with distance from the source.

\subsubsection{HH 1018}

HH~1018 (Figure~\ref{fig:hh1018}) is a microjet in the central Carina
Nebula, located within the Tr~16 cluster not far from $\eta$~Car and
HH~900 (see Figure~\ref{fig:color}).  Like HH~1017, the jet originates
from an exposed star, although here there is a small dark globule
projected nearby. HH~1018 is a one-sided jet, with no sign of features
associated with a counter jet in a bipolar flow.  The microjet is
about 2\arcsec\ (0.02 pc) long, but there is also an additional shock
structure located $\sim$10\arcsec\ (0.11 pc) along the same axis.

The central star and its microjet may resemble some proplyds at first
glance.  However, we have discounted this possibility, since the
proplyd tail would point in the wrong direction, i.e. the proplyd tail
should point {\it away} from the massive O-type stars in Tr~16, but it
points toward them.  This fact, combined with the additional emission
knot along the same axis, argues that HH~1018 is a jet.

\subsection{Jets in NGC~3324}

\begin{figure*}\begin{center}
\includegraphics[width=6.0in]{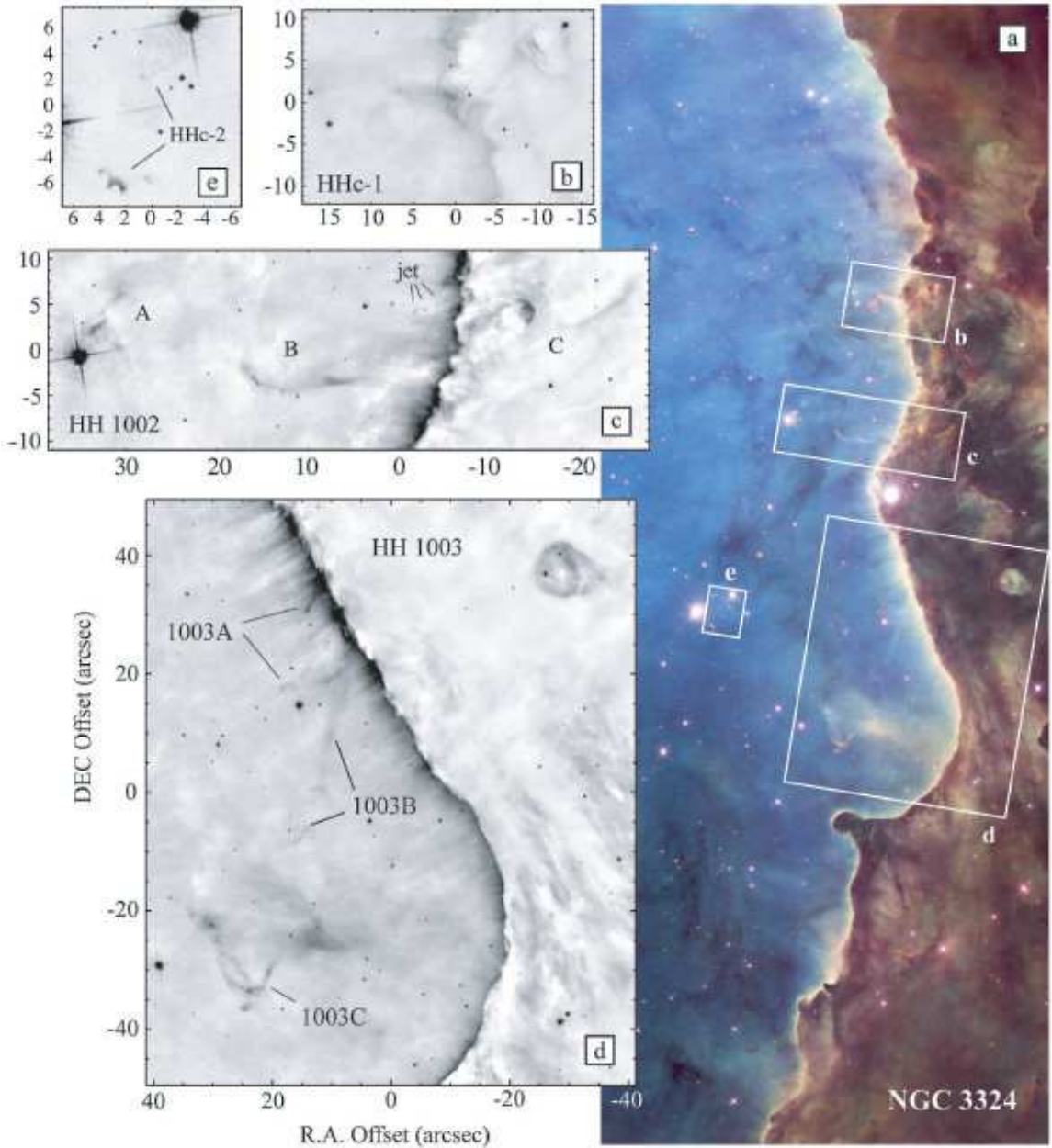}
\end{center}
\caption{Outflows in the neighboring H~{\sc ii} region NGC~3324.  (a)
  The color image is a {\it Hubble Heritage} image composed of one ACS
  H$\alpha$ footprint (green), plus the same area covered with
  additional pointings using the WFPC2 camera in the [O~{\sc iii}]
  $\lambda$5007 F502N filter (blue) and [S~{\sc ii}]
  $\lambda\lambda$6717,6731 F673N (red).  The WFPC2 images in these
  two additional filters were obtained by the {\it Hubble Heritage}
  team.  (The axes of the color image are rotated slightly with
  respect to R.A.\ and DEC.)  (b) HH candidate jet 1, a [S~{\sc
    ii}]-bright feature that apparently protrudes from the ionization
  front.  (c) HH~1002 protrudes eastward from the edge-on ionization
  front. (d) HH~1003 has several parts, not necessariy all part of a
  single flow.  HH~1003A appears to be a collimated jet plus internal
  working surfaces.  HH~1003B may be side walls of the outflow or a
  separate flow.  HH~1003C is a clear bow-shock plus Mach disk
  feature.  The structure surrounding the star in the upper right
  corner of panel (d) may be a compact H~{\sc ii} region or wind-blown
  bubble from a young star behind the ionization front.  The nebular
  filaments in (e) may be shocks in an outflow, but no clear source or
  collimated jet can be identified so they are designated as part of a
  candidate outflow HHc-2.  It is not clear if they are associated
  with any other jets in the image.}
\label{fig:n3324}
\end{figure*}

We imaged a small portion of the adjacent H~{\sc ii} region NGC~3324
at its sharp edge-on ionization front at the western side of the
nebula using the H$\alpha$ filter on ACS/WFC.  Subsequently, the same
region was imaged with {\it HST}/WFPC2 in the [O~{\sc iii}]
$\lambda$5007 F502N and [S~{\sc ii}] $\lambda\lambda$6717,6731 F673N
filters by the {\it Hubble Heritage} team, and the resulting color
image of the region is shown in Figure~\ref{fig:n3324}a.\footnote{The
  process by which these images were combined is described on the
  Hubble Heritage website: {\tt
    http://heritage.stsci.edu/2008/34/index.html}.} The resulting data
reveal multiple outflows based on their collimated or bow shock
morphology and their high [S~{\sc ii}]/H$\alpha$ ratios.  Two objects
(HHc-1 and HHc-2) remain candidate jets, while another (HH~1003) has
several parts that may be more than one outflow.  In all three cases
discussed below, the driving source is not identified at visual
wavelengths.

\subsubsection{HH~1002}

Figure~\ref{fig:n3324}c shows HH~1002, which is an irradiated
collimated jet about 50\arcsec\ long (0.56 pc). It has a somewhat
asymmetric or irregular bow shock, with features located at
(+15\arcsec,--4\arcsec) and (+33\arcsec,2\arcsec) in
Figure~\ref{fig:n3324}c.  It is possible that there may actually be
two separate flows in Figure~\ref{fig:n3324}c, since there seem to be
two separate bow shocks, and possibly also two separate collimated
jets.  For example, the feature labeled ``jet'' in
Figure~\ref{fig:n3324}c resembles a series of small shocks in a
collimated outflow, but it is off axis compared to some of the other
features.

The average jet axis is roughly along P.A.$\simeq$106\arcdeg, and
protrudes nearly perpendicular to the main ionization front, aimed
eastward into the interior cavity of NGC~3324.  The driving source is
not seen.  There is some irregular structure west of the ionization
front, including a cometary structure that may represent the position
where the jet first breaks out of the dark cloud.  If so, one would
suspect the IR driving source to be located at roughly
(--16\arcsec,+5\arcsec) in Figure~\ref{fig:n3324}c, or $\alpha_{2000}$
= 10:36:54.0, $\delta_{2000}$ = --58:37:19.  There is a red 2MASS
source at this position, but deeper IR images with better angular
resolution are needed for a confident comparison with the {\it HST}
data.  No clear shock structures are seen outside the region included
in Figure~\ref{fig:n3324}c, nor is there a likely counter-jet visible
in these images.

\subsubsection{HH~1003}

Figure~\ref{fig:n3324}d shows the region containing HH~1003.  The most
prominent structure in this outflow is a large ($\sim$10\arcsec) and
well-defined bow shock that we denote as HH~1003C in
Figure~\ref{fig:n3324}d.  It shows a clear parabolic structure common
to the main shocks in well-studied HH jets like HH~34, including
detailed morphology consistent with a Mach disk near its apex.  The
main body of the jet that powers this bow shock is less clearly
defined.  Projecting back (north) along a line that bisects the bow
shock, one sees structures consistent with a dense collimated jet body
and collimated filaments that we denote HH~1003A in
Figure~\ref{fig:n3324}d.  Bright structures at the base connect back
to the ionization front, perhaps marking the position where this flow
breaks out of the molecular cloud.  If this is the base of the jet,
then the axis that connects it to the center of the bow shock runs
along P.A.$\simeq$171\arcdeg.  The group of filaments marked as
HH~1003B may be part of the same flow (perhaps density enhancements at
the sides of the outflow), a separate outflow, or unrelated filaments
in the H~{\sc ii} region that happen to lie near the axis of HH~1003.

We also draw attention to an interesting feature in the upper right
corner of Figure~\ref{fig:n3324}d.  In ground-based images with good
seeing this structure resembles other proplyd candidates with a star
at its apex, but its morphology in the {\it HST} images is not
consistent with other proplyds.  Instead, it may be a compact young
H~{\sc ii} region or wind-blown cavity around the star at its eastern
edge.  This star is located at $\alpha_{2000}$=10:36:49.00,
$\delta_{2000}$=$-$58:38:04.9.  The surrounding diffuse emission has
strong H$\alpha$ emission compared to [S~{\sc ii}] and [O~{\sc ii}],
suggesting photoionization by a relatively soft UV field -- perhaps an
early/mid B star.  It would be interesting to obtain a classification
spectrum of the central star, as it may represent a stellar wind
bubble around a relatively isolated young B star still within its
natal cloud.

\subsubsection{HHc-1 in NGC~3324}

The structure shown in Figure~\ref{fig:n3324}b resembles the main jet
bodies of some other irradiated outflows that break out of edge-on
ionization fronts.  However, it appears to be less well-collimated
than other features associated with jet bodies, and it lacks any clear
evidence for a bow shock downstream.  Its color indicates a strong
[S~{\sc ii}]/H$\alpha$ ratio similar to other dense jets, but one
cannot be confident that this is an outflow from a young star.
Complex structures in dense photoevaporative flows at ionization
fronts can sometimes create features like this.  Note, for example,
the striated features along the ionization front in
Figure~\ref{fig:n3324}d.  Therefore, we classify it as a jet candidate
in Table 3, worthy of followup spectroscopy to study its kinematics or
IR observations to search for an embedded IR source.

\subsubsection{HHc-2 in NGC~3324}

The nebular structures shown in Figure~\ref{fig:n3324}e constitute
HHc-2 in NGC~3324.  In the lower left of this panel is a connected
group of condensations that may be knots within a flow.  In the upper
right of Figure~\ref{fig:n3324}e is a much fainter structure that is
almost lost in the diffraction pattern of a bright neighboring star,
and its structure resembles a bow shock with the opening of the
parabola pointed at P.A.$\simeq$320\arcdeg.  If this represents a jet
axis it points back roughly toward HH~1002 or HHc-1, although one
cannot unambiguously associate it with either of those jets.  The
color of HHc-2 in the 3-color image is similar to the other HH jets
and candidates in the same image, suggesting that these are
photoionized nebular structures, possibly in a jet.  Kinematics are
needed before further conclusions can be drawn, however.

\subsection{Comments on Individual Jet Candidates}

As noted earlier, we identify over a dozen candidate jets that for one
reason or another lacked structures that were clear enough to identify
them as young star outflows based on morphology in images alone.  We
list these candidates in Table 3 as they provide good targets for
followup spectroscopy to confirm or refute their nature as true jets.
HHc-1 and HHc-2 in NGC~3324 were already discussed in the previous
section; HHc-1 in Carina is discussed next.

\begin{figure}\begin{center}
\includegraphics[width=2.7in]{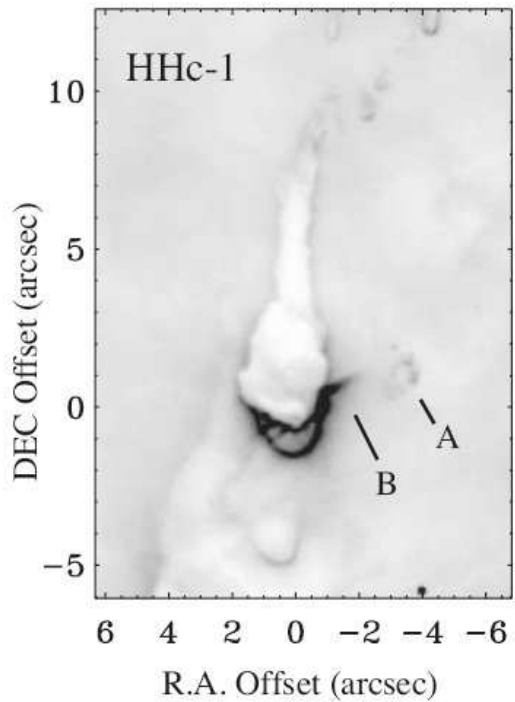}
\end{center}
\caption{Candidate jet HHc-1 in Carina (this object is different from
  HHc-1 in NGC~3324).  It is near HH~901 and 902, and can also be seen
  in Figure~\ref{fig:hh901}. The candidate jet extends to the west
  from a cometary globule that points toward Tr~14.  Feature A may be
  a bow shock or working surface, whereas feature B appears to be the
  main body of the jet.  Possible collimated features appear on the
  other side of the globule as well.}
\label{fig:hhc1}
\end{figure}

\begin{figure}\begin{center}
\includegraphics[width=3.0in]{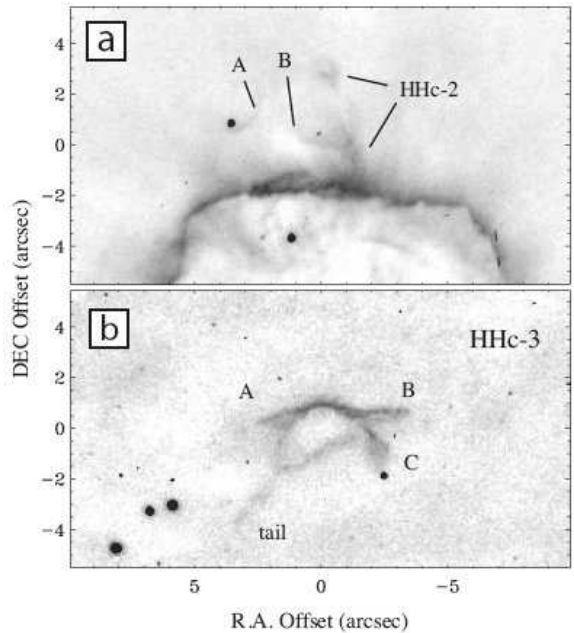}
\end{center}
\caption{(a) Candidate jet HHc-2 in the South Pillars of Carina.  This
  is a detail of the larger image seen in Figure~\ref{fig:hh1008}.
  The possible jet HHc-2 is marked, emerging northward out of the
  pillar head.  Two other features identified as wind bow shocks,
  identified as A and B, are discussed in the text.  (b) Candidate jet
  HHc-3, with three possible collimated jets labeled A, B, and C
  emerging from a small cometary cloud.  This image is just off the
  lower right of Figure~\ref{fig:c45678}.}
\label{fig:hhc23}
\end{figure}

\subsubsection{HHc-1 in Carina}

Carina HHc-1 is shown in Figure~\ref{fig:hhc1}, which is a detail of
the larger field around HH~901 and HH~902 seen in
Figure~\ref{fig:hh901}.  The evidence for a jet in this image includes
a possible collimated jet body emerging westward from the cometary
cloud (feature B in Figure~\ref{fig:hhc1}), and structures that
resemble a small bow shock or working surface in the jet (feature A).
The position angle of this western portion of the jet is
$\sim$300\arcdeg, and the jet is $\sim$4\arcsec\ (0.05 pc) long.
There are also two spikes of emission on the east side of the cometary
cloud, either of which may be a counter-jet in a bipolar flow.

Interestingly, there is a very faint point source along the putative
jet axis, within 0$\farcs$2 of the tip of the cometary cloud, located
at position (0\arcsec,0\arcsec) in Figure~\ref{fig:hhc1}.  If HHc-1 is
a true jet, then the driving source is likely to be this star embedded
within the small cometary cloud.  This is significant because, like
HH~900, this dark cloud was one of the original proplyd candidates in
Smith et al.\ (2003).  While the object does not appear to be a
photoevaporating disk like the proplyds in Orion, the jet candidate
HHc-1 suggests that it does, in fact, contain an embedded prototstar
that is still actively accreting and driving an outflow.  This adds
further evidence that the large number of tiny cometary clouds
suspended within the H~{\sc ii} region cavity in Carina, as well as
those in other regions like the similar cometary clouds in NGC~3603
(Brandner et al.\ 2000), are important potential sites of ongoing star
formation.

\subsubsection{HHc-2}

HHc-2 is shown in Figure~\ref{fig:hhc23}a; again, note that this
object is different from HHc-2 in NGC~3324.  It is a concentration of
nebulosity elongated northward from the head of a long and twisted
pillar/cometary cloud in the South Pillars of Carina.  This is the
same pillar head which we suspect may contain the driving source of
HH~1008 (the larger structure is shown in Figure~\ref{fig:hh1008}),
which is a large bow shock located $\sim$105\arcsec\ to the south.
Emerging from the same pillar head in the opposite direction, HHc-2
may therefore be the counterflow to HH~1008, although this is quite
speculative because the morphology of HHc-2 is not clearly that of a
jet.

Interestingly, Figure~\ref{fig:hhc23}a also shows two clear parabolic
shock structures, each surrounding a star (objects A and B in
Figure~\ref{fig:hhc23}a).  Their morphology and direction suggests
that they are not jet bow shocks, but rather, stand-off shocks marking
the collision between the winds of their stars and the
photoevaporative flow off the nearby pillar head.  Other examples of
such structures are seen in our data, but they are not the main topic
here.

\subsubsection{HHc-3}

\begin{figure*}\begin{center}
\includegraphics[width=5.8in]{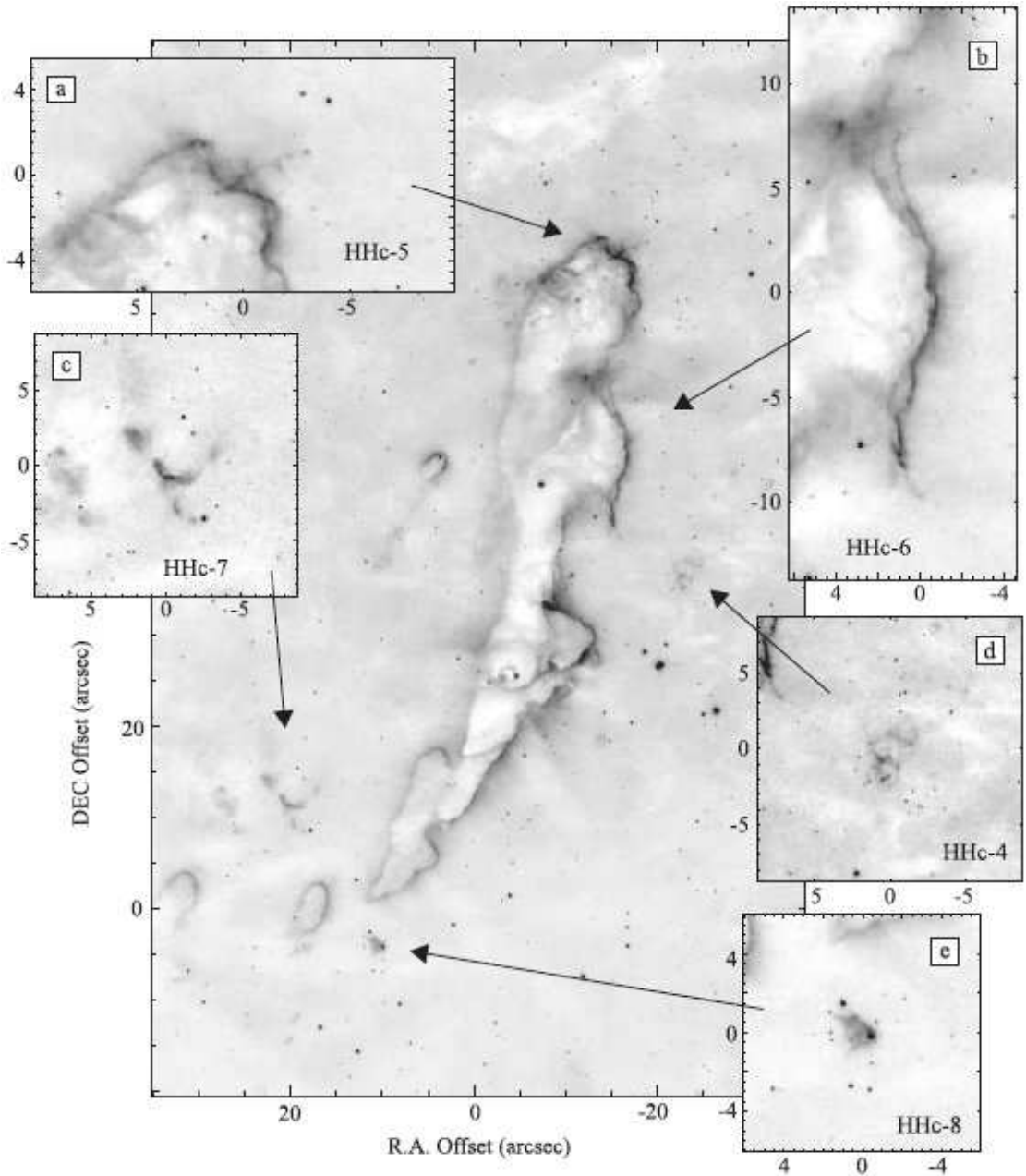}
\end{center}
\caption{Several HH jet candidates associated with a complex of dense
  molecular globules in the South Pillars.  (a) Carina HHc-5 appears
  to be a collimated chain of ionized knots protruding from the head
  of the globule.  No counter-jet is seen.  (b) HHc-6 may be a bipolar
  jet flowing north and south from a clump on the west edge of the
  globule.  The candidate jet flows nearly parallel to the edge-on
  ionization front and appears to be swept back (curved) toward the
  globule.  (c) HHc-7 is a bow shock with an unclear origin.  It may
  be associated with the proplyd-like object located 30\arcsec\
  north-west. (d) HHc-4 is also a bow-shock struture with an unclear
  origin. (e) HHc-8 is an LL-Ori or proplyd-like object with cometary
  shape surrounding a bright central star.}
\label{fig:c45678}
\end{figure*}

HHc-3 in Figure~\ref{fig:hhc23}b has an unusual appearance.  One can
clearly see a cometary cloud with a long and faint tail pointing
southeast.  There are three protrusions from the head of this cometary
cloud, marked A, B, and C, which are candidate jets.  Were it not for
the fact that there are {\it three} protrusions, one might naturally
claim that A and B (or alternatively, A and C) constitute a bipolar
jet from a star within the cometary cloud, much like the more
spectacular example of HH~900.  If A and B are a bipoalr jet, then the
jet is fairly straight running east/west, but if A and C are a pair,
then the jet is severely bent backward in a direction consisent with
being bent by the same mechanism that shapes the cometary cloud and
its tail.  Either option seems plausible, as does the possibility that
there are two separate flows.  Spectra are needed to trace the
kinematics of these features.  (N.B.: HHc-3 is located very close on
the sky to HHc-4, 5, 6, 7, and 8 discussed in the next section.  It is
just off the lower right corner of the main image in
Figure~\ref{fig:c45678}.)

\subsubsection{HHc-4, 5, 6, 7, and 8}

In ground-based images, the elongated silhouette globule shown in the
main frame of Figure~\ref{fig:c45678} is a relatively inconspicuous
dark clump.  The high angular resolution afforded by {\it HST},
however, reveals evidence for bustling outflow activity from young
stars embedded in it and immediately around it.  Altogether, this
activity suggests that this complex of silhouette globules likely
represents the last vestiges of the dense cloud core at the head of a
large dust pillar akin to others seen in the South Pillars or the
Eagle Nebula (Hester et al.\ 1996) --- although in this case, most of
the surrounding dust pillar has already been photoevaporated away
leaving only the densest parts as a cometary cloud complex amid newly
exposed young stars. While there are many indications of likely
outflow activity here, for each individual flow we prefer to await
spectroscopic confirmation of their jet nature before assigning HH
numbers, so we include likely outflows as candidate jets in Table 3.

HHc-4 in Figure~\ref{fig:c45678}d resembles the bow shocks seen in
many well-studied HH jets.  The parabola of the putative bow shock
opens to the north, but in that direction the {\it HST} images show no
clear evidence for other emission nebulosity associated with the body
of the jet, and the driving source of the outflow is not clear.  One
might suspect its driving source to be an exposed young star about
10--20\arcsec\ west of the head of the complex of dark clouds.

HHc-5 in Figure~\ref{fig:c45678}a is a collimated chain of clumps
about 4\arcsec\ (0.05 pc) in length, stretched along
P.A.$\simeq$298\arcdeg.  Similar to HH~1015, it seems to emerge from
the northernmost tip of the complex of dark clouds in
Figure~\ref{fig:c45678}. No clear bow-shock is seen at larger
distances along the same axis.

HHc-6 in Figure~\ref{fig:c45678}b may be a bipolar jet much like
HH~902 or HH~1010, bursting out from either side of the curved surface
of the head of a pillar or cometary cloud.  The putative jet has a
total length of about 18\arcsec\ (0.2 pc).  It runs nearly parallel to
the curved limb-brightened edge of the cometary cloud, and because of
this, it is difficult to be certain that it is in fact a jet and not a
sharp edge-on ionization front.  The putative jet axis is bent back
toward the cloud, reminiscent of the many bent jets in Orion (Bally et
al.\ 2006), so the P.A. listed in Table 3 is very rough.  The case
that this is in fact a jet would be strengthened if a luminous
embedded IR source were detected near the edge of the ionization front
along the jet axis near position (0\arcsec,0\arcsec) in
Figure~\ref{fig:c45678}b.

HHc-7 in Figure~\ref{fig:c45678}c is a relatively isolated bow-shock
structure similar to HHc-4, with the opening of the parabolic shock
structure pointing to the north.  However, in this case there is some
intriguing (although unclear) evidence of a possible jet and source.
Looking 30--40\arcsec to the northeast, one sees a faint chain of
aligned filaments that may lead back to a star within a curved
ionization structure that resembles some of the large proplyds in
Orion --- especially HH~668 and its source (Smith et al.\ 2005b).
This possible source of the HHc-7 jet is located at offset position
(+7\arcsec,+48\arcsec) in the main panel of Figure~\ref{fig:c45678}.

Lastly, HHc-8 shown in Figure~\ref{fig:c45678}c is an LL~Ori-like
object.  No clear jet structure is seen, but the class of LL~Ori
objects have been linked to jet outflows that are bent by a side wind,
radiation pressure, or the rocket effect (Bally et al.\ 2006).  This
cometary-shaped nebular structure implicates a radiation or wind
source located to the west, and it has a bright star at its apex.

\begin{figure}\begin{center}
\includegraphics[width=2.9in]{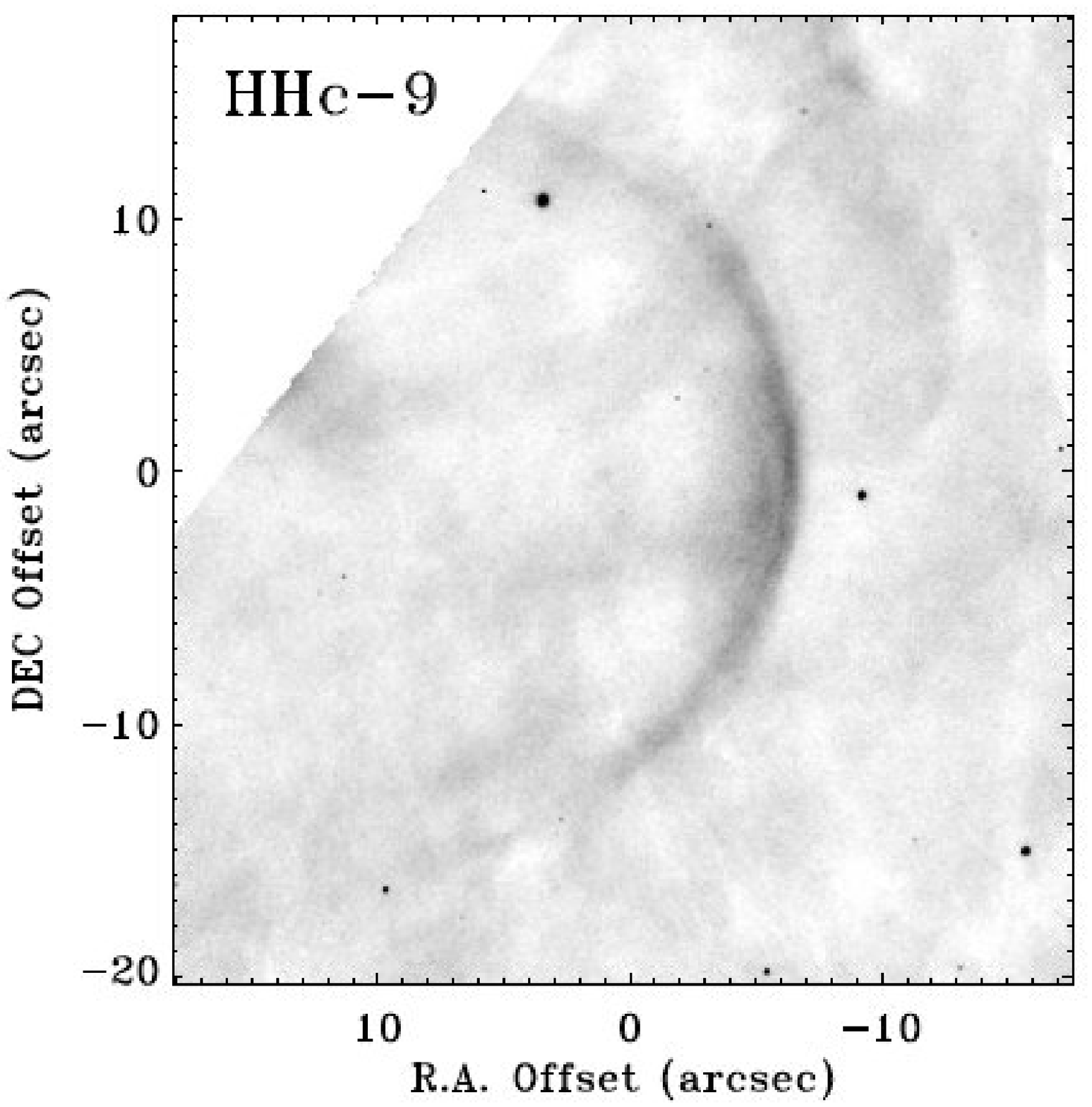}
\end{center}
\caption{Candidate jet bow shock HHc-9 located near the Tr~14 cluster,
  which is off the far right of the image.  This object fell near the
  edge of the large Tr~14 mosaic; no body of the jet or driving source
  can be seen; they would be off the image.}
\label{fig:hhc9}
\end{figure}
\begin{figure}\begin{center}
\includegraphics[width=3.1in]{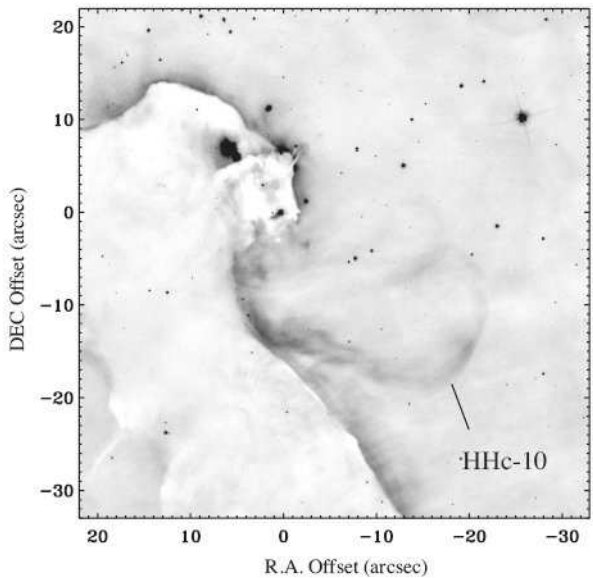}
\end{center}
\caption{Candidate outflow HHc-10, associated with the head of one of
  the most prominent dust pillars in Carina.  This is a detail of the
  field north of HH~903 (see Figure~\ref{fig:hh903}).}
\label{fig:hhc10}
\end{figure}
\begin{figure}\begin{center}
\includegraphics[width=2.8in]{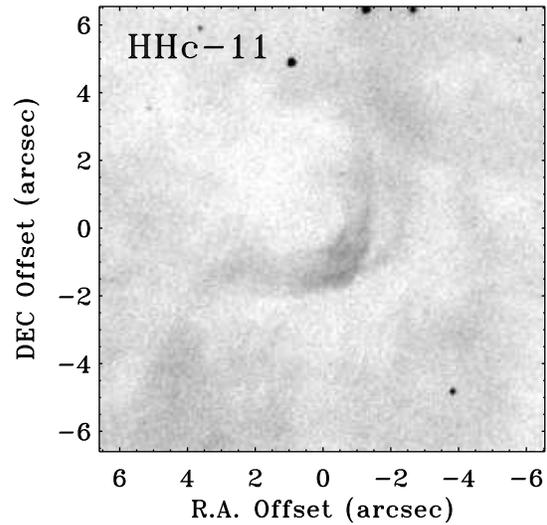}
\end{center}
\caption{Candidate jet bow shock HHc-11 amid the South Pillars in
  Carina.  No collimated jet or driving source is seen.}
\label{fig:hhc11}
\end{figure}
\begin{figure}\begin{center}
\includegraphics[width=2.8in]{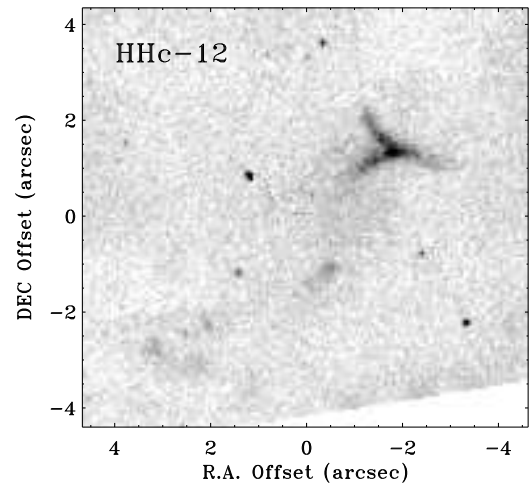}
\end{center}
\caption{Candidate shock structure HHc-12 amid the South Pillars in
  Carina.  No collimated jet or driving source is seen, although this
  structures are located about 30\arcsec\ southeast of HHc-3.  It is
  also located just off the bottom of the image in
  Figure~\ref{fig:c45678}, containing the complex of globules housing
  jet candidates HHc-4, 5, 6, 7, and 8.}
\label{fig:hhc12}
\end{figure}
\begin{figure}\begin{center}
\includegraphics[width=2.8in]{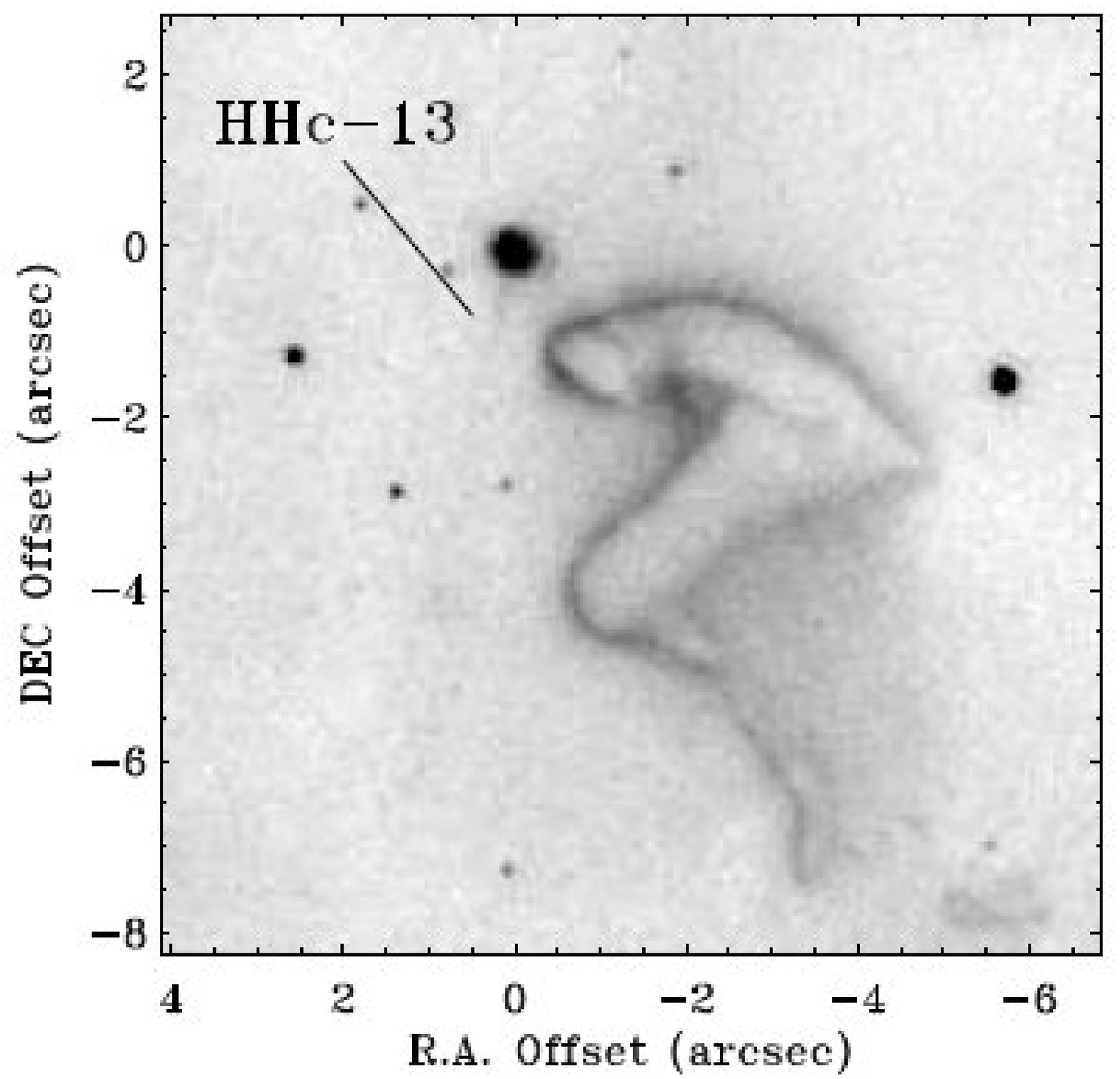}
\end{center}
\caption{Candidate microjet HHc-13, from a star located about
  1\arcmin\ NE of the origin of HH~666 (see Figure~\ref{fig:hh666}).}
\label{fig:hhc13}
\end{figure}

\subsubsection{HHc-9}

HHc-9 shown in Figure~\ref{fig:hhc9} is a large and smoothly curved
bow shock structure.  There is tenuous evidence for a Mach disk at its
apex, implying that the jet direction is due west.  This would also
imply that any hypothetical collimated jet and driving source, not
detected here, would be off the left edge of the field, since HHc-9
fell near the edge of the detector at the boundary of the large mosaic
including Tr~14.  HHc-9 could also plausibly be a density enhancement
in a stellar wind bubble that is illuminated by Tr~14.

\subsubsection{HHc-10}

HHc-10 is shown in Figure~\ref{fig:hhc10}.  It is a series of curved
shock structures associated with the head of a very large dust pillar
that also contains the driving source of HH~903, located about
90\arcsec\ south.  An image of the larger field including HH~903 is
shown in Figure~\ref{fig:hh903}. The head of this pillar appears to
have active star formation, including an embedded cluster that is just
beginning to emerge.  There is likely to be outflow activity
associated with these sources.  If HHc-10 is a group of outflows of
this type, the flows appear to be deflected toward the south by
radiation pressure or a large-scale flow in the H~{\sc ii} region
($\eta$ Car and Tr~16 are located to the north).  However, because of
this complexity and because of the lack of a clear collimated jet
body, these features could also be photoevaporative flows off the
surface of the pillar head that are deflected and shaped by the same
mechanism.  Therefore, HHc-10 is a candidate outflow in Table~3.

\subsubsection{HHc-11}

HHc-11, shown in Figure~\ref{fig:hhc11}, is another case of an orphan
bow shock structure without clear evidence for a collimated jet and no
good candidate for a driving source.  It is located several arcminutes
south of HH~1006.  While its location may be consistent with being
part of that flow if the flow has some bend in its axis, the line of
symmetry through the arc in HHc-11 does not point back toward HH~1006,
so an association with this jet seems unlikely.

\subsubsection{HHc-12}

Another orphan shock structure, HHc-12 in Figure~\ref{fig:hhc12},
shows two curved shocks plus a collection of fainter emission knots.
The curved shocks are reminiscent of two overlapping bow shocks, but
the identification of this as part of a true protostellar outflow is
uncertain.  It is not obviously associated with any collimated jet,
and the morphology of its nebular emission is not obviously that of a
parabolic bow shock.  It is however, very near on the sky to HHc-3
(about 30\arcsec\ to the upper right on Fig.~\ref{fig:hhc12}) and also
close to the complex of molecular globules and jet candidates in
Figure~\ref{fig:c45678}.  HHc-12 may be a distant shock in one of
these jets.

\subsubsection{HHc-13}

HHc-13, shown in Figure~\ref{fig:hhc13}, is a candidate microjet
flowing at P.A.$\simeq$165\arcdeg\ from a moderately bright star
located about 1\arcmin\ NE of the dust pillar from which HH~666
emerges (see Fig.~\ref{fig:hh666}). It appears to be highly
collimated, with an unresolved width, and a length that can be traced
to $\sim$2\arcsec\ (0$\farcs$22 pc) from the star where its surface
brightness falls below the detection limit.  The P.A.\ of this
elongated radial feature is different from the diffraction spikes in
the PSF of the image, which can also be seen in some brighter stars in
Figure~\ref{fig:hh666} to be at roughly 110\arcdeg\ and 200\arcdeg, so
it is a real nebular feature.  Without kinematic information it is
difficult to be sure that it is a true jet, however.

\subsubsection{HHc-14}

The candidate jet shocks comprising HHc-14 are near HH~1014 and can be
seen in Figure~\ref{fig:hh1014}.  Their curved structure may denote
partial illumination of a bow shock structure, but this is unclear.
Given the direction at which the HH~1014 jet emerges from its dust
pillar, it seems unlikely that HHc-14 is part of the same flow.  It
may come from an IR source embedded in a different pillar head.

\begin{figure*}\begin{center}
\includegraphics[width=4.8in]{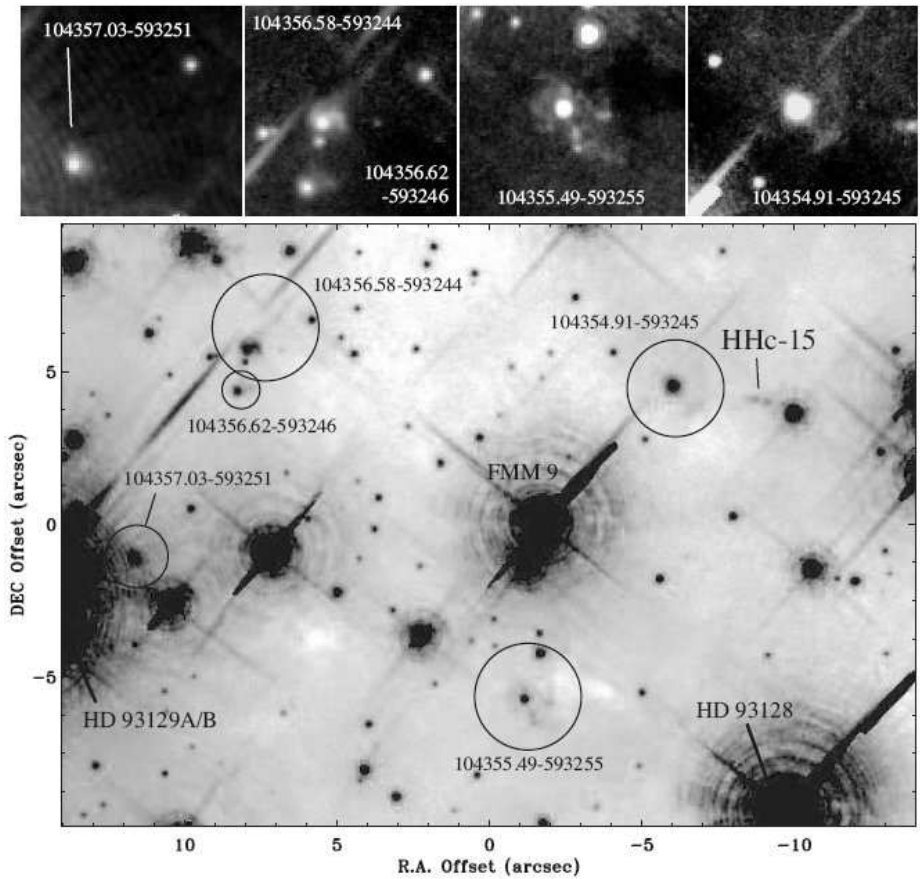}
\end{center}
\caption{The core of the Tr~14 cluster in an H$\alpha$ ACS/WFC image.
  The candidate microjet HHc-15 is in the upper right, and several
  other extended objects that are possible proplyds are circled.
  Postage stamps of these individual extended objects with a different
  intensity scale are shown in the four panels at the top.  The
  brightest cluster members with HD designations are noted, as well as
  star 9 from Feinstein, Marraco, \& Muzzio (1973) in the center of
  the image.}
\label{fig:tr14}
\end{figure*}

\subsubsection{HHc-15 and Proplyds in the Core of Tr~14}

Figure~\ref{fig:tr14} shows a detail of the {\it HST}/ACS F658N image
in the core of the Tr~14 cluster, revealing the discovery of several
extended objects seen on small scales.  HHc-15 is a candidate
microjet, made of a collimated chain of emission clumps stretching
about 2\arcsec\ (0.02 pc) eastward at P.A.$\simeq$85\arcdeg\ from a
visible star.  Without kinematic information or spectra of the stars
in this field, however, one cannot be certain that it is a jet or
which star is its driving source.

Immediately east of HHc-15 is a thin circumstellar nebula
104354.91-593245 (following the coordinate-based naming convention of
Smith et al.\ 2003).  It is shown in the detail panel in the upper
right of Figure~\ref{fig:tr14}.  This object has a limb-brightened
shell-like appearance, and may be a circumstellar shell, wind bow
shock, or a proplyd envelope.  It has a radius of 2000--3000 AU,
comparable to some of the larger proplyds in Orion (for background on
Orion's proplyds, see O'Dell 2001; Bally et al.\ 2000; Smith et al.\
2005b).  In the lower-middle portion of Figure~\ref{fig:tr14}, there
is a star surrounded by a very irregular shell structure
104355.49-593255, also shown in a detail panel second from right at
the top of Figure~\ref{fig:tr14}.  This object has rather chaotic
structure; perhaps it is a shell or proplyd envelope like
104354.91-593245, but which is being disrupted by stellar winds from
multiple nearby O-type stars in the cluster.

In the upper left of Figure~\ref{fig:tr14}, one sees two extended
nebular objects around stars that are good candidates for proplyds in
Tr14, 104356.58-593244 and 104356.62-593246.  (These are also shown in
the second-from-left panel at the top of Figure~\ref{fig:tr14}.)  The
larger object in particular, 104356.58-593244, has a long
$\sim$3\arcsec\ tail running away from the O2-type supergiant
HD~93129A (see Walborn et al.\ 2002), and is almost certainly a
proplyd.  Lastly, a bright source 104357.03-593251 is located very
close to HD~93129A in the western part of its diffraction pattern.  As
shown in the detail panel in the upper left corner of
Figure~\ref{fig:tr14}, however, this object is not a point source.  It
has extended H$\alpha$ emission within a radius of 0$\farcs$5 (1,000
AU), and is probably a marginally resolved evaporating protoplanetary
disk.

The detection of probable microjets and proplyds in the core of Tr~14
is an interesting and surprising discovery in these new F658N
narrowband {\it HST}/ACS images.  These structures were not seen in
{\it HST}/ACS images with the High Resolution Channel (HRC) by
Ma\'{i}z Appel\'{a}niz et al.\ (2005), presumably because they were
too faint in the broadband F435W ($B$-band) and F850LP filters.  This
suggests that the extended emission in Figure~\ref{fig:tr14} is
H$\alpha$ line emission and not starlight scattered by circumstellar
dust.  Jets and proplyds associated with stars in the core of Tr~14
would require that some of these stars have retained their
protostellar disks, despite their violent neighborhood in the core of
Tr~14 --- this location is far more extreme than the Trapezium, with
the O2 supergiant HD~93129A and several other early O-type stars
within only 0.1--0.2 pc.  This is even more surprising when one
considers that Tr~14 is also older than the Trapezium, with an age of
1--2 Myr (see Walborn 2009; Smith 2006a; and references therein).
These structures in Tr~14 therefore warrant intensive followup study,
since they provide information about the survival of resilient
protoplanetary disks in harsher environments and over a longer time
than in the very young Trapezium (age $\sim$10$^5$ yr).  Spectra of
the central stars associated with these extended features in Tr~14
would be valuable, and high-spatial-resolution thermal-IR observations
are encouraged to determine if these features resemble the mid-IR
emission from proplyds in the Trapezium (e.g., Smith et al.\ 2005d).

\begin{figure*}\begin{center}
\includegraphics[width=4.4in]{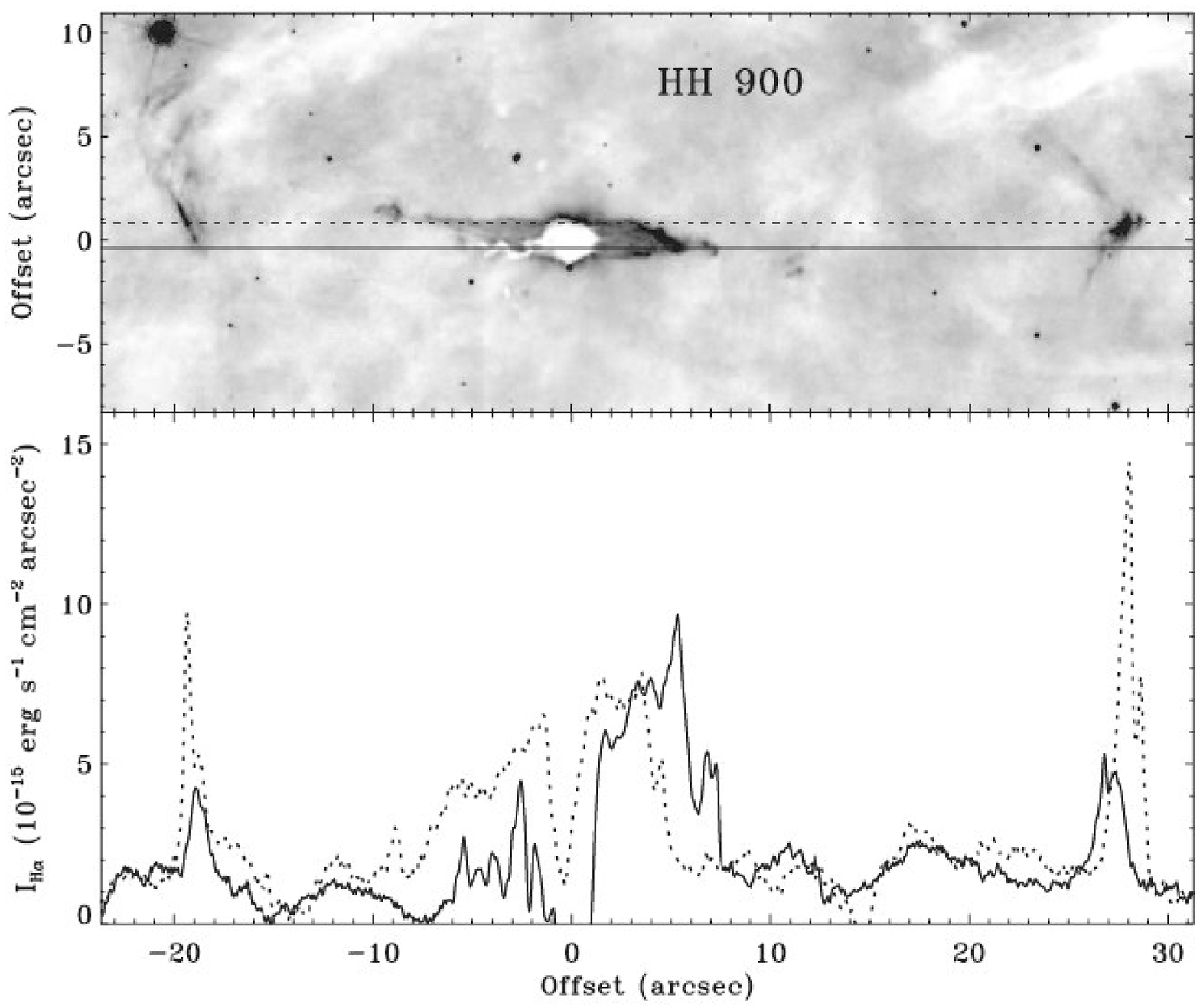}
\end{center}
\caption{A rotated image of HH~900 (top) and two intensity tracings
  along different samples of the jet (bottom).  The solid and dotted
  tracing correspond to the respective lines in the top panel.  The
  widths over which we sample the average intensity was 4 pixels, or
  0$\farcs$22.  The observed intensity, $I_{H\alpha}$, is plotted in
  units of 10$^{-15}$ ergs s$^{-1}$ cm$^{-2}$ arcsec$^{-2}$, and
  neglects possible contributions from [N~{\sc ii}].}
\label{fig:trace900}
\end{figure*}
%
\begin{figure*}\begin{center}
\includegraphics[width=4.1in]{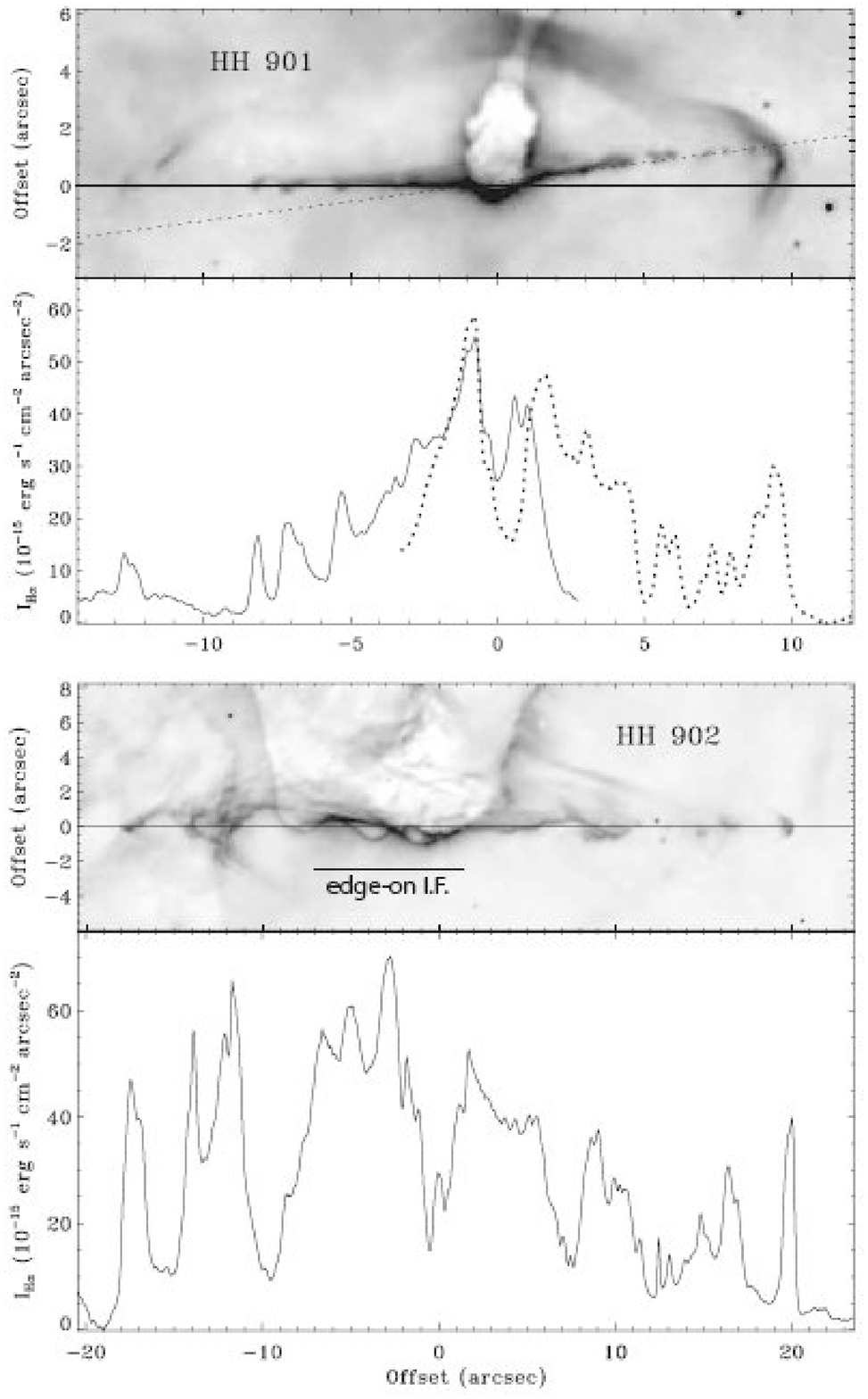}
\end{center}
\caption{Same as Figure~\ref{fig:trace900}, but for HH~901 and HH~902.
  For HH~901, we show tracings along two separate directions in order
  to follow the bend in the jet.  The solid and dashed tracings
  corrspond to the left and right sides of the jet, respectively.}
\label{fig:trace901}
\end{figure*}

\section{A CLOSER LOOK AT HH~900, 901, AND 902}

Figures~\ref{fig:trace900} and \ref{fig:trace901} show intensity
tracings ($I_{H\alpha}$ = 10$^{-15}$ ergs s$^{-1}$ cm$^{-2}$
arcsec$^{-2}$) along the jet axes of HH~900, HH~901, and HH~902.
These are intensities averaged over a few pixels, with extraction
widths corresponding roughly to the widths of the jet bodies.  (These
intensities are measured from the specific intensity in our F658N ACS
images using the pipeline calibration, integrated over the width of
the F658N filter.)  In the images, all three jets show a qualitatively
similar structure, with the inner parts showing a well-collimated
bipolar jet body, and with bow shocks at their outer extremeties.
This structure is clearly defined in HH~900 and HH~901, and somewhat
less so in HH~902 where the jet skirts along the ionization front at
the edge of its parent dust pillar and where there are multiple
overlapping bow shocks instead of two large opposing shocks.  The
implication is that in the inner parts of the jet, within a few
arcseconds of the embedded driving source, the jet body is still well
defined because it has flowed relatively unimpeded.  As time continues
and the jet material moves farther from its point of origin,
differences in velocity along the flow can become more important.
Faster material overtakes older slower material at internal shock
surfaces, forming dense clumps, so that the jet eventually takes on
the structure of a series of bullets rather than a smoother continuous
stream.  Bullets or knots may also arise from an intermittent jet
source, of course.

In the intensity plots, this progressive clumping of the jet is
manifested as spatially continuous emission within a few arcseconds of
the jet origin, followed by much larger variations in the intensity
on/off individual knots at larger separations.  (The intensity
tracings for HH~902 are complicated because some of this apparent
continuous emission at offsets of $-$2\arcsec\ to $-$7\arcsec\ arises
from emission at the ionization front.  The jet body at positive
offsets along the jet axis in Figure~\ref{fig:trace901} is less
contaminated.)  At increasing separation and increasing clumpiness,
the H$\alpha$ intensity does not drop as $r^{-2}$ as one might expect
from a conical jet.  Although the density drops in the regions between
clumps, the emission measure is dominated by material that is
piling-up at internal shocks where the density may be increasing.
Strong peaks in the H$\alpha$ intensity occur at the terminal bow
shocks along the flow, signifying a strong pile-up of jet material
there.  Note that in cases where the jet body has a substantial
neutral H fraction, one expects roughly constant $I_{H\alpha}$ since
the gas is ionization bounded (e.g., Bally et al.\ 2006).

All three jets show some deviation from a perfectly straight flow
along a single axis, and in three different ways.  As noted earlier,
in HH~900 the jet appears to be bent {\it toward} the direction of the
dominant radiation and stellar wind sources in Tr~16.  For this
source, we concentrate on the western part of the flow in
Figure~\ref{fig:trace900}.  In contrast, the HH~901 jet is clearly
bent {\it away} from the massive cluster of O-type stars in Tr~14,
with a difference of $\sim$7\arcdeg\ between the axes of its two
opposing flows.  It is not immediately clear, however, if this is
really a bend in the flow axis or a non-uniform irradiation effect
(see below).  Finally, for HH~902 in Figure~\ref{fig:trace901}, parts
of the jet body deviate from the jet axis as they seem blown back,
away from the hot O stars, whereas the densest knots seem to remain
right along a single straight jet axis.  If some of the jet material
is getting blown away while dense knots remain, this introduces
potential pitfalls in trying to deduce jet properties like the
mass-loss rate from distant knots in a flow.  In the following
section, therefore, we estimate mass-loss rates from the inner and
fairly continuous collimated parts of the jets when possible.

The H$\alpha$ intensity can be used as a direct tracer of the electron
density if the size of the emitting jet body is resolved so that one
can infer an emitting path length.  Assuming a fully ionized jet and
assuming that [N~{\sc ii}] $\lambda$6583 makes a relatively
unimportant contribution to the emission detected in the F658 filter
for these jets, one can then take the H$\alpha$ intensity to infer the
true density in the jet.  This is important for our estimates of
$\dot{M}$ for Carina's HH jets, as described in the following section.
In irradiated jets, [N~{\sc ii}] $\lambda$6583 makes a small
contribution to the emission in the filter, so neglecting it
introduces a small uncertainty compared to other sources of
uncertainty.  A more serious source of error may be the assumption
that the jet material is fully ionized.  In several irradiated HH jets
in Orion, for example, Bally et al.\ (2006) suspected significant
neutral H fractions in the jets, in some cases leading to jet bending
through the rocket effect when a photoionizing source is located to
one side.  HH~901 may be a good example of this phenomenon.  Close
examination of the inner jet structure shows a thin bright edge on the
side of the jet body that faces Tr14 (down in
Figure~\ref{fig:trace901}), whereas the other (north) side of the jet
body has lower surface brightness and is more diffuse.  This points to
a lower electron density there, possibly due to a lower ionization
fraction if this side of the jet is shielded from the UV radiation by
the neutral jet core.  The bend direction of the jet axis is in accord
with this hypothesis.  If there is a substantial neutral fraction in
the jets that are so close to a strong UV source like Tr~14, then the
densities and mass-loss rates for all jets discussed below should be
taken as lower limits.

\begin{table*}\begin{minipage}{6.5in}
\caption{Physical Paramters for HH Jets and jet candidates in Carina}\scriptsize
\begin{tabular}{@{}lccccccl}\hline\hline
HH &$L_{pc}$ &$I_{H\alpha}$ &$f$ &$V_{200}^a$ &$n_e$ &$\dot{M}_{jet}$ &Comment \\ 
   &(pc)   &(erg/s cm$^{-2}$ arcsec$^{-2}$) & &(10$^2$ km/s) &(cm$^{-3}$) &(10$^{-9}$ $M_{\odot}$ yr$^{-1}$) & \\
\hline
666 M        &0.005   &11.0$\times$10$^{-15}$   &1.0  &1.0  &705  &190  &inner jet between clumps \\
666 M        &0.005   &15.9$\times$10$^{-15}$   &1.0  &1.0  &847  &274  &bright knot \\
900          &0.02    &6.1$\times$10$^{-15}$    &0.5  &1.0  &263  &568  &wide inner jet, east \\
900 microjet &0.0025  &9.3$\times$10$^{-15}$    &1.0  &1.0  &916  &620  &bipolar microjet from star \\
901 E        &0.0035  &30$\times$10$^{-15}$     &1.0  &1.0  &1390 &184  &inner jet, east \\
901 W        &0.0035  &14.0$\times$10$^{-15}$   &0.3  &1.0  &920  &40   &clumps, west \\
902 W        &0.0031  &40$\times$10$^{-15}$     &1.0  &1.0  &1700 &176  &inner jet, west \\
903 jet      &0.0047  &11.0$\times$10$^{-15}$   &1.0  &1.0  &730  &172  &inner jet body \\
1002         &0.0034  &2.5$\times$10$^{-15}$    &...  &...  &400  &...  &no clear jet body \\
1003 A       &0.004   &3.7$\times$10$^{-15}$    &0.5  &1.0  &455  &38   &inner jet body near I.F. \\
1004 NE      &0.0067  &7.4$\times$10$^{-15}$    &0.5  &1.0  &497  &120  &inner jet body, unclear morph. \\
1005         &0.0044  &11.0$\times$10$^{-15}$   &0.8  &1.0  &750  &124  &inner jet body \\
1006 N       &0.0037  &2.5$\times$10$^{-15}$    &0.5  &1.0  &390  &28   &thin jet body, north \\
1006 S       &0.0034  &2.4$\times$10$^{-15}$    &0.4  &1.0  &403  &20   &thin jet body, south \\
1007         &0.0022  &4.4$\times$10$^{-15}$    &1.0  &1.0  &671  &35   &jet body \\
1008         &0.0078  &9.8$\times$10$^{-15}$    &...  &...  &530  &...  &no clear jet body \\
1009         &0.0028  &4.9$\times$10$^{-15}$    &...  &...  &630  &...  &no clear jet body \\
1010 SW      &0.0073  &2.9$\times$10$^{-15}$    &0.4  &1.0  &299  &68   &inner jet body \\
1011         &0.0028  &7.4$\times$10$^{-15}$    &0.8  &1.0  &769  &52   &microjet body \\
1012 microjet&0.0016  &2.5$\times$10$^{-15}$    &0.5  &1.0  &960  &13   &microjet body, not LL Ori shock \\
1013         &0.0037  &8.6$\times$10$^{-15}$    &0.2  &1.0  &723  &21   &jet body NE, but unclear \\
1014         &0.004   &7.1$\times$10$^{-15}$    &0.4  &1.0  &632  &44   &clumpy jet body \\
1015         &0.0028  &0.74$\times$10$^{-15}$   &0.4  &1.0  &244  &8.4  &faint, thin jet body \\
1016         &0.0019  &7.4$\times$10$^{-15}$    &1.0  &1.0  &934  &36   &bow shock, unclear jet body \\
1017 jet     &0.0016  &4.4$\times$10$^{-15}$    &0.5  &1.0  &787  &11   &inner jet body \\
1018 microjet&0.0011  &6.4$\times$10$^{-15}$    &1.0  &1.0  &1140 &15   &microjet  \\
HHc-1 (3324) &0.0056  &4.9$\times$10$^{-15}$    &0.5  &1.0  &444  &76   &unclear jet body \\
HHc-2 (3324) &0.0045  &8.6$\times$10$^{-15}$    &1.0  &...  &660  &...  &no clear jet body \\
HHc-1 B      &0.0028  &35.6$\times$10$^{-15}$   &1.0  &1.0  &1690 &144  &microjet body, west \\
HHc-2        &0.0022  &24.5$\times$10$^{-15}$   &0.4  &1.0  &1580 &33   &unclear jet body  \\
HHc-3 ABC    &0.0019  &3.7$\times$10$^{-15}$    &0.5  &1.0  &660  &13   &average of ABC components  \\
HHc-4        &0.0034  &6.4$\times$10$^{-15}$    &...  &...  &650  &...  &bow shock only, no clear jet body \\
HHc-5        &0.0031  &2.2$\times$10$^{-15}$    &0.3  &1.0  &400  &12   &unclear jet body  \\
HHc-6 N      &0.0028  &4.4$\times$10$^{-15}$    &0.8  &1.0  &595  &40   &curved jet body, north  \\
HHc-6 S      &0.0028  &7.4$\times$10$^{-15}$    &0.8  &1.0  &769  &52   &curved jet body, south  \\
HHc-7        &0.0037  &9.8$\times$10$^{-15}$    &...  &...  &770  &...  &bow shock only, no clear jet body \\
HHc-8        &0.0028  &6.1$\times$10$^{-15}$    &...  &...  &700  &...  &LL Ori shock, no clear jet \\
HHc-9        &0.01    &4.2$\times$10$^{-15}$    &...  &...  &306  &...  &bow shock only, no clear jet body \\
HHc-10       &0.0067  &6.1$\times$10$^{-15}$    &...  &...  &450  &...  &no clear jet body \\
HHc-11       &0.0034  &3.7$\times$10$^{-15}$    &...  &...  &490  &...  &no clear jet body \\
HHc-12       &0.0022  &3.7$\times$10$^{-15}$    &...  &...  &613  &...  &no clear jet body \\
HHc-13       &0.0012  &2.9$\times$10$^{-15}$    &1.0  &1.0  &737  &12   &microjet  \\
HHc-14       &0.0028  &3.7$\times$10$^{-15}$    &...  &...  &544  &...  &no clear jet body \\
HHc-15       &0.0012  &15.0$\times$10$^{-15}$   &0.5  &1.0  &1650 &13   &microjet  \\
\hline
\end{tabular}
$^a$Except for HH~666, $V_{200}$=1 is an assumed value in lieu of
measurements for the jet speeds.
\end{minipage}
\end{table*}

\section{JET MASS-LOSS RATES}

To estimate the mass-loss rates of the new HH jets in Carina, we first
estimate electron densities from the intensity measured in images.
The H$\alpha$ emission measure in an ionized nebulosity depends on the
electron density, $n_e$ as well as the emitting path length $L$
through the object.  For a collimated jet, we take $L$ to be the
diameter of the jet body, or $R = L/2$.  Table~4 lists estimates for
$L$ for each HH jet and jet candidate taken from spatially resolved
jet structures in our ACS images.  An expression between $n_e$, $L$,
and the emission measure can be simplified to

\begin{equation}
  n_e = 15.0 \ \sqrt{ \frac{I_{H\alpha}}{L_{pc}}} \ {\rm cm}^{-3} 
\end{equation}

\noindent where $I_{H\alpha}$ is the H$\alpha$ intensity measured in
our narrowband F658N ACS images in units of 10$^{-15}$ ergs s$^{-1}$
cm$^{-2}$ arcsec$^{-2}$, and $L_{pc}$ is the emitting path length
through the jet in pc (see Smith et al.\ 2004c; Bally et al.\ 2006;
Walawender et al.\ 2004).  Table 4 lists measured values of
$I_{H\alpha}$ and the corresponding estimates of $n_e$ calculated in
this way.  Obviously there is wide variation in both $I_{H\alpha}$ and
$n_e$ at different positions in each jet.  The values of $I_{H\alpha}$
are therefore typical average values of bright emission features in
the jet flow taken to be representative of the inner jet.  We account
for variations in $I_{H\alpha}$ along the jet (sharp drops in
intensity in Figure~\ref{fig:trace901}, for example) with a geometric
filling factor $f$, as noted below, since the observed density that
causes the observed emission measure does not uniformly fill the
column of the jet.

To provide rough estimates of the mass-loss rates for the Carina HH
jets from the electron densities in Table 4, we treat the jets as
approximately cylindrical.  Therefore, the portions of the jets where
we have chosen to measure representative values of $L$, $I_{H\alpha}$,
and $f$ correspond to inner parts of the jet that are well collimated,
such as HH~666~M and the inner parts of HH~900, 901, and 902 discussed
in the previous section.  Even so, clumping at internal shocks begins
to affect the emissivity, so to account for this effect (bright knots
separated by empty relatively spaces) we adopt a geometric filling
factor $f$, estimated from the morphology in images.  Note that for
HH~666~M, where we measure $\dot{M}$ separately for a bright knot and
a fainter region between knots, the derived mass-loss rate differs by
only $\sim$20\%.  For HH objects like HH~1003C, 1008, 1009, HHc-4, etc.,
where clear bow shocks but no clear collimated jet can be seen, we
cannot provide estimates of $\dot{M}$ in this way, although we do give
estimates of $n_e$.  Adopting cylindrical structure for the collimated
jets that we can measure, then, the average mass-loss rate flowing
through {\it one side} of the jet is given by

\begin{equation}
  \dot{M} = \mu m_H n_e V \pi (L/2)^2 f
\end{equation}

\noindent where $\mu$=1.35 is the mean molecular weight, $m_H$ is the
proton mass, $V$ is the jet speed, and $f$ is a geometric filling
factor inferred from structure in images.  To represent the total mass
carried away by both sides of a bipolar jet, this number must be
multiplied by 2.  Numerically, this can then be expressed as

\begin{equation}
  \dot{M} = 1.1 \times 10^{-5} \ \big{(} n_e \ V_{200} \ L_{pc}^2 \ f \big{)} \ M_{\odot} \ {\rm yr}^{-1}
\end{equation}

\noindent where $V_{200}$ is the jet speed in units of 200 km
s$^{-1}$.  This expression is used to calculate the values of
$\dot{M}$ in Table 4, which also lists adopted values of $V_{200}$ and
$f$.  Using only a single-epoch of images and no kinematic
information, we can make only rough estimates of the characteristic
mass-loss rates for the Carina jets.  Better estimates will come from
future studies that can provide proper motions with multi-epoch images
and Doppler shifts from high resolution spectra.  For now, we adopt
$V_{200}$ = 1.0 as a fiducial value for all.  In HH~666, spectra show
speeds of roughly 200 km s$^{-1}$ for HH~666~M and O (Smith et al.\
2004c), and this is a typical speed for the inner parts of HH jets.
Also, we will be comparing these jet mass-loss rates to those derived
in the same way for jets in the Orion Nebula, for which Bally et al.\
(2006) adopted 200 km s$^{-1}$ for the speed of most jets.

\begin{figure}\begin{center}
\includegraphics[width=2.6in]{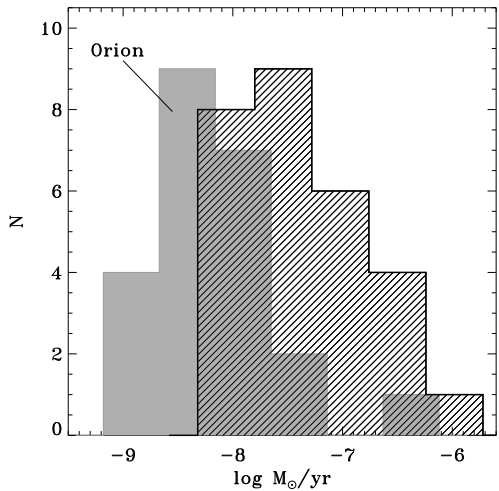}
\end{center}
\caption{A histogram of the HH jet mass-loss rates from Table 4 (black
  with hatched pattern).  For comparison, the filled gray histogram
  shows mass-loss rates for a sample of a comparable number of HH jets
  in the outer Orion Nebula, taking the value of $\dot{M}_0$ listed in
  Table 6 of Bally et al.\ (2006), but multiplied by 2 to account for
  contributions from both sides of a bipolar jet, as we have done for
  Table 4 (see text).  These Orion mass-loss rates were derived using
  the same method we adopted for Carina.}
\label{fig:mdot}
\end{figure}

There are, of course, considerable uncertainties in the estimates of
$\dot{M}$ in Table~4.  As noted above, the denser jets have an unknown
neutral H fraction, the filling factor $f$ that we adopt is rather
subjective, and the jet speed is not known for all sources except
HH~666.  Although the choice of filling factor is subjective, it is
guided by the jet structure in high resolution images, and is likely
to be reliable to within $\pm$30\%.  Judging from a wide array of
observations of HH jets, the adopted velocity of 200 km s$^{-1}$ is a
representative choice, introducing a factor of $\sim$2 uncertainty.
Fortunately, more accurate estimates of the jet speed and possibly
also the ionization fraction can be obtained with more detailed
spectroscopic followup and a later epoch of {\it HST} imaging to
measure proper motions of the jets (these observations are needed to
confirm the candidate jet objects anyway).  The estimates of $\dot{M}$
are probably good to within a factor of 2--3 or perhaps somewhat more,
and are adequate to convey the rough distribution of jet mass-loss
rates that we are sensitive to.  This level of accuracy is comparable
to expected intrinsic variability of the mass-loss rate over time for
an individual jet anyway.  For the one case where detailed
spectroscopic observations are available, HH666, it is reassuring that
the electron density derived from the emission measure in image data
(Table 4) agrees with the electron density derived from spectral
diagnotstics like the [S~{\sc ii}] doublet intensity ratio (Smith et
al.\ 2004c).

While the electron density is subject to uncertain estimates of $L$,
we note that there is not much variation in derived values of $n_e$ in
Table 4, which vary by less than a factor of 10 (200 to 2000 cm$^{-3}$).
In general, the fainter jets have $n_e$ of a few 10$^2$ cm$^{-3}$,
while the brightest jets have densities around or slightly above
10$^3$ cm$^{-3}$.  The faint end is likely a selection effect, in the
sense that lower $\dot{M}$ jets with densities near or below 10$^2$
cm$^{-3}$ would have emission measures fainter than fluctuations in
the background H~{\sc ii} region and would be difficult to detect.
Jets with densities much above 2000 cm$^{-3}$ might have dense jet
cores that remain largely neutral, despite the strong UV radiation, so
that H$\alpha$ emission does not necessarily trace the jet mass (e.g.,
Hartigan et al.\ 1994).  This is another reason to suspect that for
the densest jets, then, it is likely that the mass-loss rates and
densities in Table 4 are underestimates.

We measure a range of mass-loss rates ranging from 8$\times$10$^{-9}$
$M_{\odot}$ yr$^{-1}$ up to $\sim$10$^{-6}$ $M_{\odot}$ yr$^{-1}$.
These jet mass-loss rates trace a range of active accretion rates onto
their driving protostellar sources.  If one assumes that the mass-loss
rate of the jet is roughly 10\% of the accretion rate onto the star
(e.g., Calvet 1997), the observed jet mass-loss rates in Carina imply
relatively high mass accretion rates of 10$^{-7}$ to 10$^{-5}$
$M_{\odot}$ yr$^{-1}$.  (If the visible HH jet is only some fraction
of the wind driven from the protostellar disk, then implied accretion
rates are even higher.)  These are on the high end for the
distribution of mass accretion rates measured for nearby sources
(e.g., Calvet et al.\ 2000), so the high mass-loss rate jets in our
sample may trace more massive protostars or brief episodic
accretion/outflow events akin to FU Ori outbursts.  Indeed, mass
accretion rates around 10$^{-5}$ M$_{\odot}$ yr$^{-1}$ are comparable
to the episodic bursts of accretion inferred to occur in early
protostellar phases (Calvet et al.\ 2000).  Jets in our sample with
more typical mass-loss rates around 10$^{-7}$ $M_{\odot}$ yr$^{-1}$
are comparable to those of famous nearby HH jets like HH~34, 47, and
111 (e.g., Hartigan et al.\ 1994).

Figure~\ref{fig:mdot} illustrates a consequence of selection effects
determining the lower jet densities that we are sensitive to.  This is
a histogram of the mass-loss rates for Carina HH jets in Table 4
(black hatched) compared to a similarly sized sample of HH jets in
Orion (Bally et al.\ 2006).  Bally et al.\ list several values for
$\dot{M}$ for each jet in their Table~6 that are derived with
different assumptions; the values that we have plotted in
Figure~\ref{fig:mdot} correspond to $\dot{M}_0$ in their Table 6
(multiplied by a factor of 2 to account for both sides of a bipolar
jet, to be consistent with our estimates in Table 4).  These values
were derived using the same method adopted here (i.e., eqn.\ 2), and
therefore provide a useful comparison, subject to the same
uncertainties and potential systematic effects.  The sample discussed
by Bally et al.\ (2006) was studied with a similar {\it HST}/ACS
survey in the F658N filter.  It is also worth noting that this sample
of Orion jets represents those found in the {\it outer} parts of the
Orion Nebula, not in the brightest inner region near the Trapezium.
The contrast between jets and the background nebula is therefore
comparable to Carina.

The main lesson from Figure~\ref{fig:mdot} is that in Carina we are
likely missing a large number of HH jets at the lower end of the
$\dot{M}$ distribution, even with {\it HST}, if this distribution is
similar to that in Orion.  Both show a similar slope in the jet
mass-loss function, with very roughly $N \propto \dot{M}^{-0.5}$.  The
$\dot{M}$ distribution for Carina HH jets peaks at $\sim$10$^{-7.6}$
$M_{\odot}$ yr$^{-1}$, whereas that in Orion seems to peak at lower
values around 10$^{-8.4}$ $M_{\odot}$ yr$^{-1}$.  Orion is 5 times
closer to us than Carina, so with the same {\it HST} camera, it is
easier to detected fainter, thinner shock filaments and narrower jets
that have lower mass-loss rates.  The thin filaments of these weaker
HH jets may be lost in the bright nebular background with the lower
physical spatial resolution achieved in Carina.  If the $\dot{M}$
distribution in Orion is representative of the intrinsic distribution
in Carina, then it is likely that the HH jets we have detected in
Carina are roughly half (or less) of the total number of protostellar
outflows in the region we surveyed.  Our ACS H$\alpha$ survey did not
uniformly cover the entire Carina Nebula, however, and is not an
unbiased sample with any strict completeness criteria (see
Figure~\ref{fig:map}). Nevertheless, it seems likely that we are
missing a large number of weaker jets.  Extrapolating from the 39 jets
we detected with HST, correcting for both a factor of $\sim$2 because
of undetected jets at the low end of the $\dot{M}$ range plus the fact
that we only surveyed a portion of the nebula with {\it HST}, it is
likely that Carina currently harbors at least 150--200 active
protostellar jets.

\section{DISCUSSION}

\subsection{Jets from Tiny Cometary Globules}

A key qualitative result of this study is that the {\it HST} images of
Carina reveal several cases where HH jets emerge from tiny cometary
clouds (roughly 0.01$-$0.05 pc in size).  The most striking example
was HH~900, but several other examples are seen as well, like HH~901,
1006, 1011, 1013, HHc-1, and HHc-3.  The presence of a collimated HH
jet emerging from such a globule requires that the globule harbor an
actively accreting protostar, so it proves that even the smallest
clouds -- which appear to be only shredded remnants of a larger
molecular cloud --- are, in fact, important sites of ongoing active
star formation in an evolving H~{\sc ii} region like Carina.  Since
the phase during which an active HH outflow may be detected is quite
short (a few 10$^4$ yr) compared to the expected evaporation time of
the cloud ($>$10$^5$ yr), and because we are biased toward detecting
the most powerful outflows (see above), it is possible that more of
these small clouds also harbor young stars.

Interestingly, the clouds from which the jets emerge are small enough
and isolated enough that they may be formation sites of a {\it single}
star or binary system.  This makes them valuable objects for case
studies in star formation because their isolated nature simplifies the
complexities normally encountered in crowded massive star-forming
regions.

There is another reason that these objects may be of particular
interest, having to do with the looming deaths of several nearby
massive stars and possible connections to our own Solar System.  As
discussed above for the case of HH~900, these globules are destined to
intercept the SN ejecta that will soon result from the deaths of
$\eta$ Car and several other massive stars that are only a few parsecs
away.  This can lead to efficient injection of radioactive nuclides
into the tiny clouds, as is thought to have occurred in our own Solar
System, making these objects possible analogs of the conditions that
formed the Sun.  In the specific case of HH~900, we showed in \S 3.2.2
that the cross-sectional size of its cloud would intercept a fraction
of the SN ejecta from a hypothetical $\eta$ Carinae SN that would be
commensurate with the amount of SN ejecta thought to be injected into
the cloud from which the Solar System formed.  That estimate was for
only a single SN, but Carina harbors $\sim$70 O stars (Smith 2006a), a
few dozen of which are likely to explode in the next 1--2 Myr.  These
{\it multiple} SNe can greatly enhance the amount of young SN ejecta
that will be imparted to these clouds.  Without a second generation of
stars forming at the periphery of a giant H~{\sc ii} region, as seen
in Carina, it would be difficult to recreate the conditions necessary
for the birth of the Solar System.  The proto-solar cloud must have
been fairly close to a SN and the cloud itself must have cast a fairly
wide net in order to intercept enough fresh SN ejecta.

Orion has also been discussed as a possible analog of the conditions
under which the Solar System formed, where young low-mass stars are
seen in close proximity to an O-type star.  However, remember that
although one sees young protoplanetary disks very close to an O-type
star that will {\it eventually} explode as a SN, that O star in Orion
($\theta^1$C Ori) is still very young and not ready to explode yet.
By the time it evolves off the main sequence and ends its life 5-6 Myr
from now, the Trapezium will likely have dissolved and the proplyds we
see now will have long since had their disks evaporated.  In Carina,
on the other hand, several of the most massive stars are already 3 Myr
old and appear poised to explode in short order --- $\eta$~Car, in
particular, is thought to be near its end.  Furthermore, the region
will not witness just one SN from $\eta$ Car, but rather, in the next
1-2 Myr there will be a rapid succession of a few dozen SNe, several
of them from some of the most massive stars known.  This key
difference makes a massive evolving H~{\sc ii} region like Carina seem
like the most promising analog for the cradle of the Solar System.

\begin{figure*}\begin{center}
\includegraphics[width=4.4in]{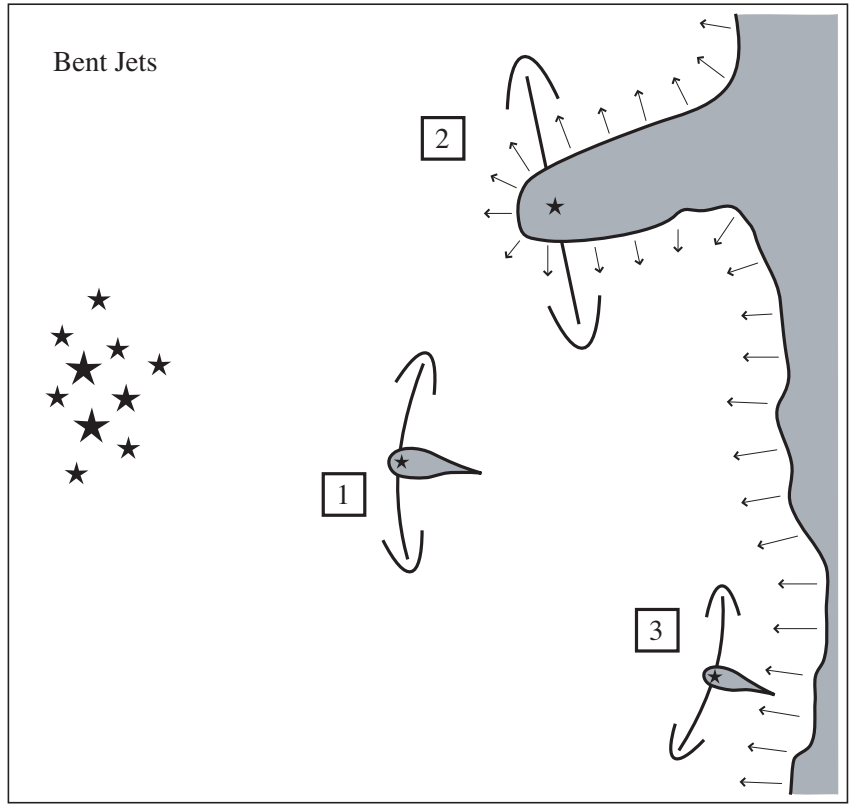}
\end{center}
\caption{A schematic illustration referring to how HH jet axes may get
  bent by external mechanisms, as discussed in \S 6.2.  A cluster of
  massive stars is located at left.  Case 1 refers to a jet that is
  bent away from the massive stars by radiation or winds.  Case 2
  refers to a jet whose axis is not bent because it is moving through
  a diverging photoevaporative flow from the surface of a dust pillar,
  which may protect the jet from winds or inhibit its bending.  Case 3
  shows a jet axis that is bent toward the massive stars because it is
  located amid a strong photoevaporative flow from the surface of an
  adjacent molecular cloud, which effectively acts as a tail wind and
  dominates over the direct radiation or wind from the massive stars.}
\label{fig:bend}
\end{figure*}

\subsection{Bending of HH Jets...or Not}

Protostellar outflows are not always straight.  HH jets with a gentle
S-shape bend can result from precession of the axis of the driving
source (e.g., Devine et al.\ 1997), whereas cases in which both
opposing flows in a bipolar jet bend in the same direction require an
external agent.  These types of bent jets, reviewed recently by Bally
et al.\ (2006), are interesting as they provide a potential diagnostic
of bulk flows of plasma in the surrounding environment.  In other
words, they serve as ``wind socks'' in H~{\sc ii} regions.  They have
been compared to the head-tail systems of jets from radio galaxies.
Several types of externally bent jets (as opposed to precessing jets)
have been discussed.

Some collimated jets show a clear bend {\it toward} the center of star
formation activity with increasing separation from the jet source,
such as HH~499 and other jets in NGC~1333 (Bally \& Reipurth 2001). In
these cases the jet bend results because the young star that drives it
is moving through a stationary ambient medium, as might occur if young
star is dynamically ejected from the active star-forming cluster
(Bally \& Reipurth 2001).

In regions where young massive stars are present near a jet, such as
in the Orion Nebula, jets tend to bend {\it away} from the massive
O-type stars.  In these cases, the bending may be caused by the
influence of winds or UV radiation from the massive stars.  This can
occur by several mechanisms (see Bally et al.\ 2006), including the
direct ram pressure of a side wind (stellar winds or a bulk flow of
plasma in the H~{\sc ii} region cavity), radiation pressure acting on
dust grains entrained by the jet, or the rocket effect if neutral gas
in the jet becomes ionized by UV radiation.  All of these forces act
to push the jet flow away from the massive stars that provide winds
and radiation.  Manifestations of this type of jet bending include the
class of LL~Ori objects (Bally et al.\ 2001), as well as smoothly
curved bipolar jets like HH~502 (Bally et al.\ 2006) and HH~555 (Bally
\& Reipurth 2003).

A protostellar jet expanding away from its driving source may also
bend suddenly if the jet has a grazing collision with a nearby cloud
or other obstacle (another jet, another star's disk, etc.).  In this
collision, the jet may be deflected in a different direction (e.g.,
Raga \& Cant\'{o} 1996; Gouveia Dal Pino 1999).  Examples of this
phenomenon include the HH~110/270 combination (Rodriguez et al.\ 1998;
Reipurth \& Olberg 1991) and the HH~83/84 complex (Reipurth et al.\
1997a).


Among the 39 known HH jets and jet candidates in Carina, we detect an
interesting variety of jet bends.  We discuss these in the context of
three different types of jet bending, illustrated schematically in
Figure~\ref{fig:bend}, numbered according to the three cases discussed
below:

1.  First, our {\it HST} images reveal several examples of jets that
bend {\it away} from nearby O-type stars, as expected in a region with
rather harsh environmental conditions driven by the $\sim$70 O-type
stars in Carina (Smith 2006a).  This is the same as the case discussed
above, exemplified by jets and LL~Ori objects in Orion (Bally et al.\
2006).  See Bally et al.\ (2006) for quantitative estimates of the
forces that bend such jets.  Examples of this phenomenon in Carina
include HH~901, 903, 1004, 1012, 1017, HHc-3, HHc-5, and HHc-8.  Two
of these, HH~1012 and HHc-8, appear to be true LL~Ori objects.  This
class of bent jets tends to be found closer to O stars, being more
directly impacted by stellar winds and radiation.

2.  Counter to expectations, several jets exhibit {\it no perceptable
  bend of their jet axis}, or very weak bends.  These jets are
surprisingly straight, considering the harsh environment in which they
reside.  A good example discussed previously is HH~666 (Smith et al.\
2004c), which maintains a very straight jet axis as each of its
bipolar flows moves more than a parsec from the point where the jet
was launched.  Other clear examples of straight bipolar jets among the
new sample of jets and jet candidates in Carina are HH~902 and 1010.
HH~902 does show some material that appears to be swept back in
between clumps, but dense knots in the jet maintain a straight
trajectory.  These jets emerge from the heads of large dust pillars
that have clearly been shaped by feedback from massive stars, so the
straight axis is somewhat surprising.  One possible explanation for
the lack of any perceptible bend in the jets has to do with the
immediate environment into which they are expanding once they exit the
obscured part of the natal dust pillar.  Namely, the bright-rimmed
dust pillars are clearly irradiated by a strong UV flux, and probably
have dense photoevaporative flows from their surfaces (depicted by
short arrows in Figure~\ref{fig:bend}).  These dense photoevaporative
flows from the head of a curved dust pillar will be radially
divergent, so they will flow in essentially the same direction as any
HH jet emerging from the pillar (i.e. normal to the surface of the
pillar).  Several dust pillars, such as the one associated with
HH~666, show clear empirical evidence for dense flows in the form of
streaks and striations proceeding radially from the irregular surface
of the cloud. These flows will therefore exert no side force on the
jet to divert its trajectory, and these dense photoevaporative flows
may even {\it shield} the HH jet from the sideward ram pressure of a
bulk flow of plasma in the larger wind-blown bubble.  Around some
pillar heads, one can see evidence for a stand-off shock between the
photoevaporative flow and winds.  Intermediate cases are also
possible, as in HH~903, where the bulk flow of H~{\sc ii} region
plasma is in the process of crossing the pillar.

3.  Lastly, we also see jets that -- counter-intuitively -- bend {\it
  toward} nearby massive stars.  The best example of this, discussed
earlier, is HH~900.  This behavior is somewhat perplexing, because it
is probably not attributable to dynamical ejection from that cluster
as in the case of bent jets in NGC~1333 (Bally \& Reipurth 2001).  (In
the case of HH~900, the cloud and its star would need to have been
ejected from Tr~16, but the cluster is much older than the young
cometary cloud, and the tail of HH~900 points away from Tr~16.)  The
explanation for this type of jet bend may be similar to the previous
case --- i.e.\ the influence of strong photoevaporative flows from the
surfaces of molecular clouds.  In this case, however, the dominant
photoevaporative flow is not from the curved surface of the pillar or
cloud from which the jet emerges, but from a neighboring cloud
surface, as depicted for case 3 in Figure~\ref{fig:bend}.  HH~900 is
located very close to the edge of a dark molecular cloud, which shows
a very bright PDR at IR wavelengths, as we noted in \S 3.2.2.  If
HH~900 is embedded within this photoevaporative flow, the bulk flow
could create the effective condition of a ``tail wind'' in its
environment, so that the ram pressure pushes its bipolar jet toward
the massive stars.  While photoionizing radiation from O stars in
Tr~16 has probably led to the initial sculpting of the cometary cloud
and its tail that points away from the massive stars, this tail wind
may also be the agent responsible for producing the unusual zig-zags
in the thin tail of HH~900's parent globule (Figure~\ref{fig:hh900})
by acting to compress it from the opposite direction.

\begin{figure}\begin{center}
\includegraphics[width=2.6in]{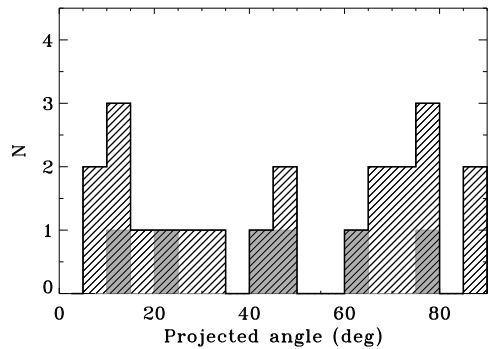}
\end{center}
\caption{A histogram of the projected angle on the sky between the HH
  jet axis and the elongation of its parent cometary cloud or dust
  pillar.  We have only plotted angles for 22 sources in which these
  angles could be determined (i.e. this does not include HH jets like
  those in NGC~3324, which emerge from a relatively straight
  ionization front).  The hatched region includes all HH jets and
  candidates, while the shaded grey histogram shows the candidates
  alone.}
\label{fig:orient}
\end{figure}

\subsection{Jet Orientations and Selection Effects}

Most of the HH jets in Carina emerge from elongated cometary clouds or
dust pillars, and some of the most striking examples are bipolar jets
whose axis is nearly perpendicular to the elongated dust pillar.  One
may naturally wonder if there is any correlation between the jet axis
orientation and the orientation of the underlying cloud (and hence,
the direction toward the massive stars whose radiation has shaped it).
In fact, as this paper was nearing completion, Lora, Raga, \& Esquivel
(2009) proposed that just such a preferential alignment should exist.
This was based on 3D simulations of the interaction between an
advancing ionization front and a neutral cloud, wherein compressed
clumps of gas that form at the unstable interface tend to collapse and
have angular momentum vectors perpendicular to the direction of the
radiation flow.  The clouds themselves tend to be elongated in the
same direction as the radiation flow, forming structures akin to
cometary clouds and dust pillars.  If cores in these clumps collapse
to form protostars and accretion disks, Lora et al.\ (2009) infer that
the resulting HH jets will have a preferential orientation
perpendicular to the long axis of their cometary cloud or dust pillar.
We can test this hypothesis with the large number of jets discovered
in a single region like Carina, many of which emerge from dust pillars
or cometary clouds that betray their dominant direction of incident
radiation.  If true, the angle between the jet axis and the long axis
of a dust pillar or cometary cloud should favor orientations near
90\arcdeg.

Figure~\ref{fig:orient} shows a histogram of this projected angle
measured in {\it HST} images.  This includes all HH jets and jet
candidates for which an approximate cloud axis and jet axis could be
measured (i.e. this does not include HH jets like those in NGC~3324,
which emerge from a relatively straight ionization front, nor does it
include HH objects like HH~1009 or HHc-4, 7, and 9 that lack clear
collimated jets; obviously HH~1017 and 1018 are not included because
they have no associated dust pillar).  We conclude from
Figure~\ref{fig:orient} that in this sample, we detect no obvious
correlation between the projected angles.  There does not appear to be
a preferred physical orientation between the jets and their parent
clouds, at least not that we can detect.  Two caveats are that the
hypothesis of Lora et al.\ (2009) requires that the formation of the
protostars driving the jets was triggered by the advancing ionization
front.  We do not know what fraction of the jet sources in Carina
formed on their own and which ones were triggered.  Furthermore, the
physical angle between the jet axis and its dust pillar is not
necessarily the same as the angle projected on the sky.  Even jets
perpendicular to their dust pillars can be oriented in such a way that
the angle between them projected on the sky is not 90\arcdeg.  One
must either select jets known to have their dust pillar axis near the
plane of the sky, or one must study the statistical distribution of
angles for random projections.

In any case, we find no evidence that selection effects like the jet
direction will bias the sample of jets detected in our images.  One
might imagine, for example, that jets emerging perpendicular to the
axis of a cometary cloud would be easier to detect than those more
closely aligned to its axis, but this is apparently not the case
(e.g., HH~1006), so we consider it unlikely that we are missing a
large number of jets whose axis is aligned with the cloud.

\subsection{Energy, Momentum, and Turbulence Injection}

Externally irradiated jets and outflows appear to be common in young
H~{\sc ii} regions.  The Orion Nebula contains dozens of examples
ranging from tiny, low mass-loss-rate jets visible only on
high-angular resolution images such as those delivered by {\it HST}
(Bally et al.\ 2006) to giant outflow lobes such as the parsec-scale
HH~400 (Bally et al.\ 2001).  The Carina Nebula also contains a
growing collection of such jets and outflows (Smith, Bally, \& Brooks
2004c; this work).  However, the total momentum and energy injected by
these outflows is unimportant in the global momentum and energy budget
of the H~{\sc ii} region.  As in the Orion Nebula, momentum injection
is dominated by the UV radiation field and stellar winds of the
massive stars.

The mass injection rate of the Carina outflows ranges from less than
$\dot{M} \ \approx \ 2 \times 10^{-9}$ $M_{\odot}$~yr$^{-1}$ to over
$2 \times 10^{-7}$ $M_{\odot}$~yr$^{-1}$ per outflow.  While the
velocities for most outflows have not been determined by either
spectroscopy or proper motions, typical values for such flows range
from under 100 km~s$^{-1}$ to over 400 km~s$^{-1}$.  Radial velocity
measurements for HH~666 indicate a range of velocities up to 250
km~s$^{-1}$ (Smith et al.\ 2004c).  Using a relatively large value for
protostellar outflows of $\dot{M} = 2 \times 10^{-7}$
$M_{\odot}$~yr$^{-1}$ and a flow velocity of 200 km~s$^{-1}$, and
assuming that at any typical time there are 100 active outflows in the
region, the total rate of momentum injection into the nebula would be
about $\dot{P} = 4 \times 10^{-3}$ $M_{\odot}$~km~s$^{-2}$.  This is
probably generaous since this estimate used the outflow parameters
near the maximum values observed, and assumes that only about 20\% of
the true number of flows have been detected.

The two dominant sources of internal motions in the nebula are likely
to be the ionized photoevaporative/photoablation flows from dense
cloud fragments trapped in the nebular interior and the impacts of
massive-star winds.  The average Lyman continuum luminosity of the
Carina Nebula is about $Q \approx 10^{51}$ ionizing photons per second
(Smith 2006a; Smith \& Brooks 2007); it may have been a bit higher
while the three current WNH stars and $\eta$ Car were main-sequence
stars (see Smith 2006a).  The total value of $Q$ may have decreased by
as much as 30\% in recent times as three massive stars evolved into
their WNH phase and $\eta$~Car became an LBV with a cooler
atmosphere/wind. The dusty Homunculus generated by the Great Eruption
in the 1840s absorbs the UV luminosity of both this massive star and
its companion, lowering the Lyman continuum luminosity by as much as
20\% since the 1840s.

Photoablation-generated momentum injection into the Carina Nebula is
determined by the incident flux of ionizing EUV ($\lambda < $ 912 \AA)
radiation.  Dense spherical clouds lose mass to EUV photo-abaltion at
a rate

\begin{eqnarray}
\dot{M}_p &\approx &\zeta \pi r_p^2 \ \mu m_H n_e \ V_{II} \nonumber \\
  &= &\zeta \mu m_H \ (3 \pi Q / 4 \alpha_B)^{\frac{1}{2}} \ V_{II} \ r_p^{\frac{3}{2}} \ D^{-1},
\end{eqnarray}

\noindent where $\zeta$ is a factor of order unity that accounts for
the illumination geometry ($\zeta \approx $ 1 for a sphere illuminated
from only 1 direction; $\zeta$=4 for isotropic illumination from all
directions), $\mu \approx$ 1.35 is the mean molecular weight, $m_H$ is
the mass of hydrogen, $\alpha _B \approx 2.6 \times 10^{-13}$
cm$^3$~s$^{-1}$ is the case-B recombination coeffficient for hydrogen
at $10^4$~K, $D$ is the distance from the ionizing source with Lyman
continuum luminosity $Q$, and $V_{II} \approx $ 10 to 20 km~s$^{-1}$
is the velocity of the photoablation flow.  The electron density in
the limit of a high-density spherical neutral condensation of radius
$r_p$ is given by 

\begin{equation}
  n_e = (3 Q / 4 \pi \alpha _B)^{\frac{1}{2}} r_p^{-\frac{1}{2}}  D^{-1}.  
\end{equation}

\noindent Assuming that there are about 100 dense globules and pillars
photoablating in the nebular interior, that their typical radii are
about $r_p$=0.2~pc, and that they are located at an average distance
of $D$=3~pc from the nearest ionizing sources, dense gas injects
plasma at a rate of order $\dot{M}$ = 0.02 $M_{\odot}$~yr$^{-1}$.
Assuming that these flows have a velocity given by the sound speed in
the plasma implies a momentum injection rate $\dot{P} \approx$ 0.25
M$_{\odot}$~km~s$^{-2}$, two orders of magnitude larger than the
injection rate by protostellar outflows.

Considering the growth of the $M_{II} \approx 10^6$ $M_{\odot}$ H~{\sc
  ii} region over the course of $\tau \simeq$ 3~Myr by ionization of
the surrounding molecular cloud complex implies $\dot{M}_{II} \approx
M_{II} / \tau \approx$ 0.3 $M_{\odot}$~yr$^{-1}$ and $\dot{P} \approx$
1.2 $M_{\odot}$~km~s$^{-2}$.  This is at least three orders of
magnitude larger than the injection rate by protostellar outflows.
The force does not scale with $\dot{M}$ due to the different geometry.

Finally, using a total stellar-wind mass injection rate of
$\dot{M}_{SW} = 1.4 \times 10^{-4}$ $M_{\odot}$~yr$^{-1}$ (Smith \&
Brooks 2007; Smith 2006a) at a typical velocity $V_{SW} $ = $10^3 $
km~s$^{-1}$ for an O-type star, implies a momentum injection rate
$\dot{P}_{SW} \, \approx \, \dot{M}_{SW} V_{SW} \, \approx \, 0.14 \
M_{\odot}$~km~s$^{-2}$, comparable to the injection rate by
photoablating clumps.  The observed protostellar outflow sources
inject momentum at a rate less than $10^{-4}$ times the rate of
combined injection rate from photoablation and massive star winds.
They are therefore globally unimportant, even if our survey has left a
large number of smaller jets undetected.  Numerical studies suggest
that HH jets can be significant in the global turbulence of a cloud
(e.g., Wang et al.\ 2009), but our study indicates that this can only
apply up to the time when massive stars are born, at which point HH
jets become globally negligible.  They may still be important locally,
within clouds in outer parts of the nebula that are shielded behind
ionization fronts.

While HH jets do not contribute significantly to the energy and
momentum budget of the region, they do have a role to play.  The
presence of outflows in Carina provides important corroborating
evidence for ongoing star formation in the dense clouds in and
surrounding the nebula.  They provide important diagnostics of
physical conditions and nebular flows.  They are the ``wind socks''
that serve to trace the motions of plasma, and the forces that act on
various plasma components.  The direction and amount of bending probes
the forces acting on the jet beams and outflow lobes.  Furthermore,
the structure and properties of the jets and their source globules may
provide diagnostics of the recent radiation history of the Carina
Nebula.  This topic will be addressed in a future paper.

\section{Conclusions}

Based on a narrow-band H$\alpha$ imaging survey of the Carina Nebula
with {\it HST}/ACS, we have reported the discovery of 38 new HH jets
and candidate jets.  Including the one jet that was previously
discovered in ground-based data, HH~666 (Smith et al.\ 2004c), there
are 22 HH jets and 17 candidate HH objects.  We described each jet and
its respective environment in detail.  With assumed outflow speeds, we
use the imaging data to derive the distribution of mass-loss rates for
this sample if jets.  A comparison with a similar distribution in
Orion, where narrow shock filaments can be resolved more easily, shows
that our survey is insensitive to jets with mass-loss rates below
10$^{-8}$ $M_{\odot}$ yr$^{-1}$.  We must have missed several other
jets, since this was a targeted study that did not cover the nebula
uniformly.  Conservatively, there may be 150--200 active protostellar
jets in Carina.

This large number of jets signifies active outflow activity in Carina,
and the short lifetime of these outflows suggests that there is a
large population of accreting young protostars and YSOs in the region.
We expect future IR studies to detect the driving sources of these HH
jets, plus numerous other embedded YSOs from which HH jets are not
seen.  YSOs that are embedded within small dark clouds in the South
Pillars offer a likely analog to the cradle of the Solar System, based
on their proximity to a large number of very massive stars that will
explode in the near future, polluting these clouds with radioactive
nuclides like the $^{60}$Fe seen in Solar System chondrites.  While we
detected a few objects that are true proplyds like the ones seen in
Orion, especially in the core of Tr~14, there are a large number of
small dark clouds that probably contain young stars.  We also
discussed the bending (or not) of HH jets exposed to feedback in giant
H~{\sc ii} regions and how the shaping of the jets depends on their
location relative to bulk flows of plasma throughout the region.  In
particular, we find that photoevaporative flows from the surfaces of
molecular clouds and dust pillars can significantly affect a jet's
morphology for sources near the periphery of a giant H~{\sc ii}
region.  Lastly, we evaluated the contribution of mass, energy, and
turbulence injection by HH jets in Carina, finding that jets are
negligible compared to winds and radiation from massive stars.  The
list of objects discovered provides the largest sample of irradiated
outflows in a giant H~{\sc ii} region, and followup studies with IR
imaging and optical spectroscopy will characterize the physics of
outflows exposed to feedback in a region like that where the Sun
formed, and where most stars are born.

\smallskip\smallskip\smallskip\smallskip
\noindent {\bf ACKNOWLEDGMENTS}
\smallskip
\scriptsize

HH catalog numbers are assigned by B.\ Reipurth in order to correspond
with the list of Herbig-Haro objects maintained at {\tt
  http://ifa.hawaii.edu/reipurth/}, and we appreciate his patience in
trying to assign consecutive HH numbers to our non-consecutive
requests. We acknowledge input during early phases of this project
from J.A.\ Morse, and we appreciate the help of the Hubble Heritage
team at STScI in generating the color images in Figures 2 and 20.
Support was provided by NASA through grants GO-10241 and GO-10475 from
the Space Telescope Science Institute, which is operated by the
Association of Universities for Research in Astronomy, Inc., under
NASA contract NAS5-26555.


\end{document}